\documentclass[a4paper,11pt]{article}
\usepackage{jcappub} % for details on the use of the package, please
\pdfoutput=1
\usepackage[T1]{fontenc} % if needed
\usepackage[utf8]{inputenc}
\usepackage{graphicx}
\usepackage{booktabs}
\usepackage{amsmath,amsbsy}
\usepackage{amssymb}
\usepackage{amsthm}
\usepackage{subfig}
\usepackage{url}
\usepackage{listings}
\usepackage[normalem]{ulem}
\usepackage[]{natbib}
\usepackage{xspace}
\usepackage[usenames,dvipsnames,svgnames]{xcolor}
\graphicspath{{./plots/}}

\newcommand{\lenstwo}{\lstinline!lenS!$^2$\lstinline!HAT!\xspace}

\newcommand{\lenspix}{\lstinline!LensPix!\xspace}
\newcommand{\healpix}{\lstinline!HEALPix!\xspace}

\newcommand{\nside}{\lstinline!NSIDE!\xspace}

\renewcommand{\vec}[1]{\mathbf{#1}}
\newcommand{\be}{\begin{equation}}
\newcommand{\ee}{\end{equation}}
\newcommand{\bea}{\begin{eqnarray}}
\newcommand{\eea}{\end{eqnarray}}

\renewcommand{\(}{\left(}
\renewcommand{\)}{\right)}
\newcommand{\la}{\langle}
\newcommand{\ra}{\rangle}

\newcommand{\vtheta}{\boldsymbol\theta}
\newcommand{\alt}{\lesssim}
\newcommand{\agt}{\gtrsim}
\newcommand{\vc}{\vec{c}}
\newcommand{\response}{\mathcal{R}}
\DeclareMathAlphabet{\pazocal}{OMS}{zplm}{m}{n}
\newcommand{\bi}{B}
\newcommand{\hard}{{\rm hardened}}
\newcommand{\lmax}{\ell_{\rm max}}

\newcommand{\vx}{{\mathbf{x}}}
\newcommand{\vl}{\vec{l}}
\newcommand{\VL}{\vec{L}}
\newcommand{\vL}{\VL}
\newcommand{\vk}{{\mathbf{k}}}

\renewcommand{\d}{\mathrm{d}}
\newcommand{\e}[1]{\mathrm e^{#1}}
\newcommand{\vnhat}{{\hat{\mathbf{n}}}}
\newcommand{\valpha}{\ensuremath{\boldsymbol\alpha}}
\newcommand\half{{\frac{1}{2}}}
\newcommand{\twopi}[1]{(2 \pi)^{ #1}} %2 pi exponents

\newcommand{\nth}{$N^{(3/2)}$\xspace}
\newcommand{\nlth}{$N_L^{(3/2)}$\xspace}

\newcommand{\kappacmb}{\kappa_{\textrm{CMB}}}

\newcommand{\Mkappa}{\pazocal{M}}
\newcommand{\Cgrads}{\tilde C^{T\nabla T}}
\newcommand{\Cgg}{\tilde C^{\nabla T\nabla T,0}}
\newcommand{\Cggtwo}{\tilde C^{\nabla T\nabla T,2}}
\newcommand{\Cexpt}{\tilde C^\expt}
\newcommand{\Aia}{A_{\rm IA}\xspace}
\newcommand{\ext}{\mathrm{ext}}

\newcommand{\Cgltwo}{C_{\text{gl},2}}

\newcommand\clo{{\cal O}}
\newcommand{\expt}{\mathrm{expt}}
\newcommand{\XY}{\mathit{XY}}
\newcommand{\ud}{{\rm d}}

\title{CMB lensing reconstruction biases in cross-correlation with large-scale structure probes}

\author[a,b]{Giulio Fabbian,}
\author[a]{Antony Lewis,}
\author[c]{Dominic Beck}
\affiliation[a]{Department of Physics \& Astronomy, University of Sussex, Brighton BN1 9QH, UK}
\affiliation[b]{Institut d'Astrophysique Spatiale, CNRS (UMR 8617), Univ. Paris-Sud, Universit\'{e} Paris-Saclay, b\^{a}t. 121, 91405 Orsay, France}
\affiliation[c]{AstroParticule et Cosmologie, Universit\'{e} Paris Diderot, CNRS/IN2P3,CEA/Irfu, Obs de Paris, Sorbonne Paris Cit\'{e}, France}

\emailAdd{G.Fabbian@sussex.ac.uk}
\emailAdd{antony@cosmologist.info}
\emailAdd{dbeck@apc.in2p3.fr}

\abstract{
The cross-correlation between cosmic microwave background (CMB) gravitational lensing and  large-scale structure tracers will be an important cosmological probe in the coming years. Quadratic estimators provide a simple and powerful (if suboptimal) way to reconstruct the CMB lensing potential and are widely used. For Gaussian fields, the cross-correlation of a quadratic-estimator CMB lensing reconstruction with a tracer is exactly unbiased if the power spectra are known and consistent analytic lensing mode response functions are used.
However, the bispectrum induced by non-linear large-scale structure growth and post-Born lensing can introduce an additional bias term (\nlth) in the cross-correlation spectrum, similar to the \nlth bias in the auto-spectrum demonstrated in recent works. We give analytic flat-sky results for the cross-correlation bias using approximate models for the post-Born and large-scale structure cross-bispectra, and compare with N-body simulation results using ray-tracing techniques. We show that the bias can be at the 5-15\% level in all large-scale structure cross-correlations using small-scale CMB temperature lensing reconstruction, but is substantially reduced using polarization-based lensing estimators or simple foreground-projected temperature estimators. The relative magnitude of these effects is almost three times higher than in the CMB lensing auto-correlation, but is small enough that it can be modelled to sufficient precision using simple analytic models. We show that \nlth effects in cross-correlation will be detected with high significance when using data of future surveys and could affect systematic effects marginalization in cosmic shear measurements mimicking galaxy intrinsic alignment.}

\begin{document}
\maketitle
\flushbottom

\section{Introduction}
Weak gravitational lensing of cosmic microwave background (CMB) anisotropies in temperature and polarization is one of the key signals exploited by current CMB experiments to obtain cosmological constraints \cite{lewis2006, planck-lensing2018, sptsz2018}. CMB lensing is sourced by gradients in the gravitational potentials between $z=0$ and the last-scattering surface ($z\approx 1100$), and is a probe of large-scale structure (LSS) at higher redshifts than those probed by galaxy surveys. It can be used to constrain the matter fluctuation amplitude, sum of neutrino masses ($M_\nu$) and properties of the dark energy and other parameters affecting the late-time geometry and growth of structure. It can also help to break degeneracies of cosmological parameters in the CMB angular power spectra ~\cite{stompor1999}, and in lensing-only parameter constraints obtained with galaxy lensing at lower redshift.\\*

Gravitational lensing induces a non-Gaussian signal in the observed CMB anisotropies. This can be used to reconstruct the distribution of the line-of-sight integrated gravitational potential that lensed the CMB, i.e. the so-called lensing potential.
The most powerful measurement of CMB lensing potential has been reported by the Planck Collaboration, who measured this signal with a significance higher than $40 \sigma$ on nearly the full sky \cite{planck-lensing2018}. The reconstruction can be further improved using high-sensitivity measurements of the CMB polarization $B$-mode on sub-degree scales: in the absence of large primordial tensor modes, the observed $B$-modes are mainly sourced by the lensing of the primordial $E$-mode polarization and hence a clean probe of the effect. The current generation of ground-based experiments (POLARBEAR, ACTpol, SPTpol) produced the first measurements of B-modes \cite{pb-bb2014, pb-bb2017, keisler2015} and managed to isolate CMB polarization lensing with good statistical significance \cite{pb-lens, story2015, sherwin2017, pb-cib,  hanson2013}. The latest SPTpol measurements \cite{wu2019} for the first time produced CMB lensing maps from CMB polarization data with less noise than those using temperature data. Next generation experiments such as CMB-S4 and Simons Observatory \cite{cmbs4,so} will improve the quality of CMB lensing maps dramatically, from both higher-resolution temperature data and higher-sensitivity polarization.\\

The matter distribution can also be probed by observations of galaxy counts, and through measurements of their shape distortion induced by gravitational lensing. Galaxy lensing measurements in particular have recently begun to provide competitive constraints on cosmological parameters on their own \cite{desy1, desy1-2, chft, kids, kids-des, hsc}.  Both galaxy counts and galaxy lensing are correlated with CMB lensing, since CMB photons pass through some of the same intervening gravitational potentials. 
The noise and systematics of CMB lensing are quite different, so the cross-correlation between the galaxy density, galaxy lensing, and CMB lensing fields, as well as their full joint analysis, make appealing cosmological probes.
A joint analysis helps to improve cosmological constraints by breaking degeneracies, for example involving galaxy bias, and by constraining nuisance parameters associated with sources of systematic error in galaxy measurements \cite{vallinotto2012, liu2016, schaan2017, cawthon2018, singh2018} (e.g. lensing multiplicative bias, photometric redshift errors). Measurements of the cross-correlation between CMB lensing and LSS tracers were initially used to obtain evidence of CMB lensing \cite{smith2007, hirata2008}. However, the first analyses that achieved a high-significance measurement of the cross-correlation between galaxy density and galaxy lensing with CMB lensing were reported by Refs.~\cite{bleem2012} and~\cite{hand2015}, respectively. Despite their limited sky coverage, current CMB polarization experiments such as POLARBEAR recently demonstrated the feasibility of carrying out these measurements using CMB polarization data alone thanks to the low noise of their CMB lensing reconstruction maps~\cite{pbXhsc, pbXherschel}. The first applications of  CMB lensing cross-correlation in cosmological analysis have been reported in Refs.~\cite{giannantonio2016, baxter2016, pullen2016} and, more recently, in Refs.~\cite{doux2018, peacock2018, bianchini2018}, while the most advanced and comprehensive studies in the field have been recently published by the DES and SPT collaborations~\cite{desXC, desXg}; for future galaxy surveys, such as Euclid and LSST, this approach is destined to become the standard baseline analysis to obtain cosmological constraints \cite{pearson2014, Merkel:2017amt, so}.\\

Since the CMB lensing signal has substantial contributions from large-scale structure at $z\gtrsim 1.5$, it is possible to make high-sensitivity measurements of cross-correlation also with density tracers at high redshift.
For example, galaxy surveys outside the optical band, such as sub-mm wavelengths, are efficient in mapping LSS in this regime. Refs.~\cite{bianchini2015, bianchini2016} were the first to perform cross-correlation analysis between CMB lensing and resolved sub-mm galaxies using data from the Planck satellite and \emph{Herschel} H-ATLAS survey. The diffuse emission at these wavelengths, the Cosmic Infrared Background (CIB), is generated by cumulative unresolved dusty star-forming galaxies. In addition to being a sensitive probe of star formation history, this emission has a remarkably high correlation ($\approx 80\%$) with CMB lensing \cite{planckCIBxlensing} and is thus an ideal proxy for CMB lensing itself. CIB data can be combined with CMB reconstruction data to improve the sensitivity of CMB lensing measurements at small scales \cite{planck-lensing2018, Yu:2017djs}, and to improve subtraction of the lensing-generated CMB polarization B-modes on degree scale (delensing). For low noise levels, delensing allows better constraints on the primordial B-mode signal
generated by the stochastic background of primordial gravitational waves produced during inflation \cite{sz1997, kamionkowski2016}. The first demonstrations of CIB delensing have been recently reported by Refs.~\cite{Larsen:2016wpa, Manzotti:2017net}. Although delensing with external galaxy surveys is also an appealing option \cite{Smith:2010gu, manzotti2018}, CIB will be the most faithful large-scale structure lensing tracer at the scales relevant for B-mode delensing for the next few years, and is potentially able to achieve 40\% delensing on 60\% of the sky. Internal delensing (using CMB data alone) becomes most efficient for very low noise levels, but in the intermediate regime a joint analysis with CIB can significantly improve the delensing efficiency.\\

In this context of increasing experimental sensitivity and extended scientific applications, theoretical modelling of CMB lensing and its cross-correlation with LSS probes has to keep up. In particular, any biases in the cross-correlation estimators could lead to spurious correlations or miscalibration of the tracer maps used for delensing.
On small scales lensing becomes a large effect, dominating the unlensed signal which is washed out by Silk damping, so  corrections to low-order perturbative calculations can become substantial.\\*
The power spectrum estimators typically used to measure CMB lensing potential spectra also assume Gaussianity of the CMB lensing field.
This approximation has been shown to be valid to high accuracy for modelling the power spectra of both CMB lensing potential and the lensed CMB power spectra \cite{merkel2011, namikawa2016, pratten2016, Lewis:2016tuj, fabbian2018, marozzi2016, marozzi2016pol}. However, the non-linear LSS evolution and beyond-Born effects generate significant non-Gaussianity in the CMB lensing field that will be observable by future surveys \cite{pratten2016}. Recent work showed that the bispectrum of  the CMB lensing field can lead to a biased estimate of the CMB lensing power spectrum if standard quadratic estimators are used without correction~\cite{bohm2016, bohm2018, beck2018}. Such bias, known as \nlth,  could in turn affect cosmological parameters estimation if left unmodelled~\cite{beck2018}. \\

In this paper we evaluate how \nlth propagates to the cross-correlation of CMB lensing with other density tracers such as galaxy lensing and galaxy counts, and assess the relevance for future CMB experiments and galaxy surveys. Since the non-Gaussianity of the matter distribution increases to lower redshift, the bias in correlation with low-redshift tracers could become important. Furthermore, for the CMB auto-spectrum there is a near cancellation between the biases from the bispectrum from large-scale structure and post-Born lensing. Low redshift tracers have a smaller post-Born signal due to the reduced path length, so there is much less cancellation, potentially making the \nlth bias relatively much more important in cross-correlation~\cite{bohm2018}.\\

We show that the \nlth cross-correlation bias can be modelled quite well theoretically, and hence should be relatively straightforward to include in future analyses. However detailed numerical results are somewhat sensitive to the details of the fully-nonlinear LSS bispectrum shape, which can only be modelled rather approximately analytically, and residual accuracy is hard to assess. We therefore also use numerical simulations for comparison, where the non-linear effects can be measured from the simulation non-perturbatively.\\

In Sec.~\ref{sec:theory} we review the relevant theoretical background. In Sec.~\ref{sec:sims} we review the details of the modelling implemented in the simulations and the assessment of their level of realism in both the CMB lensing potential and LSS tracers. In Sec.~\ref{sec:results} we show the results of our numerical experiments and their impact for future surveys, including the estimation of cosmological and systematics parameters. Conclusions are drawn in Sec.~\ref{sec:conclusions}.

\section{Theory}\label{sec:theory}
\subsection{Gravitational lensing}
In the weak lensing the effect of deflections of light rays coming from a source plane is described by the lens equation. This maps the source position $\boldsymbol\beta$ of a ray originating at comoving radial distance $\chi$ to the observed angular position $\boldsymbol\theta$. Assuming General Relativity and using natural units
\be
\beta_i (\boldsymbol\theta, \chi) = \theta_i -
2\int_0^{\chi}\frac{D_A(\chi -
  \chi')}{D_A(\chi)D_A(\chi')}
\Psi,_{\beta_i}\left(\boldsymbol\beta(\boldsymbol\theta,
\chi'), \chi' \right) {\rm d}\chi',
\label{eq:beta}
\ee
where $\Psi(\boldsymbol\beta, \chi)$ is the Weyl gravitational potential located on the photon path, $\Psi,_{\beta_i}(\boldsymbol\beta, \chi),$ its angular derivatives\footnote{The derivatives in the small angle limit should be computed using a coordinate
system orthogonal to the current light ray's direction of travel. Numerical tests have shown that using angular derivatives causes a negligible
error (see e.g. \cite{becker2013} and references therein).}, and $D_A(\chi)$ is the comoving angular diameter distance.\\*
In the Born approximation, the photon path is approximated by the unperturbed photon geodesic $\mathbf{x}(\chi) \approx \boldsymbol\theta \chi$, such that the line integral of the Newtonian potential $\Psi$ simplifies and the geodesic equation becomes

\be
\beta(\boldsymbol\theta,\chi^*) = \boldsymbol\theta + \nabla\phi(\boldsymbol\theta), \label{eq:remapping}
\ee
where $\nabla\phi(\boldsymbol\theta)$ is the deflection field and $\phi$ the lensing potential
\be
\phi(\boldsymbol\theta) =
-2\int_0^{\chi_s}\frac{D_A(\chi_s -
  \chi')}{D_A(\chi_s)D_A(\chi')}
\Psi\left(\boldsymbol\theta, \chi' \right) {\rm d}\chi'.
\label{eq:lenseq}
\ee
Here, $\chi_s$ is the distance to the source plane, i.e. the comoving distance to the last scattering surface for the CMB, or the distance to source galaxies in the case of galaxy lensing.  If we want to evaluate the lens equation at higher order, i.e. beyond the Born approximation (post-Born), we have to account for the fact that photons do not travel along the unperturbed background geodesics. Higher-order corrections are typically introduced perturbatively in Eq.~\eqref{eq:beta} by Taylor expanding the potential $\Psi$ around the unperturbed geodesic position \cite{cooray2002, pratten2016, krause2010, marozzi2016}. The post-Born corrections only affect the lensing potential power spectrum at sub-percent level on angular scales $L\lesssim 10000$ for CMB lensing and galaxy lensing alike.
However, as discussed in Ref.~\cite{pratten2016, fabbian2018}, the lensing potential bispectrum and higher-order correlation functions are more affected as discussed further below. An additional characteristic signature of post-Born corrections is the appearance of curl-like deflections such that
\be
\beta(\boldsymbol\theta) = \boldsymbol\theta + \nabla\phi(\boldsymbol\theta)  + \nabla\times\Omega(\boldsymbol\theta).\label{eq:full-remapping}
\ee
Here we define $(\nabla\times\Omega)_i \equiv \epsilon_{ij}\partial_j\Omega$, where $\epsilon_{ij}$ is the  Levi-Civita symbol in two dimensions and $\Omega$ is a pseudo-scalar field \cite{hirata2003}. Curl modes are correlated at higher-order with gradient-like deflections so that the $\phi\phi\Omega$ or $\phi\Omega\Omega$ bispectra are non-null\footnote{Following the consistency relations of Eq.~\eqref{eq:conskappa}, \eqref{eq:rotationdisplacement}, we will refer to these higher order correlations as $\kappa\kappa\omega$ or $\kappa\omega\omega$ to conform to the most common convention adopted in the literature.}. However, the effects of this higher-order correlation is negligible for the analysis presented in the following. We finally note that post-Born corrections depend strongly on the length of the photon path and are thus progressively less important for source planes located at lower redshift.

\subsection{CMB lensing cross-correlation}
The lensing potential can be related to the lensing convergence $\kappa$ in the weak lensing regime through the Poisson equation $\kappa = -\nabla^2\phi/2$, so that in the harmonic domain%\footnote{Despite being derived in the Born approximation, these relations hold in the post-Born regime at sub-percent accuracy \cite{fabbian2018}.}
\bea
\kappa_{LM} &=& \frac{L(L+1)}{2}\phi_{LM}.\label{eq:conskappa}
\eea
Gravitational lensing directly probes the Weyl gravitational potential, but in General Relativity (and after matter domination) the potential can be related directly to the comoving density perturbation $\delta$ via the Poisson equation. Observed angular galaxy densities as a function of redshift depend on a variety of effects (including redshift distortions, magnification bias, velocity and potential effects), but can also be approximated at some level as biased tracers of the comoving density perturbation.
It is therefore convenient to rewrite the lensing observables in terms of convergence field so that the cross-correlation between CMB lensing and LSS tracers in a given redshift bin can be written in the Limber approximation as
\be
C_{L}^{AB} \approx \int\frac{d\chi}{\chi^2}W_A(\chi)W_B(\chi)P_\delta\left(k=\frac{L+1/2}{\chi},z(\chi)\right), \qquad A,B \in\{g,\kappa_{\textrm{CMB}},\kappa_{z}\}
\label{eq:limbercls}
\ee
where $P_\delta$ is the comoving density matter power spectrum, $g$ is the galaxy density, $\kappa_{\rm CMB}$ the CMB lensing convergence and $\kappa_z$ the lensing convergence of galaxies located at redshift $z$. The window function $W$ determines the redshift distribution so that
for galaxy density
\be
W_g(\chi) = b_g(z)\frac{1}{n}\frac{dn}{dz}\frac{dz}{d\chi} \qquad n = \int dz \frac{dn}{dz}.
\label{eq:w_g}
\ee
where $b_g$ is the galaxy bias and $dn/dz$ the redshift distribution of the observed galaxies. For lensing of a source at a comoving distance $\chi_s$
\be
W_\kappa(\chi) \equiv W_\kappa(\chi,\chi_s) = \gamma(\chi)  \chi^2\left(\frac{1}{\chi}-\frac{1}{\chi_s}\right) \Theta(\chi_s-\chi),
\ee
where $\Theta$ is the step function and the potential $\Psi$ and density are related by $k^2\Psi \approx -\gamma(z)\delta$, where $\gamma$ is approximately independent of $k$.
In the case of the CMB, $\chi_s$ is the comoving distance to the last scattering surface, which can be well approximated as a single source plane. In the case of the lensing convergence of galaxies located at $\chi_z$, the lensing efficiency has to be integrated over the source distribution used to estimate the convergence field
\be
W_{\kappa_z}(\chi) = \frac{1}{n}\int d\chi \frac{dn}{dz}\frac{dz}{d\chi}W_\kappa(\chi,\chi_z).
\ee
\subsection{CMB lensing quadratic estimators and cross-correlation}\label{sec:quest}
\label{subsec:estimators}
The CMB lensing potential can be extracted statistically from noisy observations of the CMB using lensing reconstruction. The simplest estimators are quadratic in the lensed CMB fields and fast to evaluate~\cite{Hu:2001kj,okamoto-hu}, though more optimal (and costly) alternatives have been proposed~\cite{hirata2003, carron2017, millea2017}. On the full sky, the quadratic estimator takes the form
\be
\hat{\phi}_{LM}^{XY}=\sum_{\ell_1 m_1 \ell_2 m_2} \left[\frac{A^{XY}_L}{L(L+1)} (-1)^M
\begin{pmatrix}
L & \ell_1 & \ell_2 \\
-M & m_1 & m_2
\end{pmatrix}
g^{XY}_{L\ell_1\ell_2}\right]\tilde X^{\expt}_{\ell_1 m_1}\tilde Y^{\expt}_{\ell_2 m_2}, \label{eq:estimator}
\ee
where X, Y represent temperature (T) or polarization (E or B-modes) fields and $\tilde X^\expt_{\ell m}$ are the harmonic coefficients of the noisy, beam-deconvolved data.
The weight functions $g^{XY}_{L\ell_1 \ell_2}$ can be freely chosen, for example to approximately minimize the variance, and the $A_L^{XY}$ normalization is defined so that the estimator is unbiased. All the different reconstruction channels $\hat\phi_{LM}^{XY}$ involving temperature and polarization can then be combined using an inverse variance weighting to obtain the so-called minimum variance (MV) reconstruction $\hat\phi^{\rm MV}_{LM}$.
For theoretical calculations we use the flat-sky approximation for convenience,
following the notation of~\cite{bohm2016} with Fourier conventions
\begin{equation}
	\label{FourierConvention}
		f(\vl)=\int \d^\mathrm{2} \vx\, f(\vx) \e{-i\vl\vx},
		\hspace{1 cm}
		f(\vx)=\int \frac{\d^\mathrm{2} \vl}{\twopi{\mathrm{2}}}\, f(\vl) \e{i \vl\vx}.
\end{equation}
For statistically isotropic fields the angular power spectrum $C$ and bispectrum $B$ are then defined by
\begin{align}
	\label{Spectra}
		\langle X(\vl)Y(\vl')\rangle&= (2\pi)^\mathrm{2}\delta(\vl+\vl')C_{l}^{XY}\\
		\langle X(\vl)Y(\vl')Z(\vl'')\rangle&= (2\pi)^\mathrm{2} \delta(\vl+\vl'+\vl'')B^{XYZ}(l,l',l''),
\end{align}
where $l\equiv |\vl|$ and we will use tildes to denote lensed quantities.
For brevity we also write
\begin{equation}
  \label{eq:5}
  \int_{\vl} \equiv \int\frac{\d^2 \vl}{(2\pi)^2}.
\end{equation}
The lensing potential quadratic estimator can then be written as~\cite{Hu:2001kj}
  \begin{equation}
    \label{eq:quadest}
    \hat\phi^\XY(\VL) = A_L^{\XY} \int_\vl g_\XY(\vl,\VL)\tilde X_\expt(\vl)\tilde{Y}^*_\expt(\vl-\VL),
  \end{equation}
where $A_L^{XY}$ is the normalization.\\*
In this paper we consider the cross-correlation of the lensing reconstruction $\hat\phi^\XY(\VL)$ with an external tracer $\phi_\ext(\vL)$ (e.g. galaxy lensing or tracer of the galaxy density). Assuming Gaussian CMB fields and that the lensing potential and $\phi_\ext$ are only weakly non-Gaussian, for uncorrelated CMB and lensing fields we can write the cross-correlation to leading order in the non-Gaussianity as
 \begin{multline}
  \la \phi_\ext(\vL') \hat\phi^\XY(\VL)\ra =
  A_L^{\XY} \int_\vl g_\XY(\vl,\VL) \la \phi_\ext(\vL')\tilde X_\expt(\vl)\tilde{Y}^*_\expt(\vl-\VL)\ra
  \\
  \quad = A_L^{\XY} \int_\vl g_\XY(\vl,\VL) \la \phi_\ext(\vL')\tilde X_\expt(\vl)\tilde{Y}^*_\expt(\vl-\VL)\ra_G \hfill
 \\
+\frac{A_L^{\XY}}{2} \int_\vl g_\XY(\vl,\VL) \int \ud^2\vL'' \ud^2\vl_1 \ud^2\vl_3 \left\la \frac{\delta^3\left(\phi_\ext(\vL')\tilde{X}(\vl)\tilde{Y}(\vL-\vl)\right)}{\delta \phi_\ext(\vL'')\delta\phi(\vl_1)\delta\phi(\vl_3)} \right\ra_G
\la \phi_\ext(\vL'')\phi(\vl_1)\phi(\vl_3)\ra \\
+ \dots,
\label{eq:xcorrphi}
  \end{multline}
where $G$ subscripts denote expectations evaluated for Gaussian lensing potentials and unlensed CMB fields.
This expression can be obtained by evaluating the expectation on the left hand side by using the leading-order multivariate Edgeworth expansion for the distribution of the fields about a Gaussian (for small non-Gaussianity)\footnote{Similar to the result of Ref.~\cite{Lewis:2016tuj} for the impact of lensing bispectrum on the CMB power spectra.}.

\begin{table}
\centering
\resizebox{\textwidth}{!}{
\begin{tabular}{|l|l|}
\hline\rule{0pt}{2.6ex}XY & $f^{XY}(\vl_1,\vl_2)$ \\[2mm]
\hline\rule{0pt}{2.6ex}TT & $\vl_1\cdot \vL \Cgrads_{l_1} + \vl_2\cdot \vL \Cgrads_{l_2}$  \\[2mm]
EE & $\cos(2(\varphi_{l_1}-\varphi_{l_2})) \left[\vl_1\cdot \vL \tilde C^{E\nabla E}_{l_1} + \vl_2\cdot \vL \tilde C^{E\nabla E}_{l_2}\right]
+ \frac{1}{2}\sin(2(\varphi_{l_1}-\varphi_{l_2}))\left[ \vl_1\times \vL \tilde C^{PP_\perp}_{l_1} - \vl_2\times\vL \tilde C_{l_2}^{PP_\perp}\right] $  \\[2mm]
EB & $\sin(2(\varphi_{l_1}-\varphi_{l_2}))\left[ \vl_1\cdot \vL \tilde C^{E\nabla E}_{l_1} + \vl_2\cdot \vL \tilde C_{l_2}^{B\nabla B}\right] - \frac{1}{2}\cos(2(\varphi_{l_1}-\varphi_{l_2}))\left[ \vl_1\times \vL \tilde C_{l_1}^{PP_\perp} + \vl_2\times \vL \tilde C^{PP_\perp}_{l_2}\right]$\\[2mm]
BB & $\cos(2(\varphi_{l_1}-\varphi_{l_2})) \left[ \vl_1\cdot \vL \tilde C_{l_1}^{B\nabla B} + \vl_2\cdot \vL \tilde C_{l_2}^{B\nabla B}\right] -\frac{1}{2}\sin(2(\varphi_{l_1}-\varphi_{l_2}))\left[ \vl_1\times \vL \tilde C_{l_1}^{PP_\perp} - \vl_2\times \vL \tilde C_{l_2}^{PP_\perp}\right]$ \\[2mm]
TE & $\cos(2(\varphi_{l_1}-\varphi_{l_2}))\vl_1\cdot \vL \tilde C_{l_1}^{T\nabla E} + \sin(2(\varphi_{l_1}-\varphi_{l_2}))\vl_1\times \vL \tilde C_{l_1}^{TP_\perp}+ \vl_2\cdot \vL \tilde C_{l_2}^{T\nabla E} $ \\[2mm]
TB & $\sin(2(\varphi_{l_1}-\varphi_{l_2})) \vl_1\cdot \vL \tilde C_{l_1}^{T\nabla E}
-\cos(2(\varphi_{l_1}-\varphi_{l_2}))\vl_1\times \vL \tilde C_{l_1}^{TP_\perp} - \vl_2\times \vL \tilde C_{l_2}^{TP_\perp}$ \\
\hline
\end{tabular}}
\caption{Exact flat-sky non-perturbative mode response functions for the lensing quadratic estimators for Gaussian fields assuming no primordial $B$ modes (as defined in Eq.~\eqref{responsedef}). Here we used $\vec a \times \vec b \equiv \epsilon_{ij}a^i b^j$ and the various lensed gradient spectra defined in the appendix of Ref.~\cite{lewis2011}. To a $\sim 0.5\%$ approximation the $\tilde C_{l}^{TP_\perp}$ and $\tilde C_{l}^{PP_\perp}$ terms can be neglected for all but the BB estimator (which we do not use), since they are much smaller than $\tilde C_l^{E\nabla E}$.
Approximating the gradient spectra with equivalent lensed CMB spectra (and taking $\tilde C_{l}^{TP_\perp}=\tilde C_{l}^{PP_\perp}=0$) is a good approximation for $\lmax \alt 2000$ (Ref.~\cite{Hanson:2010rp}, see Fig.~\ref{fig:Cgrads} for an example), but leads to biased results using smaller scales as shown in Fig.~\ref{fig:response} below.
%Further replacing lensed spectra with unlensed spectra gives the leading-order perturbative result.
}
\label{table:nonpertresponse}
\end{table}
\noindent
The Gaussian expectation in the first term of Eq.~\eqref{eq:xcorrphi} can be evaluated exactly. Assuming no correlation between the unlensed CMB fields and the lensing potential, it becomes (after integration by parts, see e.g. Ref.~\cite{Lewis:2011au}),
  \begin{eqnarray}
  \la \phi_\ext(\vL')\tilde X_\expt(\vl)\tilde{Y}^*_\expt(\vl-\VL)\ra_G  &=&
   (2\pi)^2C^{\phi_\ext\phi}_{L'}\left\la \frac{\delta}{\delta \phi(\vL')^*}\left( X_\expt(\vl)\tilde{Y}^*_\expt(\vl-\VL)\right)\right\ra_G\nonumber\\
   &=& (2\pi)^2\delta(\vL+\vL') C^{\phi_\ext\phi}_{L} f^{XY}(\vl, \vL-\vl).
  \end{eqnarray}
 Here we defined the non-perturbative mode response functions $f^{XY}$ by
  \be
  \left\la \frac{\delta}{\delta \phi(\vL)}\left( \tilde X(\vl_1)\tilde{Y}(\vl_2)\right)\right\ra_G
  = \delta(\vl_1+\vl_2-\vL) f^{XY}(\vl_1,\vl_2),
  \label{responsedef}
  \ee
  which can be calculated analytically in terms of the lensing potential power spectrum and the unlensed CMB power spectra following Ref.~\cite{lewis2011}. Explicit expressions in terms of the lensed gradient spectra are given in Table~\ref{table:nonpertresponse}, and numerical results for the gradient spectra can be calculated using the public CAMB\footnote{From version 1.0.5, \url{https://camb.info}} code~\cite{Lewis:1999bs}.
   In the simplest case of temperature reconstruction, the exact result is equivalent to replacing the unlensed spectrum with the lensed gradient spectrum in the leading perturbative expressions for $f(\vl_1,\vl_2)$ given by Ref.~\cite{Hu:2001kj}\footnote{The expressions in Table~\ref{table:nonpertresponse} reduce the leading perturbative results of Ref.~\cite{Hu:2001kj} after fixing typos (a factor of 2 in angular argument for $f^{TE}$ and a sign in second term for $f^{EB}$). Note that the gradient spectra like $\tilde C^{T\nabla T}_l$ by definition reduce to the unlensed spectra for no lensing, and $\tilde C^{PP_\perp}_l$ is just notation for the spectrum defined in Ref.~\cite{lewis2011} that reduces to zero for no lensing.}.

For Gaussian fields the quadratic estimator correlation is non-perturbatively unbiased if the normalization is given by
  \be
  [A_L^{XY}]^{-1} = \int_\vl g_{XY}(\vl,\vL)f^{XY}(\vl,\vL-\vl),
  \ee
  since we then have
  \be
    \la \phi_\ext(\vL') \hat\phi^\XY(\VL)\ra_G  = (2\pi)^2\delta(\vL+\vL') C^{\phi_\ext\phi}_{L}.
  \ee
The $g_\XY$ weight functions can be freely chosen, but if they are chosen to minimize the (non-lensing) variance, they can be  related directly to the mode response functions $f^{XY}$~\cite{Hu:2001kj}. For comparison with simulations we use the often-used but slightly-suboptimal weights where the lensed gradient spectra are approximated by the lensed CMB spectra and  $\tilde C_{l}^{TP_\perp}=\tilde C_{l}^{PP_\perp}=0$. However, it remains important to calculate the normalization using the lensed gradient power spectra in the mode response functions $f^\XY$  to obtain results that are unbiased at the percent level for future high-resolution data (as shown in Fig.~\ref{fig:Cgrads} the temperature lensed gradient spectrum is substantially smaller than the lensed spectrum on small scales). We also use the separable (and hence fast but somewhat suboptimal) $g_{TE}$ weights where $\tilde C^{TE}_\ell$ is set to zero.\\*
The non-Gaussian contribution to Eq.~\eqref{eq:xcorrphi} gives a bias term proportional to the mixed tracer-CMB lensing potential bispectrum,
 \begin{multline}
    \la \phi_\ext(\vL') \hat\phi^\XY(\VL)\ra  = (2\pi)^2\delta(\vL+\vL') C^{\phi_\ext\phi}_{L} + \\
  + (2\pi)^4\frac{A_L^{\XY}}{2}
\int_{\vl_1}  B^{\phi_\ext\phi\phi}(L',l_1,|\vL'+\vl_1|)
  \int_\vl g_\XY(\vl,\VL)
  \left\la \frac{\delta^2 \left(\tilde{X}(\vl)\tilde{Y}^*(\vl-\vL)\right)}
{ \delta\phi(\vl_1)\delta\phi^*(\vl_1+\vL')} \right\ra_G + \dots.
\label{eq:bispectrumresponse}
  \end{multline}
By statistical isotropy the remaining expectation must be $\propto \delta(\vL+\vL')$ and we define the bias $N^{(3/2), XY}_L$ proportional to the bispectrum so that~\cite{bohm2016}
\be
    \la \phi_\ext(\vL') \hat\phi^\XY(\VL)\ra  = (2\pi)^2\delta(\vL+\vL')\left(
     C^{\phi_\ext\phi}_{L} + N^{(3/2),XY}_L\right).
\ee
The remaining Gaussian expectation value in Eq.~\eqref{eq:bispectrumresponse} can be calculated analytically in terms of the unlensed and lensing potential fields.
Explicit expressions for the  \nlth biases in terms of the bispectrum are given in Appendix~\ref{sec:Nthreetwo} (to lowest perturbative order), and non-perturbatively in Appendix~\ref{sec:nonpert}. The $N^{(3/2)}$ biases turn out to be relatively small, and hence do not need to be calculated to high precision, and we use the general perturbative expressions but substitute unlensed spectra for lensed gradient spectra to capture the main non-perturbative correction (see Appendix~\ref{sec:nonpert}).
The bispectrum has contributions from non-linear structure growth, and from post-Born lensing. The theoretical calculation of the relevant bispectra is discussed in detail in Appendix~\ref{sec:bispectrum}.\\*
The detailed shape and sign of \nlth depends on the bispectrum, where there can be a partial cancellation between the post-Born and non-linear matter evolution (LSS) terms as explained in Ref.~\cite{pratten2016}.  For the CMB lensing auto-spectrum, the cancellation can be significant, substantially reducing the size of the overall bias~\cite{bohm2018,beck2018}. However, the LSS bispectrum can become larger relative to the cross-correlation power spectrum for lower-redshift tracers, and the post-Born contribution smaller; so there is substantially less cancellation for cross-correlation with low-redshift tracers, giving a larger net bias.

\section{Simulations}\label{sec:sims}
To test the bias in the cross-correlation of LSS tracers with CMB lensing reconstruction we need consistent simulations of non-Gaussian evolution of the matter distribution and its tracers, and post-Born lensing.
 We use $N$-body simulations analysed using a multiple-lens raytracing technique. We focus on galaxy lensing and galaxy density LSS tracers, where for simplicity we assume the latter directly probe the matter distribution (leaving a detailed study of non-linear biasing effects to future work). In this section we describe the details of our procedure and the validation tests performed on the simulated observables. For the power spectrum tests in this section we computed the theoretical predictions in the flat-sky approximation using the first-order Limber approximation with $k\approx \frac{L+1/2}{\chi}$ to achieve a more reliable comparison with full-sky simulation results. In the subsequent sections we use the zeroth-order Limber approximation for the analytic flat-sky fractional \nlth\ bias results.

\subsection{N-body simulations and lensing modelling}\label{sec:nbody}
To model the realistic non-linear LSS matter distribution we used a $\Lambda$CDM simulation from the DEMNUni suite \cite{carbone2016, castorina2015}. This simulation uses a Planck 2013 cosmology
with massless neutrinos and parameters
\begin{equation}
\{ \Omega_{\rm dm},  \Omega_{\rm b}, \Omega_{\Lambda}, n_{\rm s}, \sigma_8, H_0,M_{\nu},\tau
  \} = \{ 0.27, 0.05, 0.68, 0.96, 0.83, 67 \: \rm{Km / s /
    Mpc}, 0, 0.0925 \},
\end{equation}
\noindent
with $2048^3$ dark matter particles in a box size of 2 Gpc/$h$. The simulation stored 62 snapshots of the box at logarithmically-spaced redshifts between  $z=99$ and $z=0$. The mass resolution of the simulation at $z = 0$ is $M_{\rm CDM} = 8.27 \times 10^{10} M_{\odot} / h$ and the gravitational softening length is set to $\epsilon_s = 20$ kpc$/h$, corresponding to 0.04 times the mean linear inter-particle separation.
Using the snapshots at our disposal we reconstructed the full-sky past lightcone of the observer over the whole redshift range covered by the simulation following Refs.~\cite{carbone2008, calabrese2015}. This approach minimizes the replication of structures along the line of sight using randomization techniques. The lightcone was then divided into spherical shells of comoving thickness $\Delta\chi\approx150\ {\rm Mpc}/h$. The particles located inside each of these volumes were projected onto spherical shells of surface mass density $\Sigma^{\theta}$ sampled on an \healpix grid at $\textrm{\nside}=4096$ \cite{fosalba2008} so that each pixel $p$ the surface density is $\Sigma^{\theta}_p = \sum_{0}^{n} m / \Delta\Omega_{\rm pix}$ where $n$ is the number of particle per pixel, $\Delta\Omega_{\rm pix}$ its area in steradians, and $m$ the mass of one dark matter particle.
 The $k$-th surface mass density plane was converted into surface mass overdensity $\Delta^{(k)}_{\Sigma} = \Sigma^{\theta(k)} /\bar{\Sigma}^{\theta(k)} -1$, and then to lensing convergence $\kappa^{(k)}_{\chi_s}$ for a source plane at $\chi_s$, using
\be
\kappa^{(k)}_{\chi_s} = 4 \pi  G\frac{D_A(\chi_s-\chi_k)}{D_A(\chi_s)}\frac{(1+z_k)}{D_A(\chi_k)}\Delta^{(k)}_{\Sigma}.
\label{eq:kappa1}
\ee
The $\Delta^{(k)}_{\Sigma}$ maps will also be used as proxy of the galaxy density distribution as described in Sec. \ref{subsec:galaxy}. The $\kappa^{(k)}_{\chi_s}$ maps are used to model gravitational lensing for the different source planes discussed in Sec. \ref{subsec:kappa} using the multiple-lens-plane raytracing algorithm (ML) \cite{becker2013,hilbert2009} implemented in the \lenstwo code \cite{fabbian2013, fabbian2018}. To do this the $\kappa^{(k)}_{\chi_s}$ maps were first converted to lensing potential maps using Eq. \eqref{eq:conskappa} and then used to propagate the light-ray trajectories between subsequent matter shells, as well as the lensing magnification matrix $A$. The components of the magnification matrix read
\be
A =
\begin{pmatrix}
  1 - \kappa -\gamma_1 & -\gamma_2 + \omega \\
  -\gamma_2 - \omega & 1-\kappa+\gamma_1
\end{pmatrix},
\ee
where $\gamma_1, \gamma_2$ are the two components of the shear and $\omega$ the lensing rotation angle which can be connected to the curl potential $\Omega$ in harmonic space as
\be
\omega_{LM} = \frac{L(L+1)}{2}\Omega_{LM}.
\label{eq:rotationdisplacement}
\ee
Our ML algorithm also propagates and outputs the components of the $A$ matrix in the Born approximation, which are used in the following to isolate the impact of post-Born effects on the cross-correlation signals.
For all the results of this work we performed the raytracing on ECP (equi-cylindrical project) grids \cite{muciaccia1997} with an approximate resolution of 3 arcsec and resampled the final products on \healpix pixelization at 1.7$^\prime$ resolution ($\textrm{\nside}=4096$) through a subsequent direct and  inverse SHT. In this last operation we did not correct for any effect of quadrature and signal smoothing due to the pixelization window as these are negligibly small given the resolution of the internal ECP grid adopted in the raytracing.

\subsection{Galaxy lensing}\label{subsec:kappa}
Future surveys will use galaxy weak lensing as a tomographic probe of the matter distribution, measured using the lensed shapes of background galaxies located mainly at $z\lesssim 2$. For the purpose of this paper we produced 5 simulated maps of convergence $\kappa_{z}$ of sources having a redshift distribution described by a Dirac delta distribution around specific redshifts $z\in\{0.2,0.35,0.6,1,2\}$ and no shot-noise using the pipeline of Sec.~\ref{sec:sims}. This is of course a simplifying assumption that neglects complications connected to conversion between observed shear estimators and convergence maps, as well as all the instrumental and theoretical systematic effects that complicate these kind of measurements (e.g. redshift estimation and distribution of background galaxies, multiplicative biases in the cosmic shear estimates, intrinsic alignments, and the impact of the baryons in the signal modelling). Within our assumptions, we validate the accuracy of the simulated convergence maps compared to theoretical results for the power spectrum  based on different models for the non-linear evolution of the matter. Comparisons of the skewness are given in Appendix~\ref{app:skewness}.\\*
Fig.~\ref{fig:convergence} shows a comparison between the power spectra of the simulated convergence maps $\kappa_{z}$ $C_L^{\kappa\kappa, z}$ in the Born approximation compared to analytic predictions computed using Eq.~\eqref{eq:limbercls} using approximations to model the non-linear matter power spectrum. The theoretical predictions obtained using $P_\delta(k,z)$ of the snapshots of the DEMNUni simulation
agree to sub-percent precision with the results of the simulation for $z\geq 0.6$\footnote{Since we measure $P_\delta$ of the N-body simulation on a grid of binned $k$ values for a finite number of snapshots in redshift, we used a spline interpolation to perform the integration.}. The agreement is degraded to about 3\% for $z\leq 0.6$ due to residual numerical effects from the finite spatial resolution of the raytracing grid. This can be compared to using semi-analytic fitting formula such as the Halofit models of Ref.~\cite{Takahashi:2012em} or \cite{Mead:2015yca}, and adding to these predictions an uncorrelated shot-noise component at each redshift given by $P_\delta^{\rm shot-noise}=1/\bar{n}$, where $\bar{n}(z)$ is the average number of particles in a given surface-mass density plane of the lightcone. The Halofit models agree to better than 7\% at all redshifts (and much better than this at $z>0.2$ and non-extreme multipoles), within the expected intrinsic accuracy of the fitting formulae for the non-linear evolution.

\begin{figure}[t]
\centering
\includegraphics[width=\textwidth]{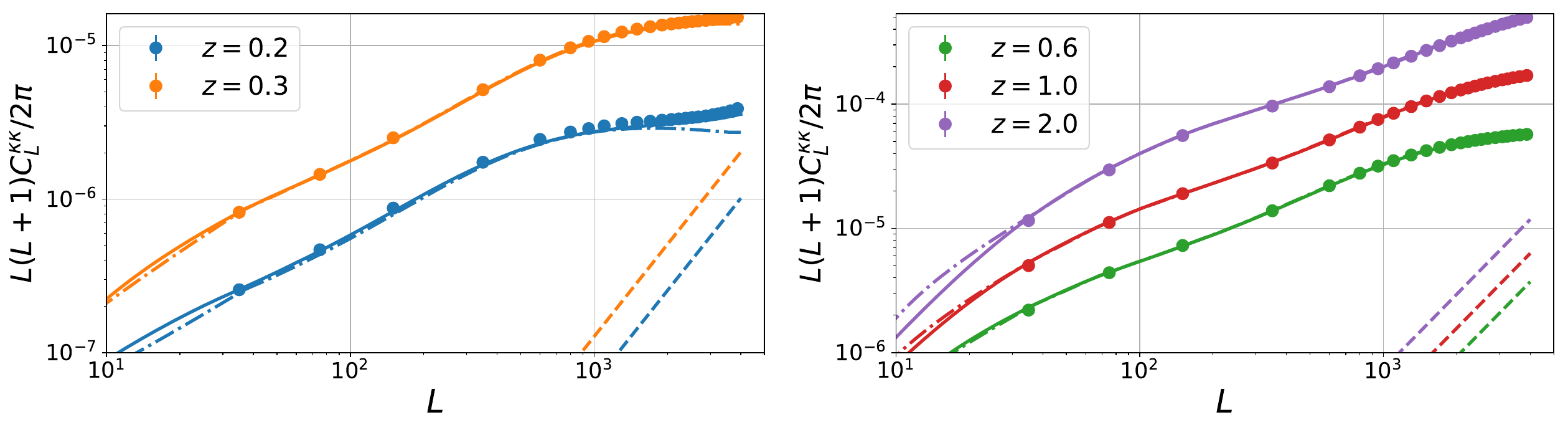}
\caption{Lensing convergence power spectrum in the Born approximation extracted from the \lenstwo simulations (dots) for the redshift bins considered in this work. Predictions for the signal computed in Limber approximation and using the measured matter power spectrum from the DEMNUni simulations are shown as solid line, while those using the Halofit fitting formula of Ref.~\cite{Takahashi:2012em}  are shown as dot-dashed lines. The discrepancy seen at small scales (in particular at $z=0.2$) is due to numerical shot noise in the N-body simulation (dashed line). Results based on post-Born lensing simulations give consistent results.
}
\label{fig:convergence}
\end{figure}

\begin{figure}[]
\centering
\includegraphics[width=\textwidth]{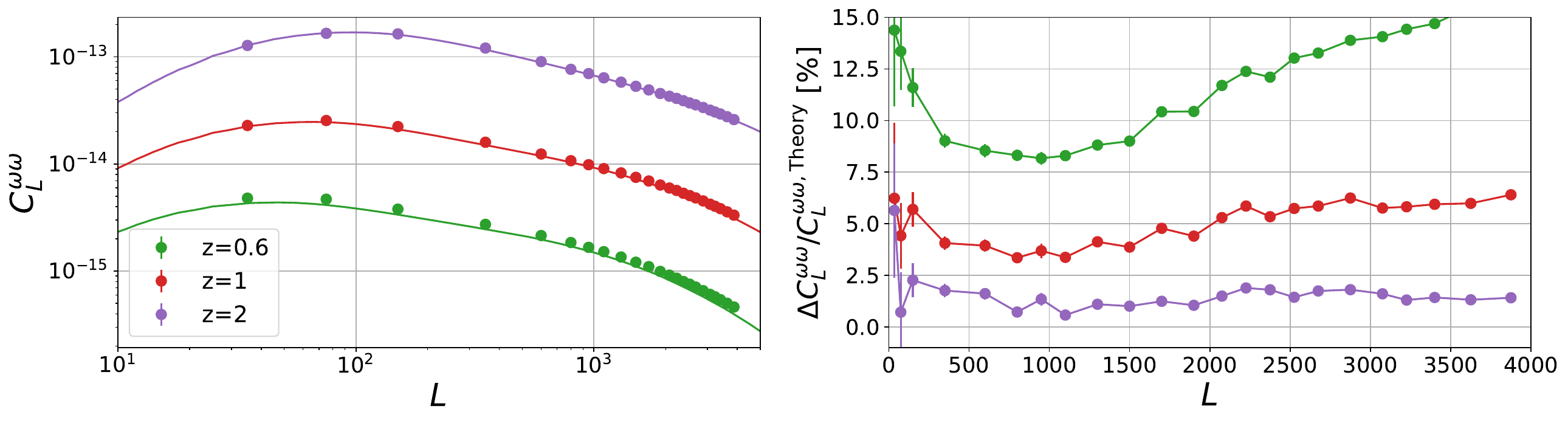}
\caption{Lensing rotation power spectra extracted from the \lenstwo simulations compared with analytic predictions (left). The binned relative difference between simulation and theory results is shown in the right panel. The error bars show the error on the mean within a given multipole bin.
}
\label{fig:rotation}
\end{figure}
\noindent
The power spectrum of the post-Born simulated $\kappa_{z}$ are consistent with the Born results, showing differences to the Born result of about 0.05\% for $z=2$ on all angular scales and consistent with zero to within sample variance. Such differences are negligible for all practical purposes and are in rough agreement with analytic estimates based on the results of Ref.~\cite{krause2010,pratten2016}, which predicts them to be lower than $0.01\%$
We found that the measured post-Born corrections grow slightly with decreasing redshift as a result of numerical artefacts and power aliasing (as an additional shot-noise bias), but stay below 3\% at $z=0.2$, which is the worst observed difference. Post-Born lensing also introduces a non-zero rotation component to the magnification matrix (which is identically zero at first order~\cite{stebbins96}). Its power spectrum comes entirely from second-order lens-lens coupling terms and provides a stringent test of the accuracy of the post-Born modelling in our observables. In Fig.~\ref{fig:rotation} we show the agreement between the power spectrum of the simulated rotation fields and the analytic predictions from Ref.~\cite{pratten2016} using Halofit for the non-linear matter power spectrum. For the bins $z\geq 0.6$, where the signal is stronger and the r.m.s. rotation angle has a magnitude much higher than the raytracing resolution ($\simeq$ 50, 20, 7 arcsec respectively), we obtain agreement at the few-percent level. To our knowledge this represents the most accurate numerical validation of analytic predictions of the rotation power spectrum for weak lensing, improving earlier estimates of Ref.~\cite{hilbert2009} and extending CMB lensing result of Ref.~\cite{fabbian2018} to lower redshift. However, at $z=0.35$ and $z=0.2$ the magnitude of the rotation angle is well below our spatial resolution (2 and 0.5 arcsec respectively) and thus the agreement is degraded to $\simeq 20\%$. As the post-Born effects are very small at these low redshifts, this accuracy level is more than enough for our purposes.

\subsection{Density field} \label{subsec:galaxy}

The second observable we considered for the cross-correlation is the projected density. This is usually reconstructed through the observation of galaxy overdensities $g$ and its use as a cosmological and astrophysical probe depends on our ability to model how galaxies trace the true underlying matter field, i.e. to the knowledge of their bias parameter $b$ as a function of scale and redshift. For the sake of simplicity we ignored these complication here and assumed that the field is perfectly reconstructed. This is equivalent to set $b_g=1$ and constant over redshift in Eq.~\eqref{eq:w_g}. This simplifies the modelling of $g$ in the simulation as we can directly use the $\Delta_{\Sigma}^{(k)}$ field described in Sec.~\ref{sec:nbody} without resorting to the creation of mock galaxy catalogues using abundance matching or Halo Occupation Distribution (HOD) techniques\footnote{This approach also does not account for the effect of lensing magnification bias. Thus, we do not include lensing magnification in the theoretical modelling for consistency.}. We will report results for \nlth in terms of relative bias with respect to the true cross-correlation power spectrum, so the impact of changes in overall normalization is minimized. Any redshift and scale dependence of the bias can easily be incorporated in the formalism presented in Appendix ~\ref{sec:bispectrum}. However, detailed modelling of additional bispectrum terms arising from non-linear biasing or redshift distortions (and other effects) will require future work as all these effects might become important. We produced simulated density fields for 7 different redshift bins with an approximately constant width of 1 Gpc. The details of the bin boundaries are shown in Table~\ref{table:deltabounds}.

\begin{table}
\centering
\begin{tabular}{|c|c|c|c|}
\hline
$\bar{z}$&$z_{\textrm{min}}$&$z_{\textrm{max}}$&$\Delta\chi$ [Gpc]\\
\hline
0.2&0.06&0.33&1.1\\
0.35&0.25&0.46&0.8\\
0.6&0.41&0.78&1.2\\
1&0.75&1.3&1.5\\
2&1.19&2.85&2.5\\
5&2.85&7.04&2.5\\
9.45&7.04&11.85&1.2\\
\hline
\end{tabular}
\caption{Details of the redshift bins used to model the density field and its cross-correlation with CMB lensing, where $\Delta\chi$ gives the comoving thickness of the each top-hat bin.}
\label{table:deltabounds}
\end{table}
\noindent

\begin{figure}[htp]
\centering
\includegraphics[width=\textwidth]{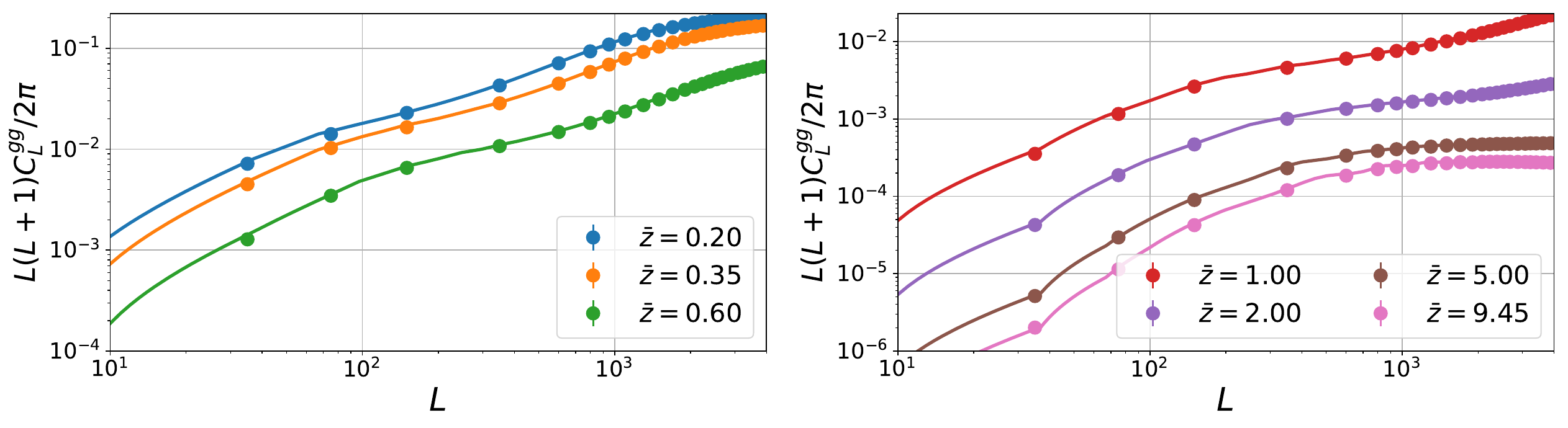}\\
\includegraphics[width=\textwidth]{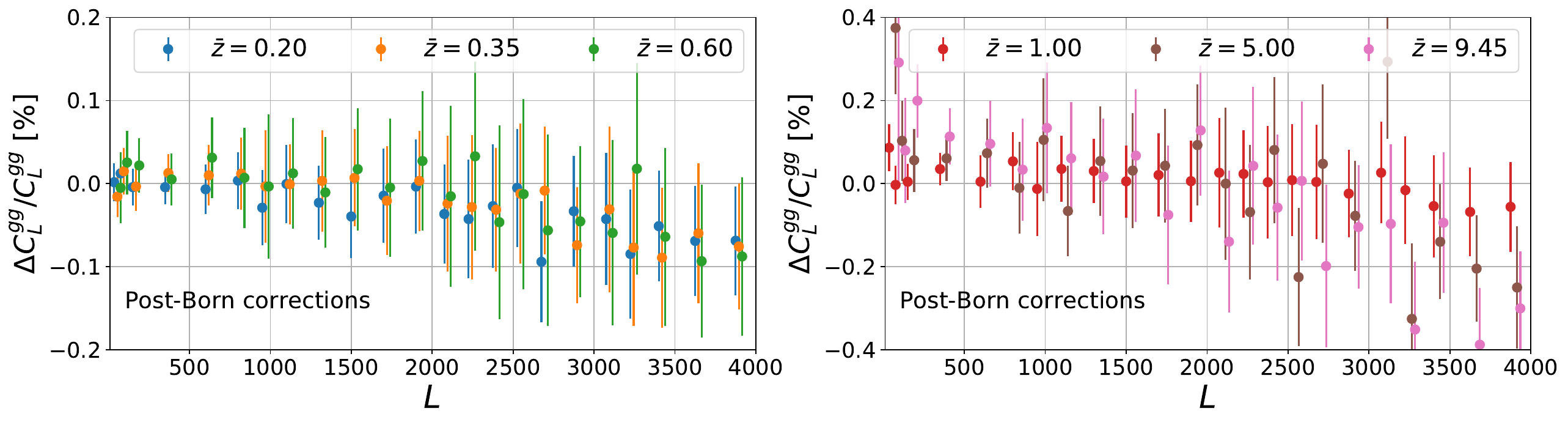}
\caption{Top: power spectra of the surface mass density of all the redshift bins used in this work. The first order simulation results extracted from the DEMNuni N-body simulation are shown as dots and analytical prediction using the Halofit fitting formulae are shown as solid lines. Analytic and simulation results agree at better than 4\%. Bottom: post-Born corrections to $C_{\ell}^{gg}$ for different redshift bins.}
\label{fig:surfacemass}
\end{figure}
\noindent
We will be correlating simulated density fields with Born and post-Born CMB lensing, so
for each redshift bin we produced lensed and unlensed density fields so that the density field is modelled to consistent perturbative order (so the signals observed in a given direction on the sky are probing densities at consistently perturbed locations along the line of sight). The first-order unlensed results are obtained by simply summing together the $\Sigma^{\theta(k)}$ planes falling within the boundaries of each redshift bin and then computing the surface mass overdensity $\Delta^{\bar{z}}_{\Sigma}$ of this map. To model the lensed surface mass density we first created an integrated convergence map $K^{(k)}$ for each $\Sigma^{\theta(k)}$ falling within a redshift bin in the Born approximation, i.e.
\begin{equation}
 K^{(k)} = \sum_{j<k} \kappa^{(j)} _{\chi_k}.
 \end{equation}
We lensed each $\Sigma^{\theta (k)}$ with a displacement field extracted from $K^{(k)}$\footnote{This operation requires solving the Poisson equation in the harmonic domain and for this purpose we set a maximum multipole $L_{\textrm{max}}=6144$.}  using a modified version of the \lenspix code, and then sum together each lensed $\Sigma^{\theta (k)}$  within a redshift bin to get the lensed surface mass density $\Delta^{\bar{z},{\rm ML}}_{\Sigma}$. This procedure is necessary to take into account the fact that at the next perturbative order, the density field is observed at the lensed position.
Neglecting this would lead to a loss of correlation, since the background and perturbed light ray paths would then see lensing by different small-scale structure (e.g. the perturbative cross power spectrum would be missing a $2\times2$-order term that is important on small scales).
However, to the order that we are working it is not required to lens the density field using the post-Born displacement. In Fig.~\ref{fig:surfacemass} we show the power spectrum $C_{\ell}^{gg,\bar{z}} = C_{\ell}^{\Delta\Delta,\bar{z}}$ of the simulated galaxy density maps compared with analytic predictions using Eq.~\eqref{eq:limbercls} and the Halofit fitting formula. The two results agree at better than $4\%$ level. We found that the difference between the lensed and unlensed $C_{\ell}^{gg,\bar{z}}$ is below the $0.2\%$ level for the highest redshift bin and below $0.05\%$ for the lowest one, and is therefore negligible. Note that this does not mean that the lensing correction is always negligible, for example the cross-power spectrum between the unlensed and lensed density fields does differ significantly on small scales, due to decorrelation between the densities observed at inconsistent positions along the line of sight. For this reason we consistently use the lensed densities when cross-correlating with the post-Born lensed CMB, and the lensed density is consistently modelled when calculating the theoretical bispectrum predictions (see Appendix~\ref{app:postborn}).

\subsection{CMB lensing and its cross-correlations with LSS tracers}

\begin{figure}[!htbp]
\centering
\includegraphics[width=\textwidth]{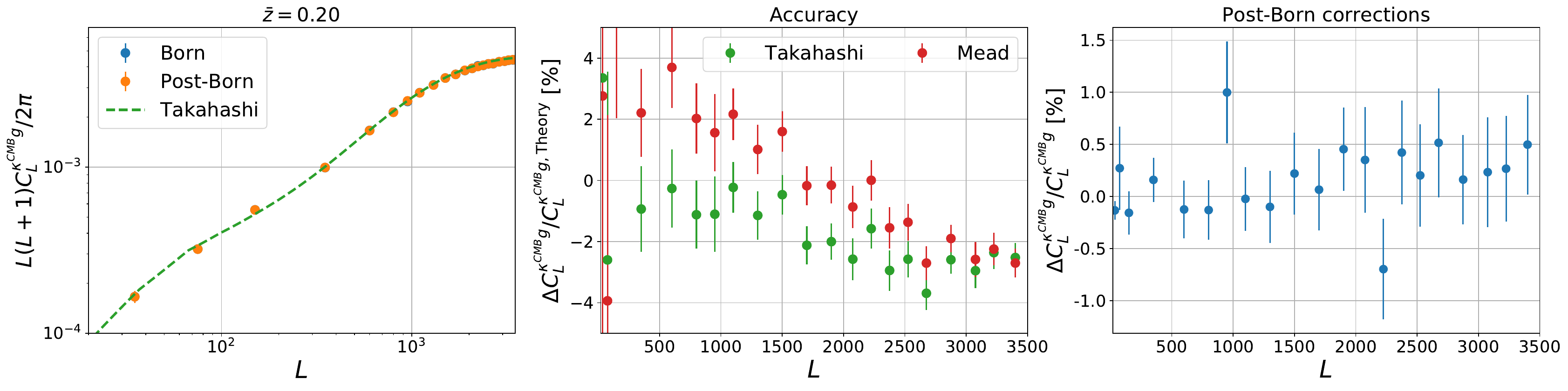}\\
\includegraphics[width=\textwidth]{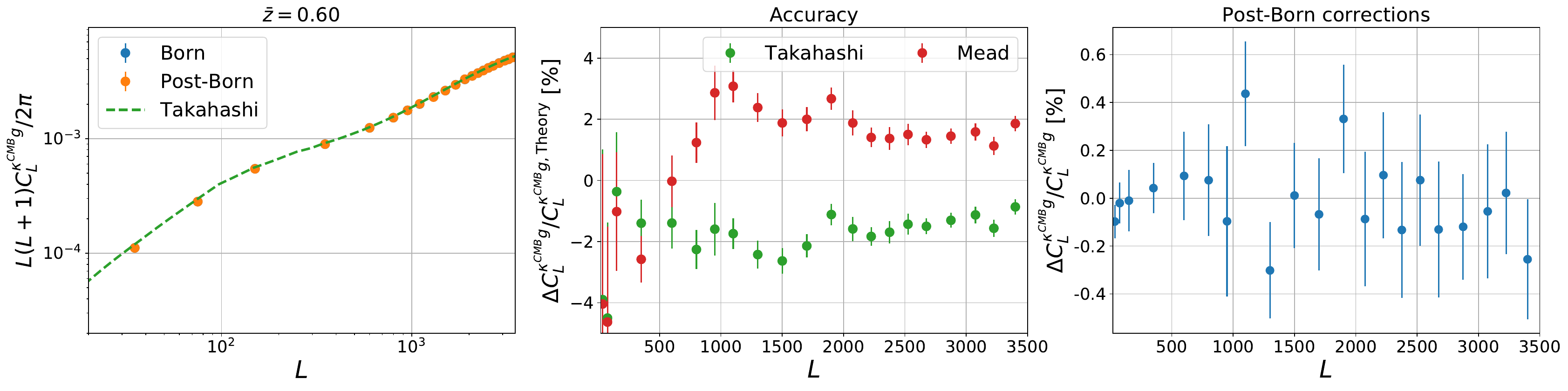}\\
\includegraphics[width=\textwidth]{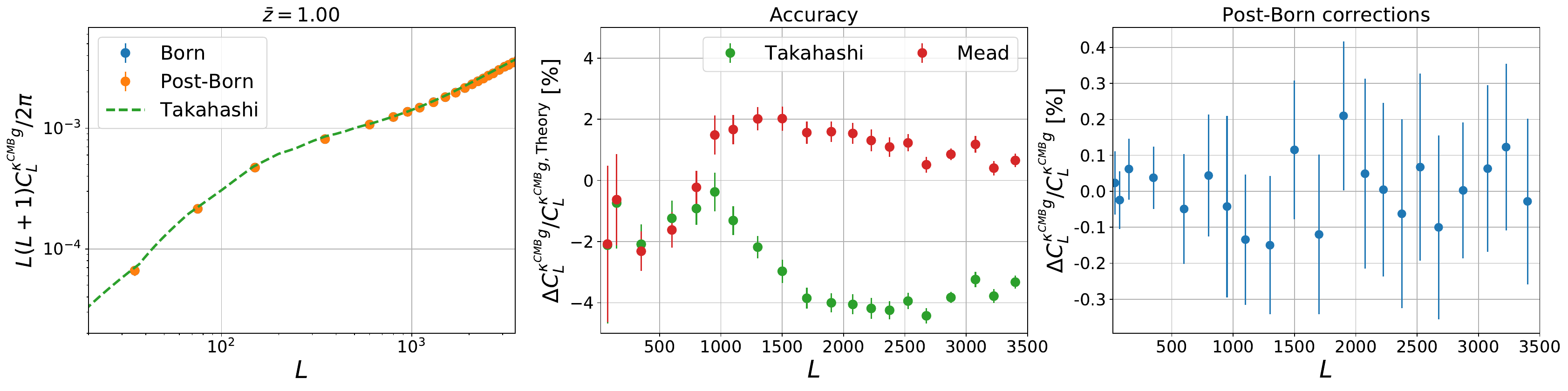}\\
\includegraphics[width=\textwidth]{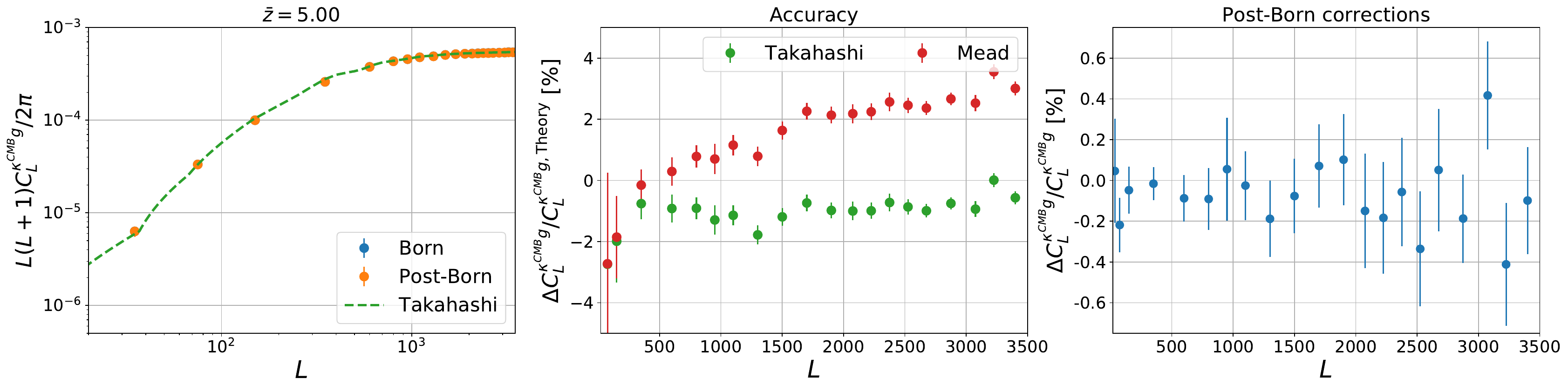}
\caption{Cross-correlation power spectrum between matter density and CMB lensing convergence. From top to bottom we show results for different density redshift bins. Results from first order (blue) and post-Born (orange) simulations are shown as dots and theoretical predictions based on Halofit (green solid) are shown in the left panels. The middle panel shows the fractional difference between simulation results and predictions obtained with the fitting formulae of Refs.~\cite{Takahashi:2012em} (green) or \cite{Mead:2015yca} (red) for the non-linear matter power spectrum. The left panel shows the difference between simulated results in Born and post-Born regime.}
\label{fig:cmbXg}
\end{figure}

\begin{figure}[!htbp]
\centering
\includegraphics[width=\textwidth]{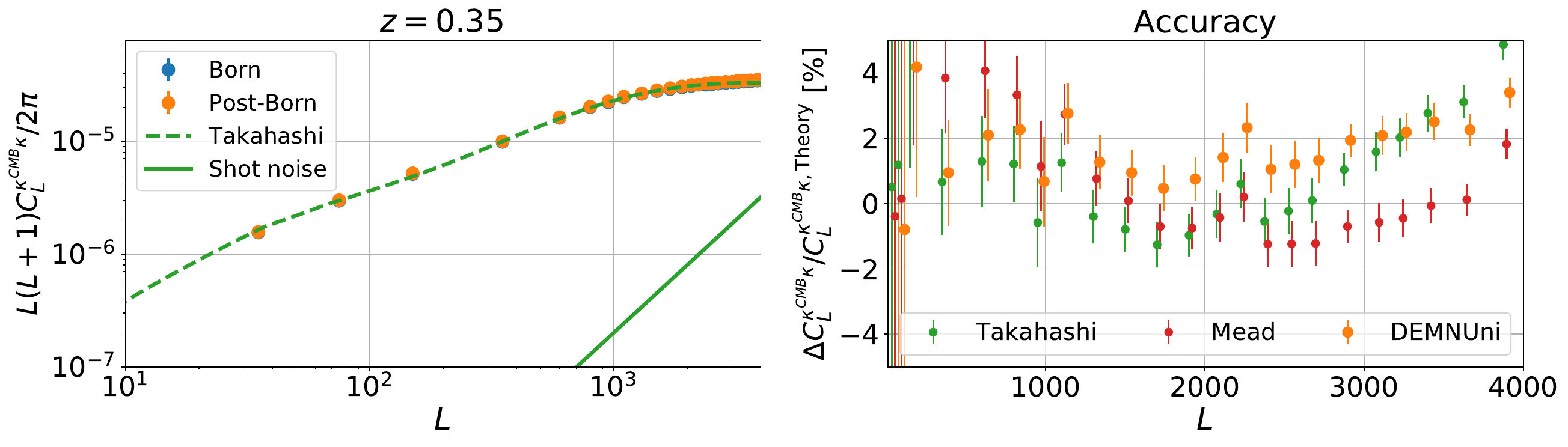}\\
\includegraphics[width=\textwidth]{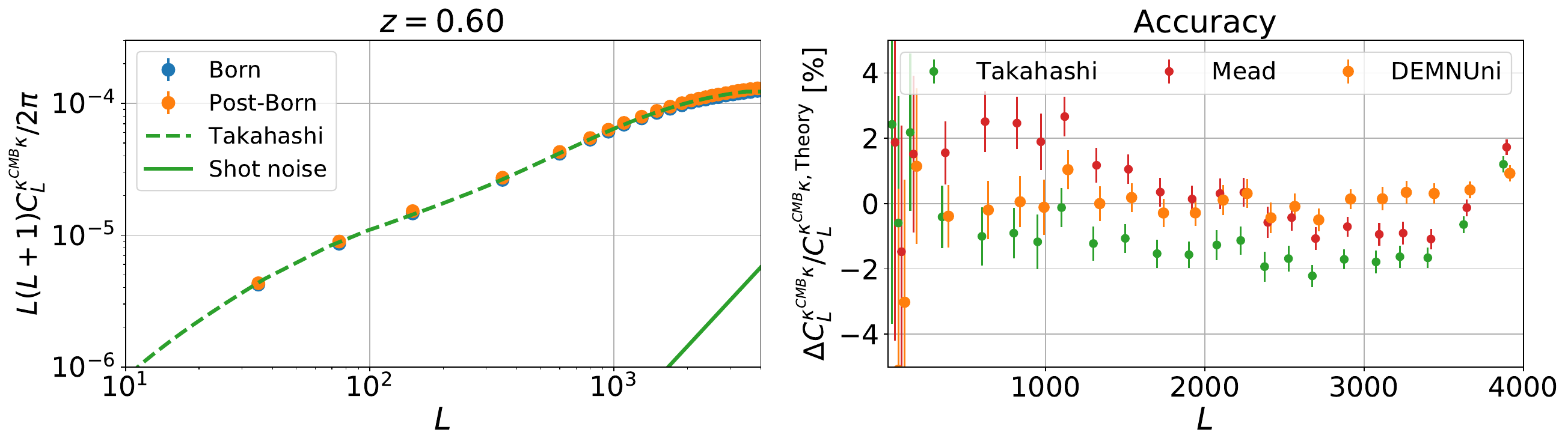}\\
\includegraphics[width=\textwidth]{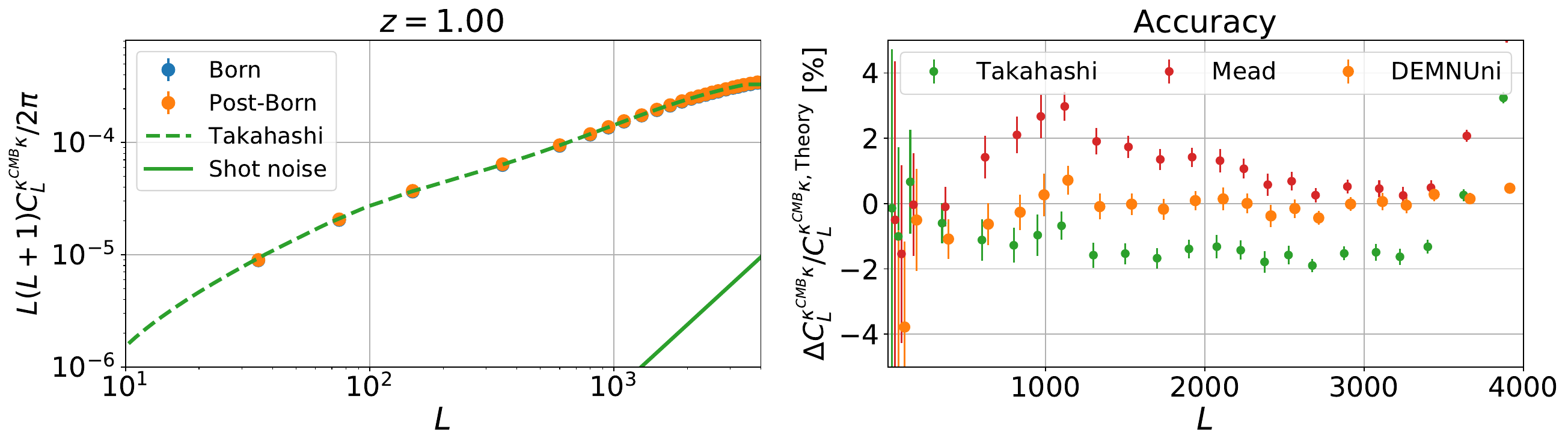}\\
\includegraphics[width=\textwidth]{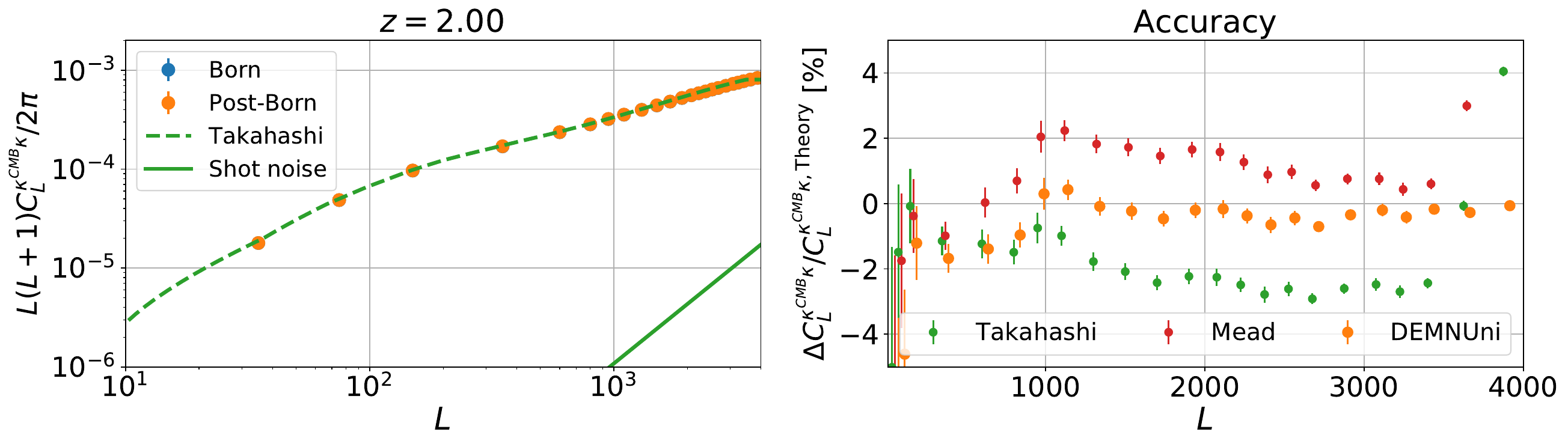}
\caption{Cross-correlation power spectra of CMB lensing and galaxy weak lensing convergence. From top to bottom we show results for different redshift bins. Results from first order (blue) and post-Born (orange) simulations are shown as dots, and theoretical predictions based on Halofit (dashed green) are shown in the left panels. The N-body shot noise contribution is shown in solid green. The right panels show the fractional difference between simulation results and theoretical predictions obtained with the fitting formulae of Ref.~\cite{Takahashi:2012em} (green) or Ref.~\cite{Mead:2015yca} (red) for the non-linear matter power spectrum, or the results obtained with the DEMNUni simulation matter power spectrum (orange). When not dominated by the numerics the difference between Born and post-Born results is $\simeq 0.2\%$.}
\label{fig:cmbXk}
\end{figure}

To model CMB lensing including all higher-order effects we used the simulated $\kappacmb$ maps in Born and post-Born regime from Ref.~\cite{fabbian2018}. These maps were derived using the same algorithm and N-body simulation products described in Sec.~\ref{sec:nbody}. We refer the reader to Ref.~\cite{fabbian2018} for a detailed summary of the validation tests and analysis performed on these maps. They reproduce the analytic predictions for the post-Born and non-linear LSS evolution on $C_L^{\kappacmb\kappacmb}, C_L^{\omega_{\rm CMB}\omega_{\rm CMB}}$ as well as all the contribution of $\kappacmb\kappacmb\kappacmb$ bispectrum to the skewness $S_3$ and lensed CMB power spectra with high precision. Here we show a few additional validation tests done to assess their accuracy in reproducing cross-correlation signals between CMB lensing and the simulated LSS tracers described in the previous sections. In Fig.~\ref{fig:cmbXg} and \ref{fig:cmbXk} we show a comparison between the analytic predictions of the cross-correlation power spectra between the CMB lensing convergence and density or galaxy convergence with the same quantities obtained from our simulations.

\noindent
In Fig.~\ref{fig:cmbXg} we show theoretical predictions in Limber approximation for the cross-correlation power spectrum with the galaxy density using different recipes to model the non-linear evolution. The difference between the Mead (HMcode)~\cite{Mead:2015yca} and Takahashi~\cite{Takahashi:2012em} Halofit versions used for the non-linear matter power spectrum has some affect on the agreement with simulation, but the agreement in both cases is at the $2-4\%$ level for all redshift bins and angular scales. In the same figure we also show for the first time an upper limit on the impact of the post-Born corrections on these observables, that amounts to $\lesssim 0.2\%$.
These results should be considered as indicative, since an accurate estimation of these effects (both at the simulation and theoretical level) should account for realistic modelling of the galaxy bias, which is beyond the scope of this paper.  Similarly to Fig.~\ref{fig:cmbXg}, in  Fig.~\ref{fig:cmbXk} we show that the Limber prediction for the cross-correlation power spectrum between CMB lensing convergence and galaxy lensing convergence agree with analytical predictions to better than 2\% for most cases, irrespective of the non-linear model used for the theory prediction. We show however that using the $P_\delta(k,z)$ from the DEMNuni simulation improves the agreement to sub-percent level, in particular for $z\geq 0.6$ where the signal is stronger and the effect of the shot-noise and power aliasing are reduced (at low-redshift we have a low number of particles per mass shell).  Shot-noise and power aliasing affects are correlated with the CMB lensing and galaxy lensing convergence, as both have been derived performing the same operations on an overlapping set of the mass planes. However, the numerical impact is more important in the galaxy convergence auto power spectrum than the cross-correlation, which has less contribution from low redshift to the total signal. The cross-correlation power spectrum with density at a given source redshift is even less affected since it has little contribution from lower redshifts.  The post-Born lensing correction has an effect that is $\simeq 0.2\%$ on the cross-correlation power spectrum with both galaxy density and galaxy lensing.\\
The tests we have outlined show that the accuracy of our numerical modelling of the cross-correlation signal, including higher-order effects, is more than sufficient to measure \nlth effects quite precisely, and the expected theoretical and numerical uncertainties are well under control.

\section{Results}\label{sec:results}

\subsection{Lensing reconstruction and \nlth extraction}\label{sec:sim-reconstruction}
\begin{figure}[!htbp]
\centering
\includegraphics[width=\textwidth]{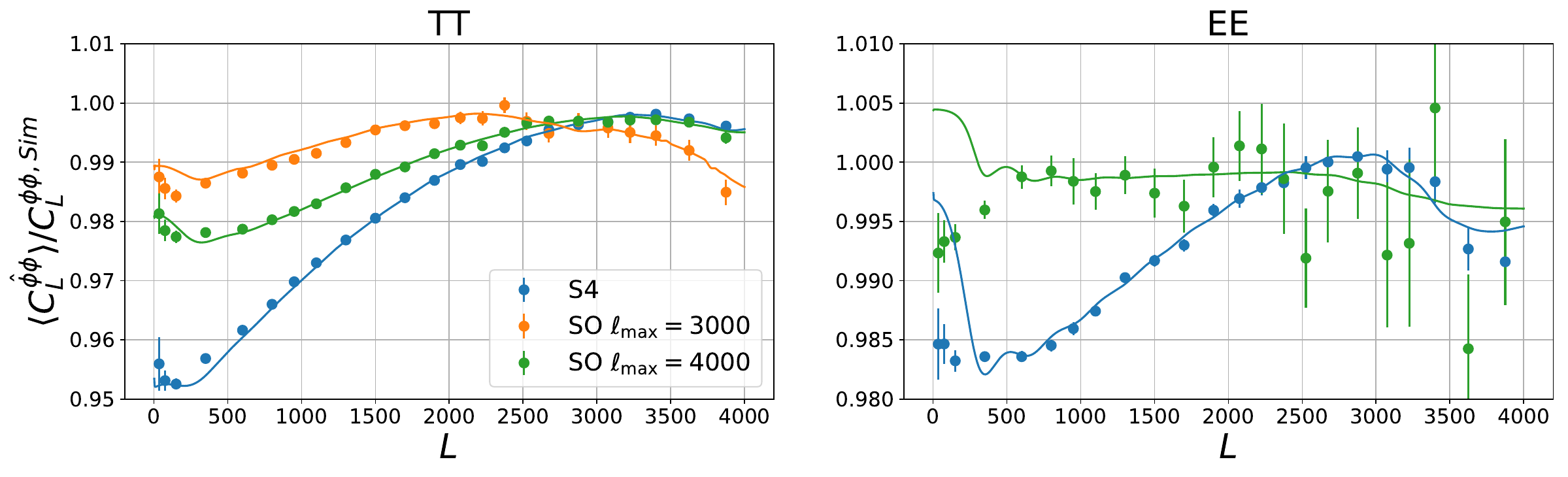}
\caption{Bias on the quadratic estimator normalization in the high signal-to-noise regime induced by using the standard perturbative response functions (plus simulation and approximation artefacts). Different reconstruction setups are shown as different colours.  The bias is evaluated as the average ratio of the cross-spectrum between the true lensing potential and the reconstructed lensing potential (normalized using lensed spectra in the estimator mode response functions), divided by the true lensing potential power spectrum for each specific realization of the $\kappa^G$ simulation suite. The solid line shows the ratio between the quadratic estimator normalization computed with the lensed power spectra in the mode response functions and the normalization computed with the non-perturbative response functions involving the gradient CMB power spectra, both calculated assuming the fiducial theoretical lensing potential power spectrum.}
\label{fig:response}
\end{figure}

To extract \nlth we followed a strategy similar to Ref.~\cite{beck2018}. We lensed the same set of 100 unlensed Gaussian CMB realizations having Planck 2013 cosmology described in Sec.~\ref{sec:nbody}, using a deflection field $\mathbf{d}$ extracted from the Born ($\kappacmb$ ) or post-Born ($\kappacmb^{\rm ML}$) convergence. The deflection can be represented by a spin-1 field and E/B harmonic decomposition, where in the Born-approximation
\be
_1d_{LM}^E = -\frac{2\kappa_{LM}}{\sqrt{L(L+1)}} \qquad _1d_{LM}^B = 0.
\ee
Since the curl mode of the deflection field of CMB lensing is related to the gradient mode at higher-order via the non-zero $\kappa\kappa\omega$ bispectrum, we also produced a set of lensed CMB simulations using the full deflection field as in Eq.~\eqref{eq:full-remapping}.  In this case, the deflection field acquires a non-zero $B$-mode component so that
\be
_1d_{LM}^E = -\frac{2\kappa^{\rm ML}_{\rm{CMB},LM}}{\sqrt{L(L+1)}} \qquad _1d_{LM}^B = -\frac{2\omega_{LM}}{\sqrt{L(L+1)}}.
\ee
where $\omega$ is the field rotation of Ref.~\cite{fabbian2018}.
In the following we use notation like $\hat{C}_L^{\hat{\phi}\phi^{\ext}}[\kappa]$ for any auto or cross-correlation power spectrum with an external LSS tracer $\phi^{\rm ext}$ that involves at least one reconstructed CMB lensing field estimated from the simulation set obtained with a deflection field extracted from a specific convergence field $\kappa$. \\*
The \nlth bias strongly depends on the experimental CMB noise level and resolution. For this analysis we considered two main idealized experimental configurations representative of future high-sensitivity CMB polarization experiments: the Simons Observatory \cite{so} (SO) and CMB-S4 (S4) \cite{cmbs4}. We add to each set of lensed CMB maps a white noise realization consistent with the beam-deconvolved noise power spectrum $N_{\ell}$ for each experiment. For SO we considered the so-called ``goal'' $N_{\ell}$ observed over 40\% of the sky, which is tabulated numerically to include the effects of component separation on the final sensitivity based on an internal linear combination (ILC) of frequencies (called \textit{Standard ILC} in Ref.~\cite{so}). For S4 we assumed an observed sky fraction of 40\%, an $N_{\ell}$ corresponding to $1$ $(1.4)\mu$K-arcmin white noise in intensity (polarization) after component separation and a Gaussian beam of 1 arcmin for S4.  For both S4 and SO we neglect uncertainties from modelling of foreground residuals. \\*
For all the simulated lensed CMB maps we performed lensing reconstruction including CMB multipoles $40\leq\ell\leq 3000$ for SO and $2\leq\ell\leq 4000$ for S4.
As described in Sec.~\ref{subsec:estimators}, we used slightly-suboptimal lensed power spectra in the quadratic estimator weights, but calculate the normalization using the non-perturbative lensed gradient spectra in the mode response functions to obtain unbiased results. We found that using lensed power spectra in the response functions would lead to biases of up to the order of 5\% (1\%) for S4 (SO) when including temperature reconstruction at high multipoles with low noise levels (see Fig.~\ref{fig:response}).
There was also a bias in the polarization reconstruction estimators, but the effect is less important, of the order of 1\% for EE reconstruction in S4 and sub-percent for SO.

\noindent
Despite using the non-perturbative mode response functions in the computation of the estimator normalization we observed small residual biases in the normalization of the estimators. This can be observed in the solid lines of Fig.~\ref{fig:response}, where we see a discrepancy between the expected value of the normalization bias (solid lines) and the measurements of the simulations, in particular at large scales. This could be due to several factors, such as the adoption of the flat-sky approximation in the computation of the gradient spectra as well as neglecting the $P_{\perp}$ terms in Table \ref{table:nonpertresponse}. In addition, the presence of N-body shot-noise leads to an excess of lensing power on the small scales causing an additional lensing effect that is not included in our analytic modelling of the mode response functions.
We corrected for these effects introducing a Monte Carlo (MC) correction. For this purpose, we produced an additional set of simulations where we lensed each of the same 100 unlensed CMB realizations using an independent Gaussian deflection field. Each of these are computed from an independent Gaussian realization of a convergence field having the same angular power spectrum of $\kappacmb$ as extracted from the raytracing simulations\footnote{We neglected the fact that $\kappacmb$ and $\kappacmb^{\rm ML}$ have different power spectrum due to post-Born corrections as these effect are non-detectable for the experimental configurations we considered here.}. We refer to this set of simulations as $\kappa^G$ set.
On each of the lensed CMB realizations in $\kappa^G$ we reconstructed the lensing potential $\hat{\phi}^G$ following Sec.~\ref{subsec:estimators}.
We then quantified the correction to the estimator normalization $A_L^{{\rm MCbias}, XY}$ by correlating $\hat{\phi}^G$ for each realization with the simulated Gaussian realization $\phi^G$
\begin{equation}
A_L^{{\rm MC bias}, XY} = \left(\frac{\langle \hat{C}_L^{\hat{\phi}^G\phi^G, XY}[\kappa^G]\rangle_{\textrm{CMB}}}{C_L^{\phi\phi}}\right)^{-1},
\label{eq:mcbias}
\end{equation}
where $\langle\cdot\rangle_\textrm{CMB}$ denotes an average over CMB realizations and $XY$ the specific lensed CMB fields used for the lensing reconstruction. Here $C_L^{\phi\phi}$ is the lensing potential spectrum computed from $C_L^{\kappacmb\kappacmb}$ using Eq.~\eqref{eq:conskappa}. We found this MC correction to be small and different depending on the reconstruction channel used. Its magnitude is scale dependent but has an amplitude  $\lesssim 1\%$ for TT and less than $0.2\%$ for the EB estimator for S4.  We then multiplied all of the lensing reconstruction of the $\kappacmb$ and $\kappacmb^{\rm ML}$ set of simulations by the corresponding $A_L^{{\rm MCbias}}$ before performing any \nlth measurement.

\subsection{\nlth Theory and simulation comparison}\label{sec:simcomparison}
With the sets of simulations described in the previous section we measured the different sources of \nlth for the cross-correlation by combining simulations in the following way:
\begin{align}
&\text{LSS:} &  N_L^{(3/2), \textrm{XY}}&=\left\langle \hat{C}_L^{\hat{\phi}^{\textrm{XY}}\phi^\ext}[\kappacmb]-C_L^{\phi\phi^\ext} \right\rangle_\textrm{CMB}  \\
&\text{Post-Born:} &  N_L^{(3/2), \textrm{XY}}&=\left\langle \hat{C}_L^{\hat{\phi}^{\textrm{XY}}\phi^\ext}[\kappacmb^{\rm ML}]-\hat{C}_L^{\hat{\phi}^{\textrm{XY}}\phi^\ext}[\kappacmb] \right\rangle_\textrm{CMB} \\
&\text{Total:} &   N_L^{(3/2), \textrm{XY}}&=\left\langle \hat{C}_L^{\hat{\phi}^{\textrm{XY}}\phi^\ext}[\kappacmb^{\rm ML}]-C_L^{\phi\phi^\ext} \right\rangle_\textrm{CMB}\\
&\text{Curl:}  &   N_L^{(3/2),\textrm{XY}}&=\left\langle \hat{C}_L^{\hat{\phi}^{\textrm{XY}}\phi^\ext}[\kappacmb^{\rm ML},\omega]-\hat{C}_L^{\hat{\phi}^{\textrm{XY}}\phi^\ext}[\kappacmb^{\rm ML}]\right\rangle_\textrm{CMB},
\end{align}
\noindent
where $\phi^{\ext}$ are all the density or galaxy convergence tracers described in Sec.~\ref{sec:sims}.

\begin{figure}[!htbp]
\centering
\includegraphics[width=\textwidth]{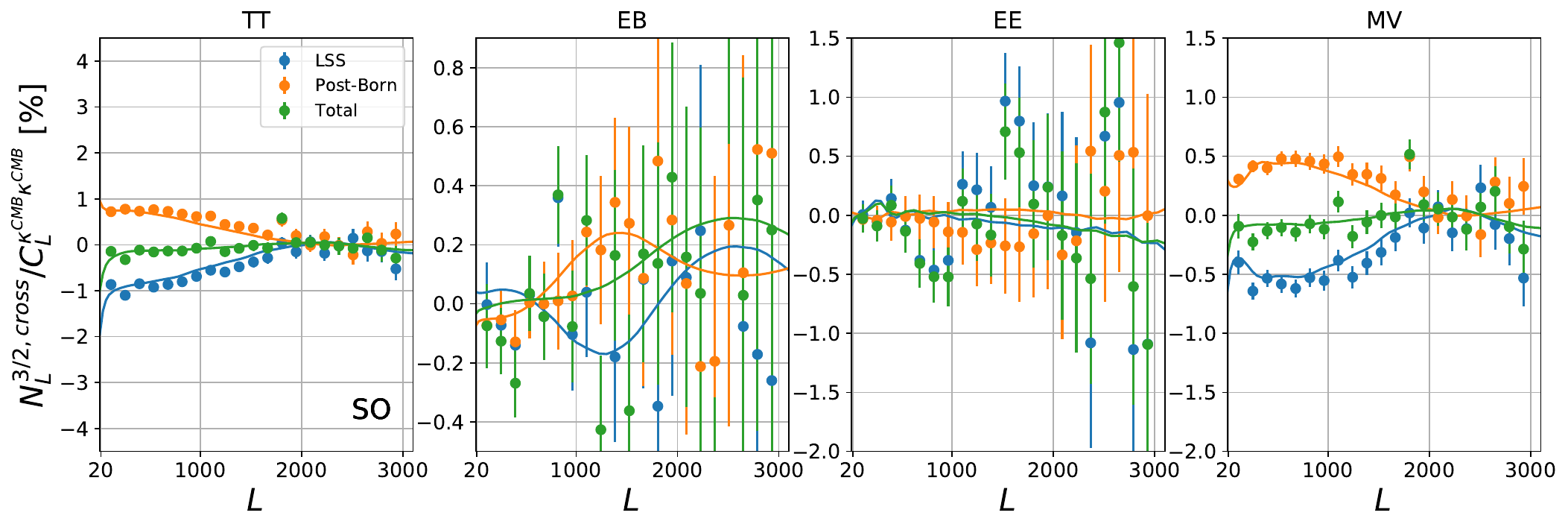}\\
\includegraphics[width=\textwidth]{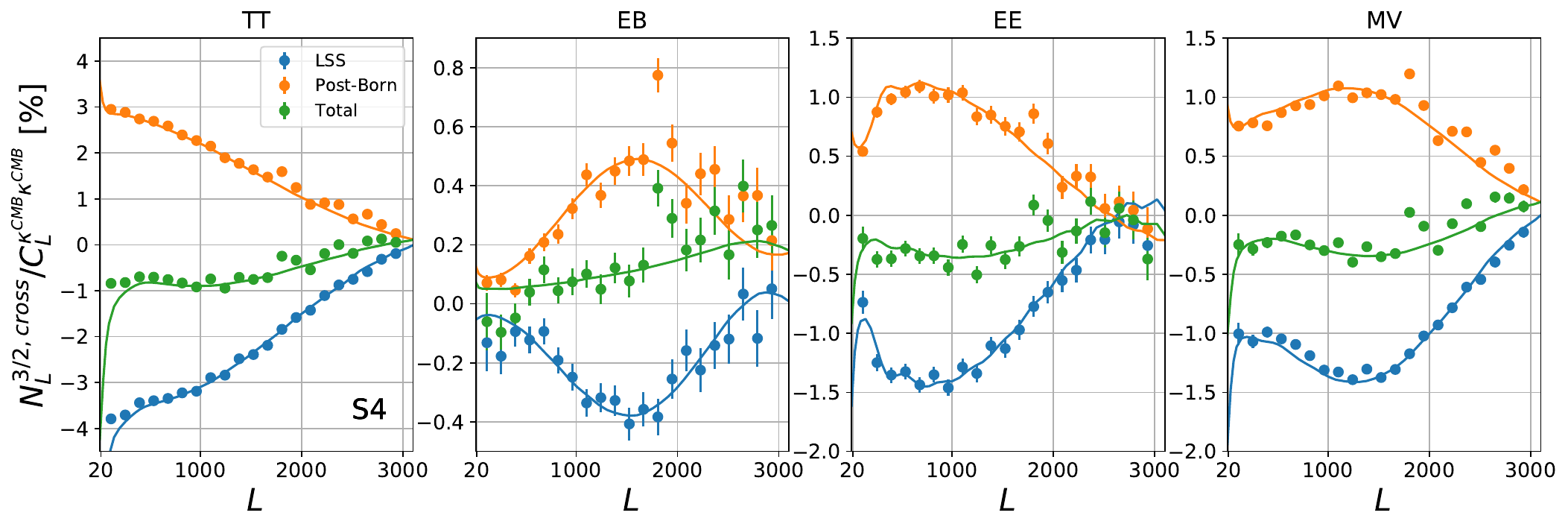}
\caption{Fractional \nlth bias in the cross-correlation power spectrum between the reconstructed CMB lensing potential and the true one measured from simulation for SO (top) and S4 (bottom). Theoretical predictions based on results of this work and GM fitting formulae for the matter bispectrum are shown as solid lines while binned simulation measurements are shown as dots. Different contributions to the \nlth signal are shown in different colours, showing the substantial cancellation between post-Born and LSS contributions. The error bars account for the sample variance of CMB alone.
}
\label{fig:cmbn32x}
\end{figure}
\noindent
We also correlate with the true simulated CMB lensing $\phi_{\rm input}$ as an interesting case study where the post-Born cancellation is largest and uncertainties due to the LSS modelling are less important. Moreover, it also provides an estimate for the \nlth bias on the CMB lensing autospectrum $C_L^{\phi\phi}$ as this is has an amplitude that is roughly $\simeq 2$ times $C_L^{\hat\phi\phi_{\rm input}}$~\cite{bohm2016}. In Fig.~\ref{fig:cmbn32x} we show the analytical predictions for \nlth compared to simulations when $\phi^{\ext}=\phi_{\rm input}$, extending the early results of Ref.~\cite{bohm2018} to the estimators including polarization and the minimum-variance estimator. The agreement between our theoretical predictions and simulations is very good for all the reconstruction channels at both SO and S4 sensitivity levels.
The TT lensing reconstruction estimator gets a large contribution from smaller angular scales where the non-linear effects are relatively more important, so we find a larger bias than with the estimators using polarization (as for the auto-spectrum case discussed in Ref.~\cite{beck2018}). The size of the bias depends on the $\lmax$ used in the temperature reconstruction, and may also be reduced by foreground mitigation strategies as discussion in more detail in Appendix~\ref{sec:foregrounds}.
The EB estimator is more robust as there is less signal from the very small scales. Note that even at S4 sensitivity, the higher order coupling due to the $\kappa\kappa\omega$ bispectrum is negligible, so we neglected this term.

\begin{figure}
\centering
\includegraphics[width=\textwidth]{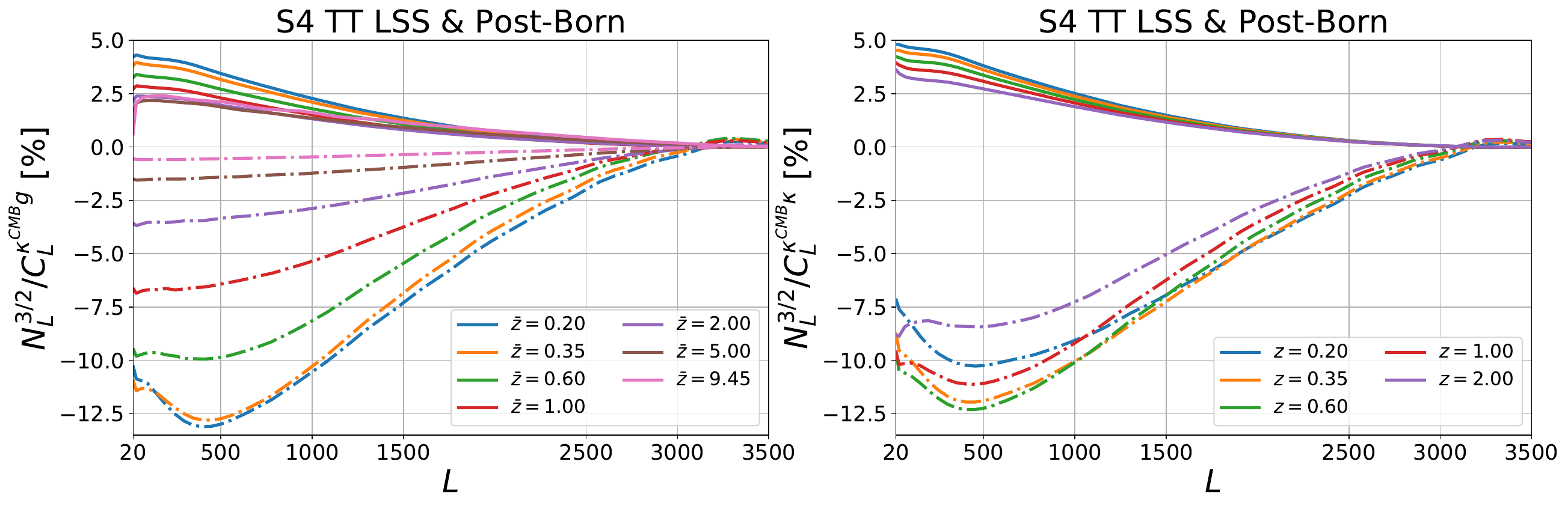}\\
\includegraphics[width=\textwidth]{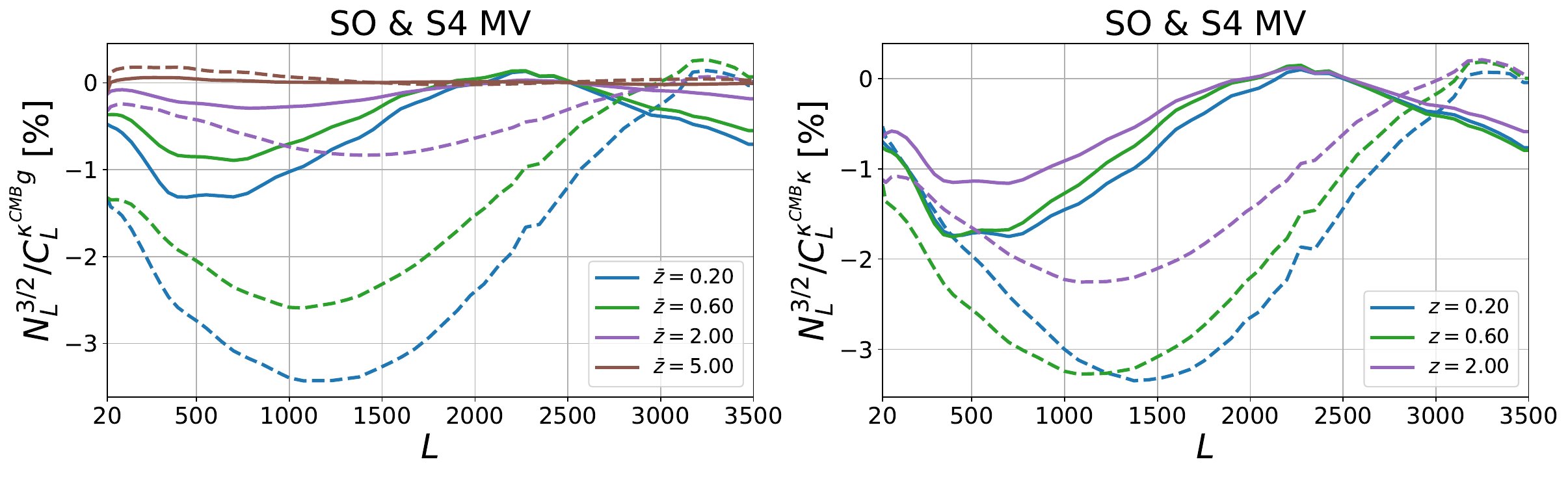}
\caption{Top: redshift dependence of the LSS (dot-dashed) and post-Born (solid) contributions to the fractional \nlth bias when using TT lensing reconstruction with S4. \nlth for cross-correlation with density is shown on the left panel while the galaxy lensing is shown on the right panel. Bottom: redshift dependence of \nlth for the minimum-variance lensing reconstructions. The SO curves are shown in solid line while S4 as dashed lines. Density is shown on the left panel and galaxy lensing is shown on the right panel. Different redshift are shown as different colours.
}
\label{fig:n32-trends}
\end{figure}
\begin{figure}
\centering
\includegraphics[width=\textwidth]{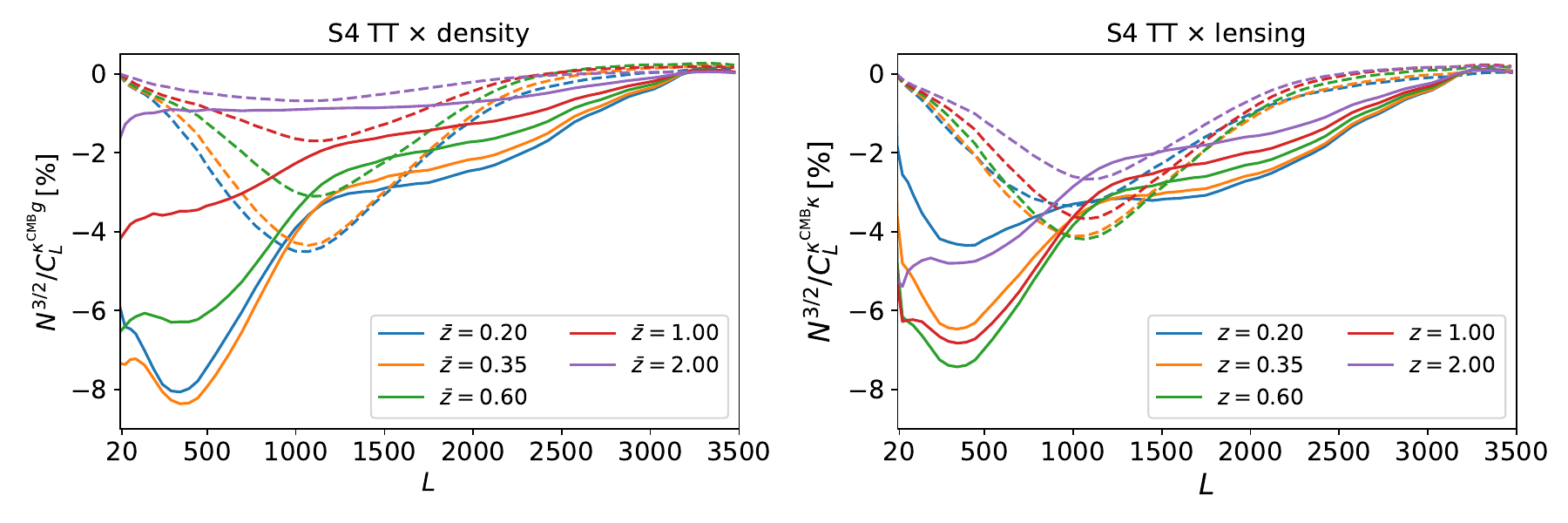}\\
\caption{Fractional \nlth bias from different configurations of the cross-bispectrum for S4 TT lensing reconstruction cross-correlated with galaxy density (left) and lensing (right). Solid lines show the contribution from semi-squeezed bispectrum triangles with short modes $L_1, L_2 > 2L$ where $L$ is largest-scale mode in the triangle, and dashed lines show the remaining contributions.
}
\label{fig:n32-squeezed-trends}
\end{figure}

\noindent
Fig.~\ref{fig:n32-trends} shows the analytic predictions for the \nlth bias in the cross-correlation with low redshift tracers for SO and S4 using the TT and MV estimators.  \nlth becomes a progressively larger fraction of the signal when using tracers probing the matter distribution at lower redshift. In the case of density cross-correlation, the LSS contribution to \nlth increases by factor of 3 going from $z=2$ to $z=0.35$ and similar trend, though less strong, is observed for the cross-correlation with galaxy lensing. The fractional post-Born bias has weaker variations of the order of $\lesssim 50\%$, since the size of the post-Born contribution and the signal itself both scale roughly the same way with path length. The stronger scaling of the LSS density cross-correlation bias means that post-Born is less important at low redshifts, leading to a net increase of the total bias because there is less cancellation between the two terms.
 It is interesting to note that the amplitude of the large-scale \nlth for galaxy lensing tends to saturate for sources located at $0.35<z<0.6$ and gets reduced for the $z=0.2$ case that mainly traces matter at $z\approx 0.1$, as the sharp decrease in the lensing kernel of CMB lensing overcomes the effect of the increased non-linearities of the LSS\footnote{
 The \nlth bias depends on the cross-bispectrum which in the Limber approximation scales with two powers of the CMB lensing window function at each contributing redshift, but the cross power spectrum scales with one power, so the ratio tends to zero as $z\rightarrow 0$ where the window function goes to zero.}. Lensing of galaxies at $z\approx 0.35 (0.6)$ mainly probes the matter distribution halfway between the source and the observer. We see the same saturation effect in the cross-correlation with the density at lower redshift with respect to the galaxy lensing case, i.e. $z<0.35$.
The largest contribution to the fractional bias on large scales comes from semi-squeezed bispectrum triangles coupling large-scale tracer modes to the small-scale lensing potential modes that enter the small-scale lensed temperature that dominates the large-scale lensing reconstruction.
 This is illustrated in Fig.~\ref{fig:n32-squeezed-trends}, which also shows the same trend of growth to low redshift followed by suppression at very low redshifts that give little contribution to the small-scale CMB lensing potential.

\begin{figure}[!htbp]
\centering
\includegraphics[width=\textwidth]{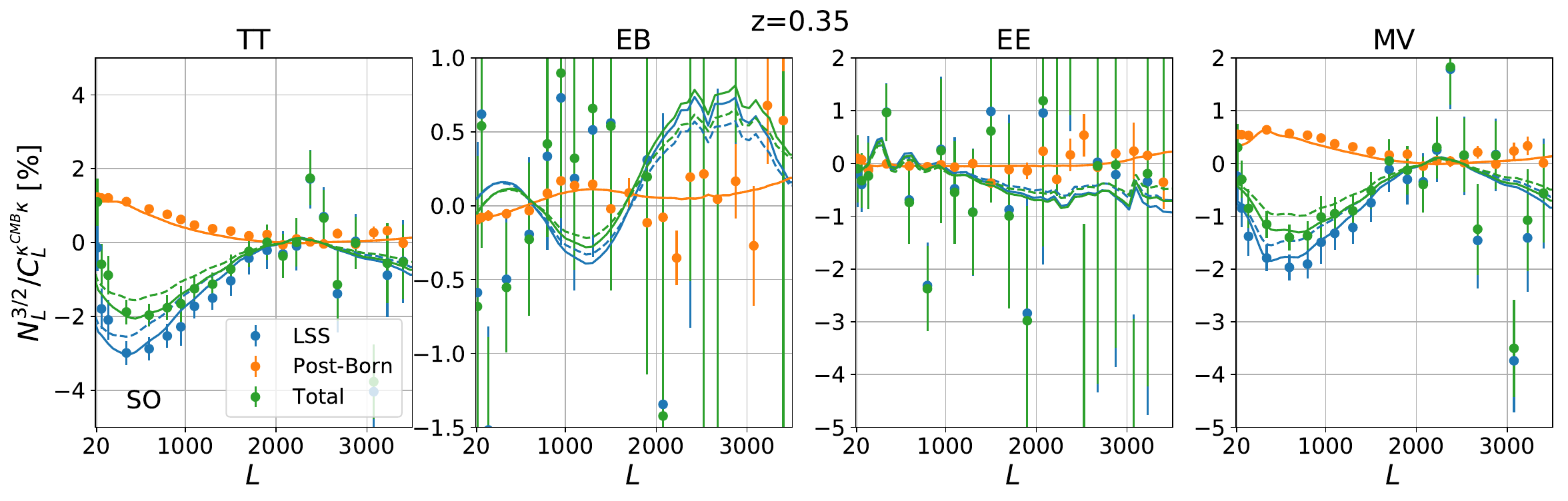}\\
\includegraphics[width=\textwidth]{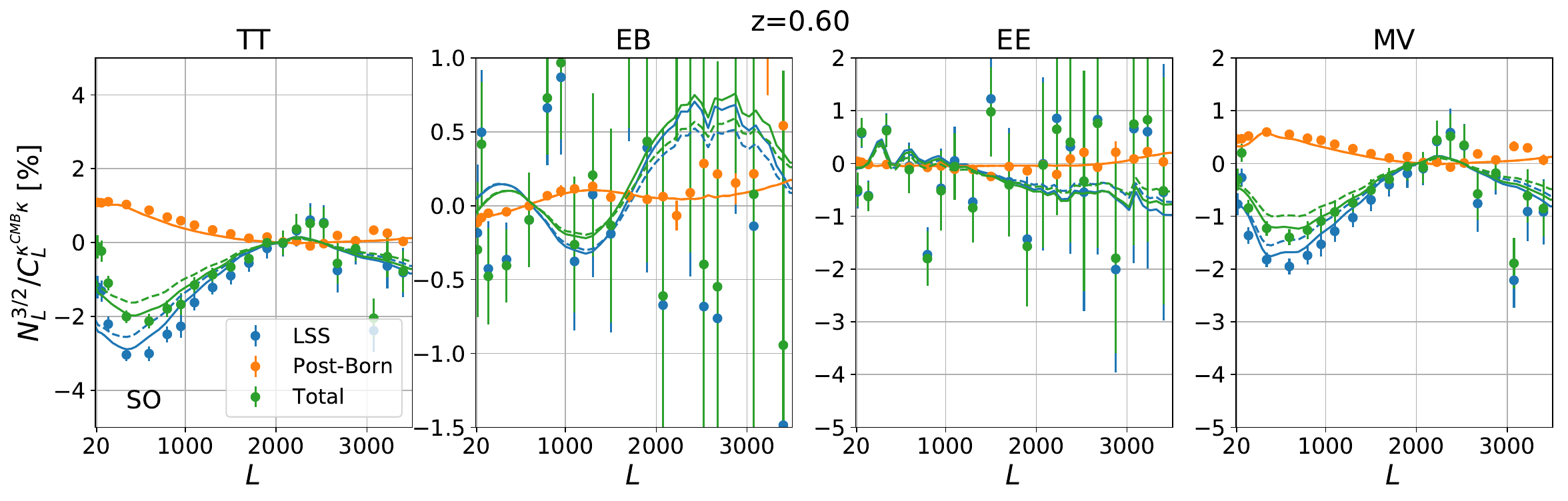}\\
\includegraphics[width=\textwidth]{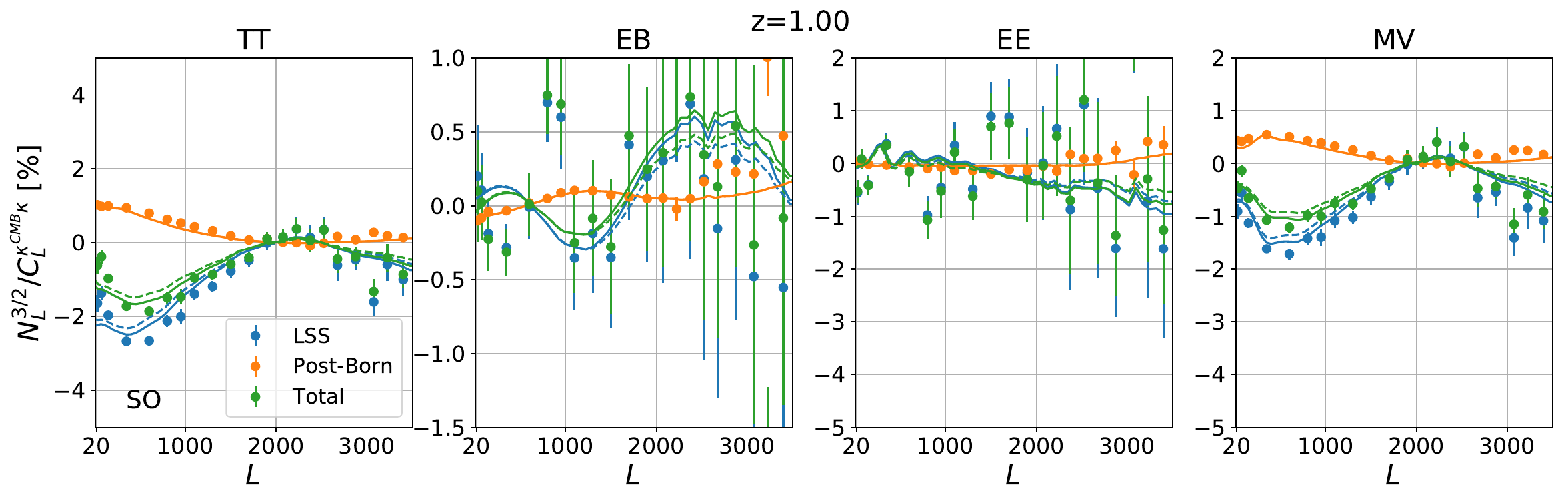}\\
\includegraphics[width=\textwidth]{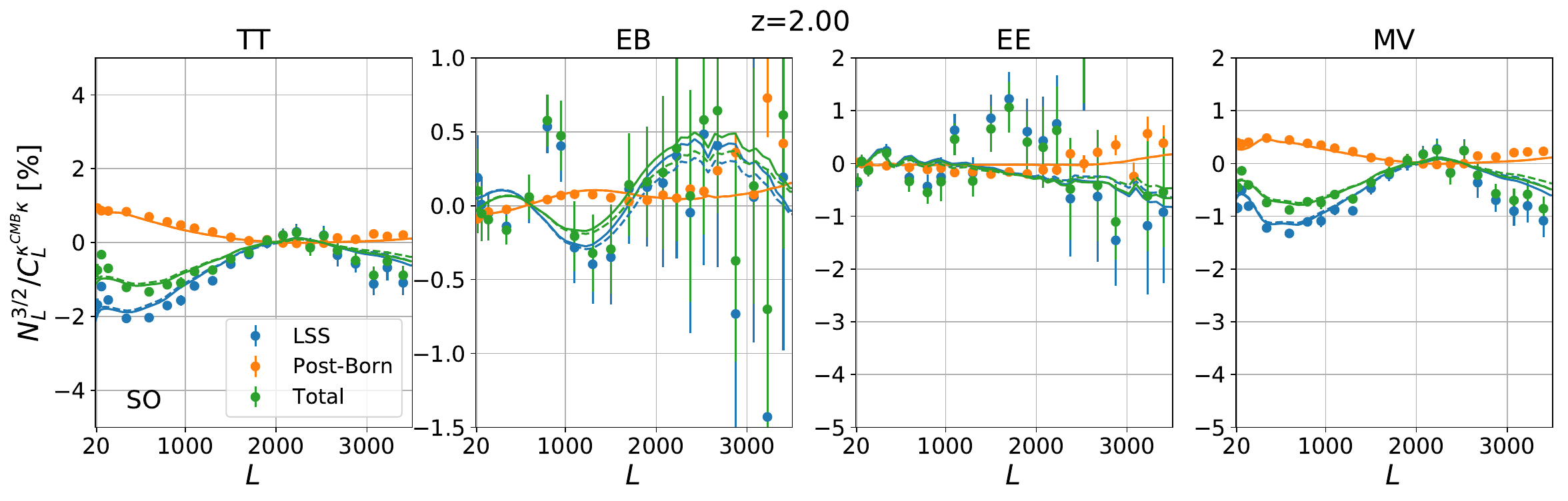}\\
\caption{Fractional \nlth bias for the cross-correlation power spectrum between the reconstructed CMB lensing potential of the Simons Observatory and galaxy lensing at different redshift bins. The redshift increases moving from top to bottom. Theoretical predictions using GM fitting formulae for the matter bispectrum are shown as solid lines while those based on SC fitting formulae are shown as dashed lines. Different contributions to the \nlth signal are shown in different colours. The error bars accounts for the sample variance of CMB alone.}
\label{fig:results_kappaSO}
\end{figure}

\begin{figure}[!htbp]
\centering
\includegraphics[width=\textwidth]{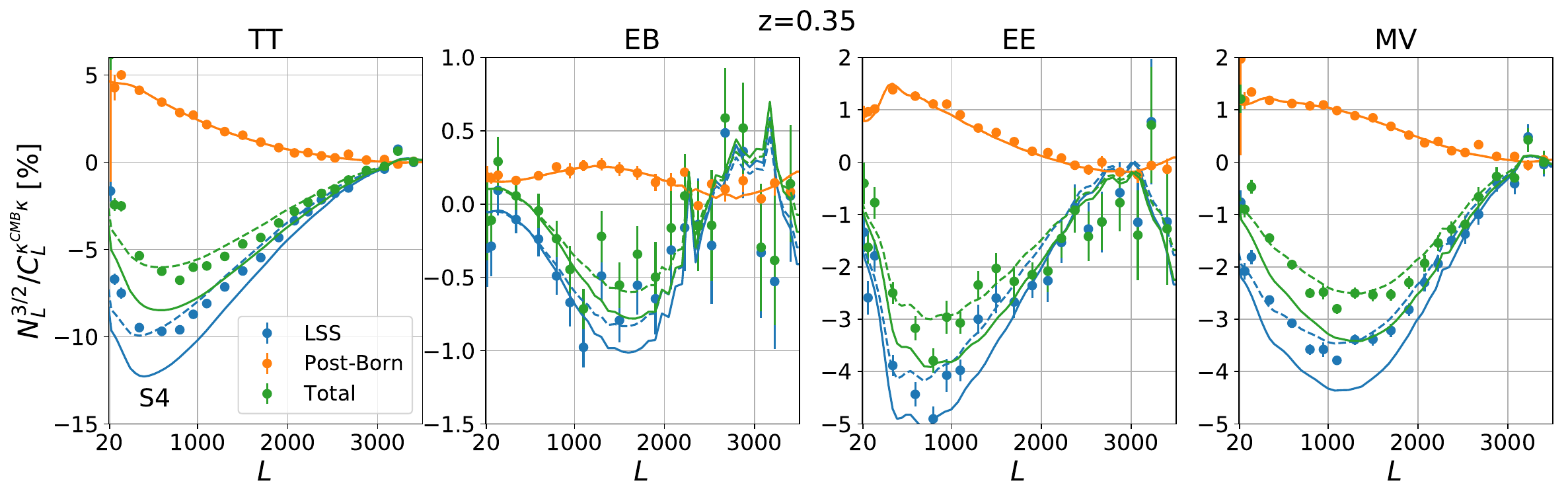}\\
\includegraphics[width=\textwidth]{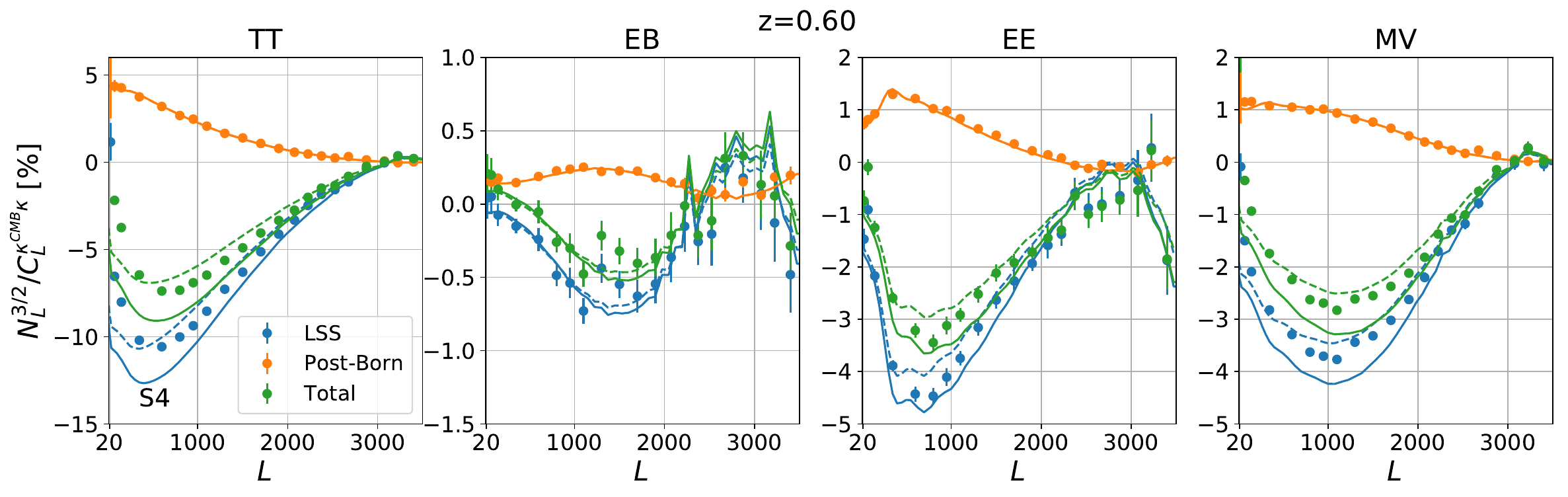}\\
\includegraphics[width=\textwidth]{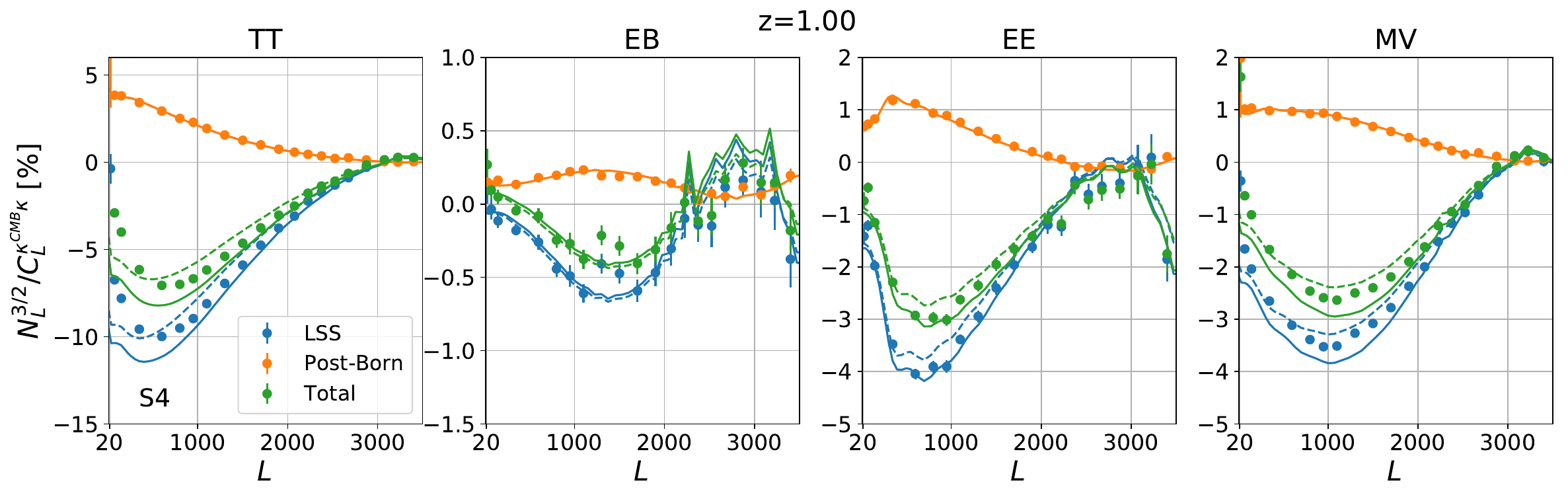}\\
\includegraphics[width=\textwidth]{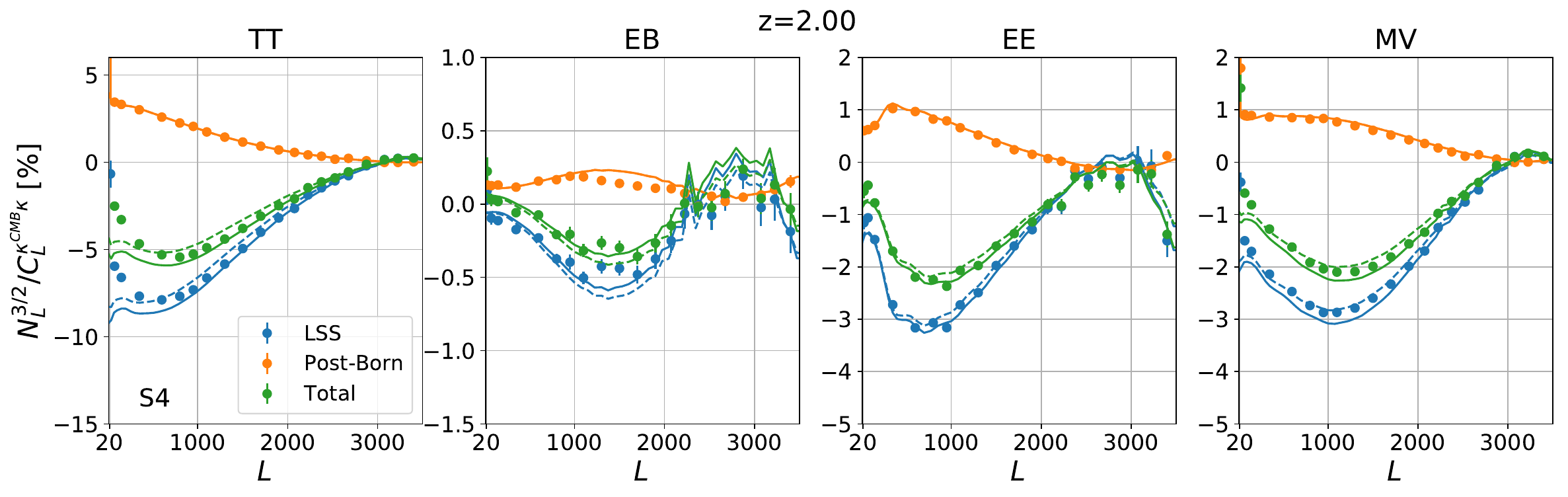}\\
\caption{Same as Fig.~\ref{fig:results_kappaSO} for S4}
\label{fig:results_kappaS4}
\end{figure}

\begin{figure}[!htbp]
\centering
\includegraphics[width=\textwidth]{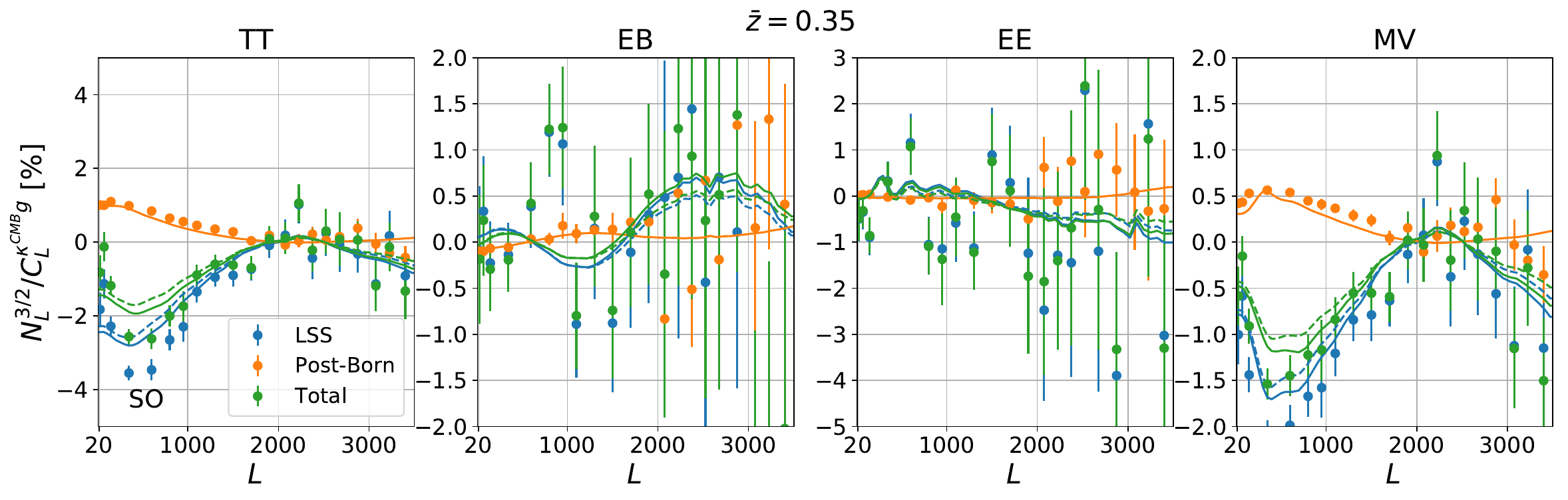}\\
\includegraphics[width=\textwidth]{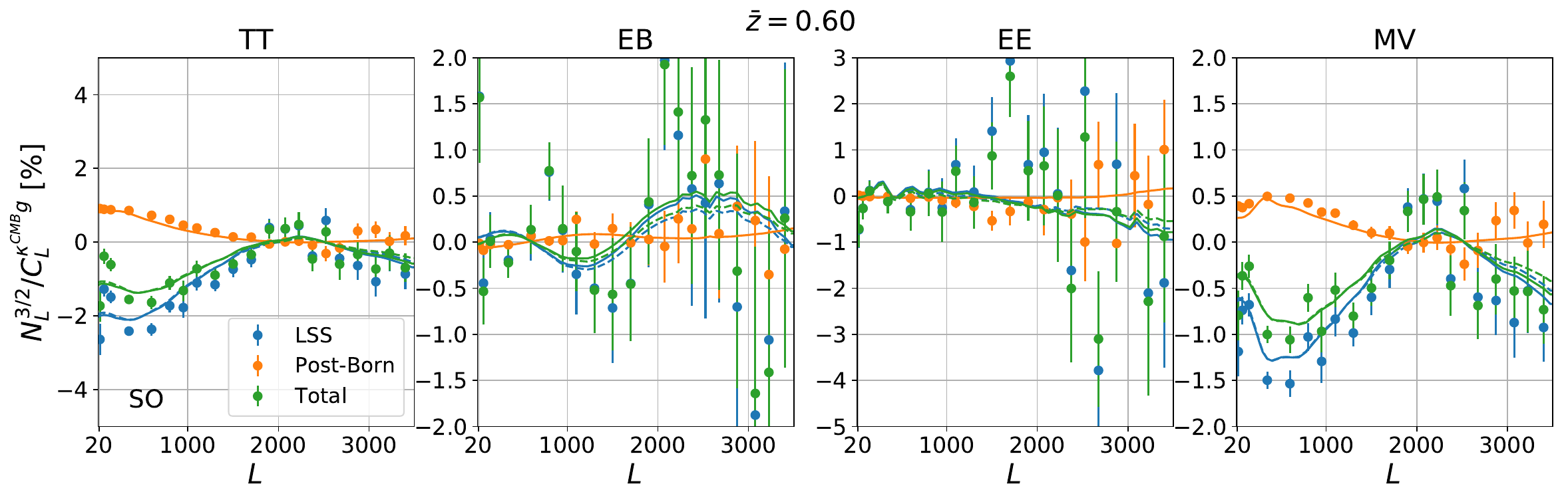}\\
\includegraphics[width=\textwidth]{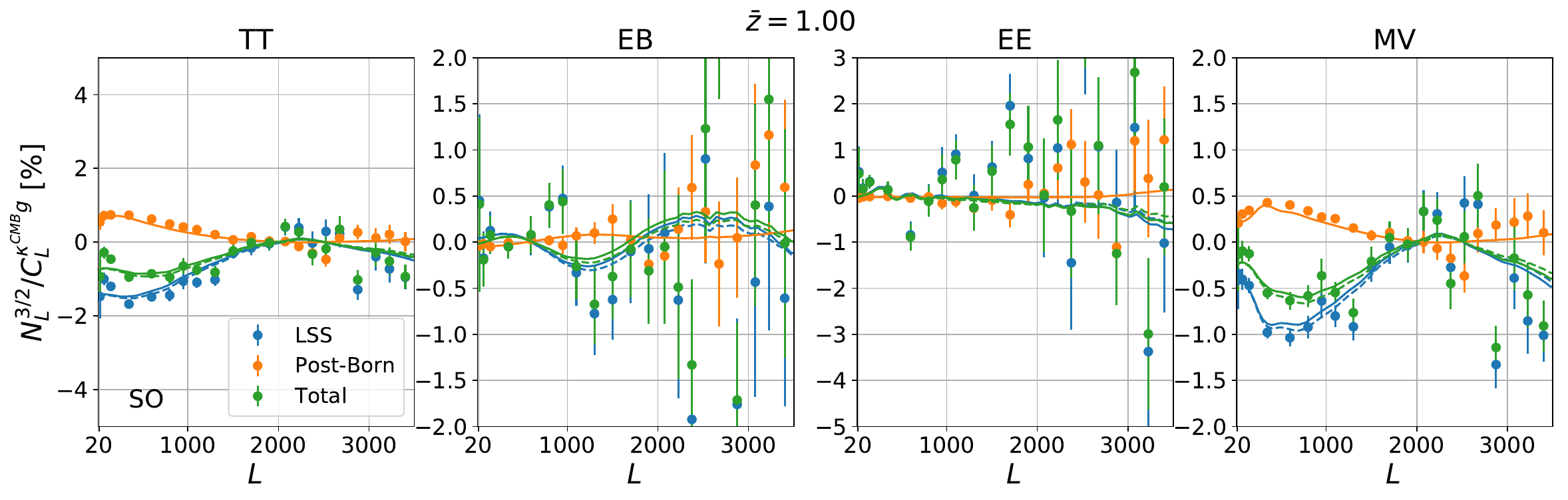}\\
\includegraphics[width=\textwidth]{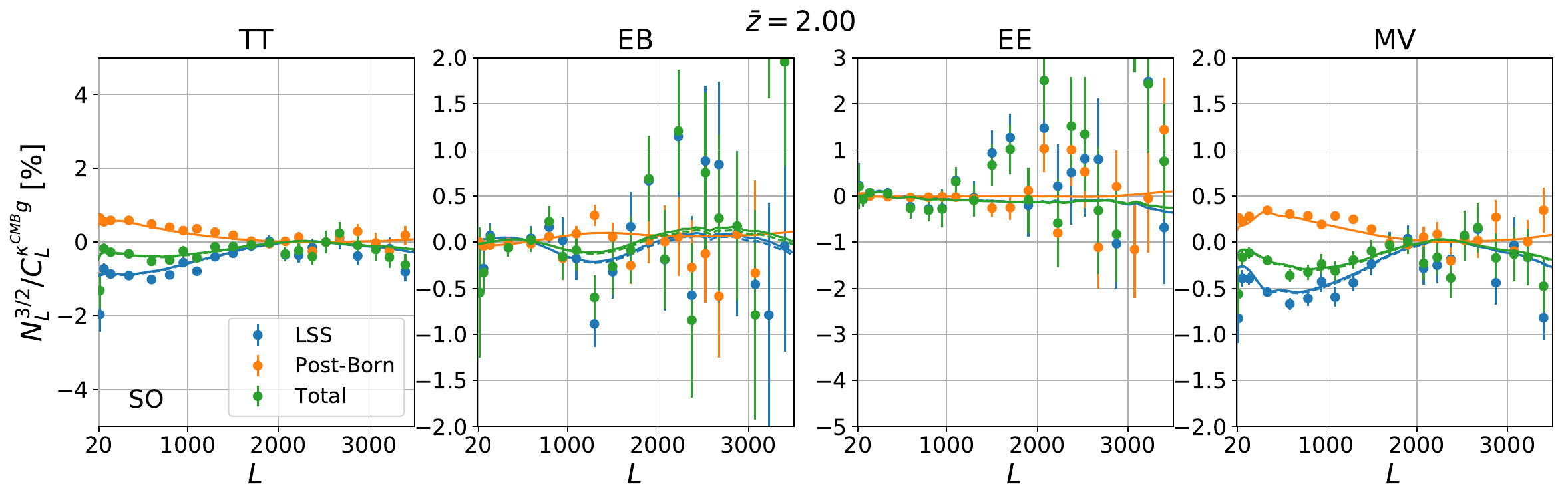}\\
\caption{Fractional \nlth bias for the cross-correlation power spectrum between the reconstructed CMB lensing potential of SO and galaxy density at different redshift bins. The redshift increases moving from top to bottom. Theoretical predictions using GM fitting formulae for the matter bispectrum are shown as solid lines while those based on SC fitting formulae are shown as dashed lines. Different contributions to the \nlth signal are shown in different colours. The error bars accounts for the sample variance of CMB alone. }
\label{fig:results_countsSO}
\end{figure}

\begin{figure}[!htbp]
\centering
\includegraphics[width=\textwidth]{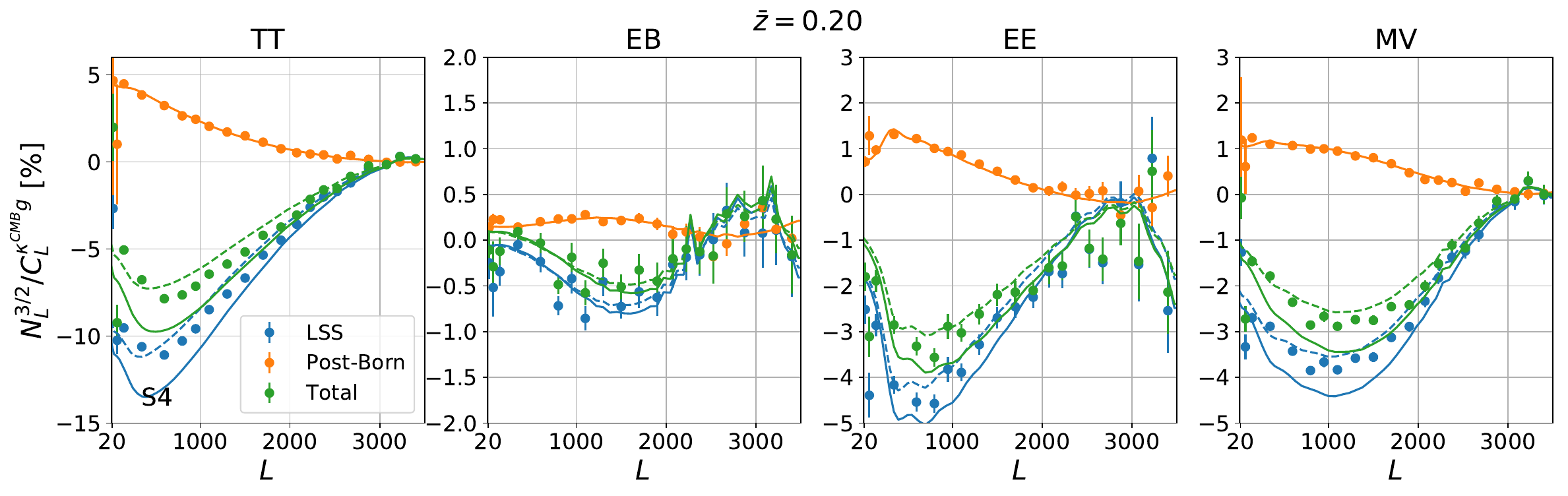}\\
\includegraphics[width=\textwidth]{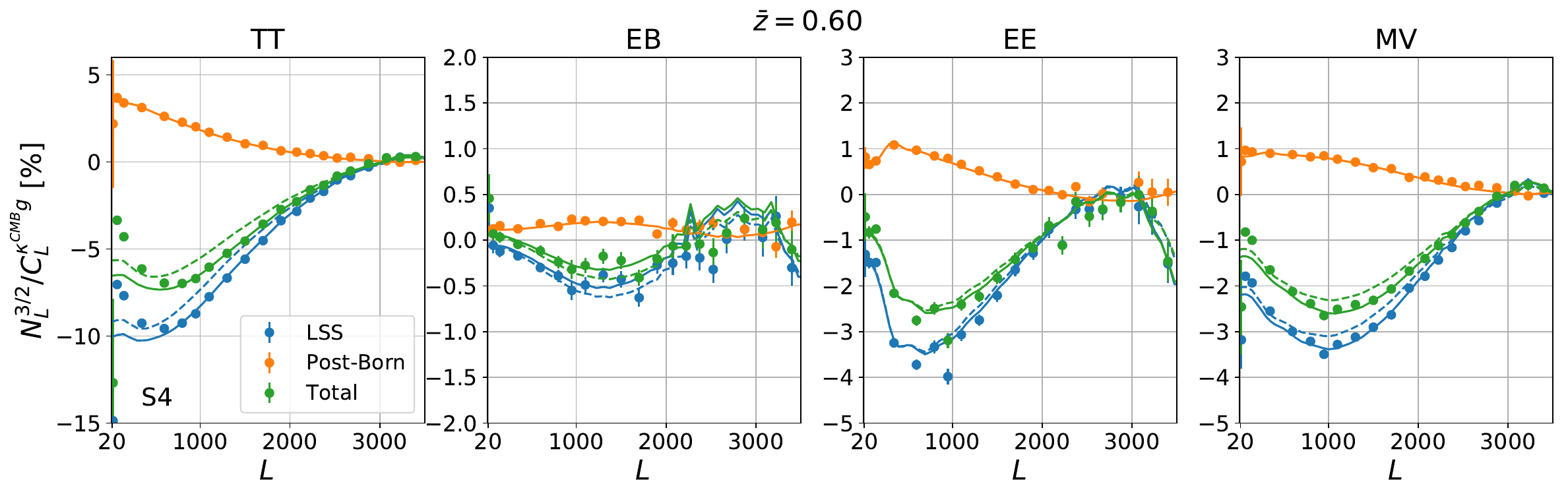}\\
\includegraphics[width=\textwidth]{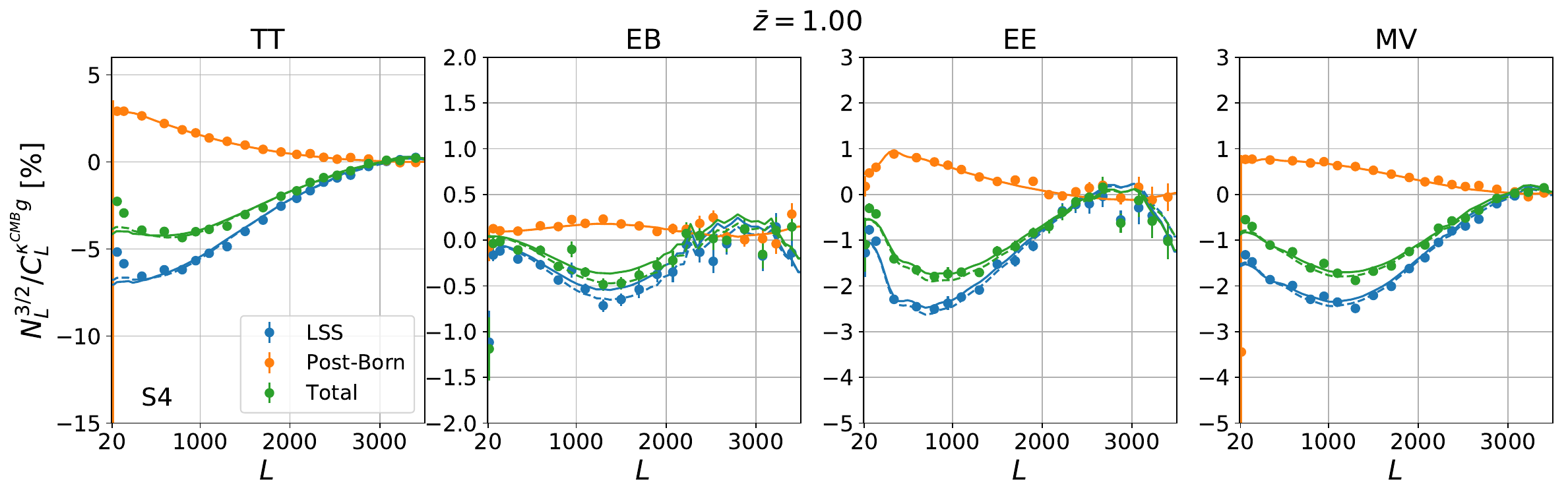}\\
\includegraphics[width=\textwidth]{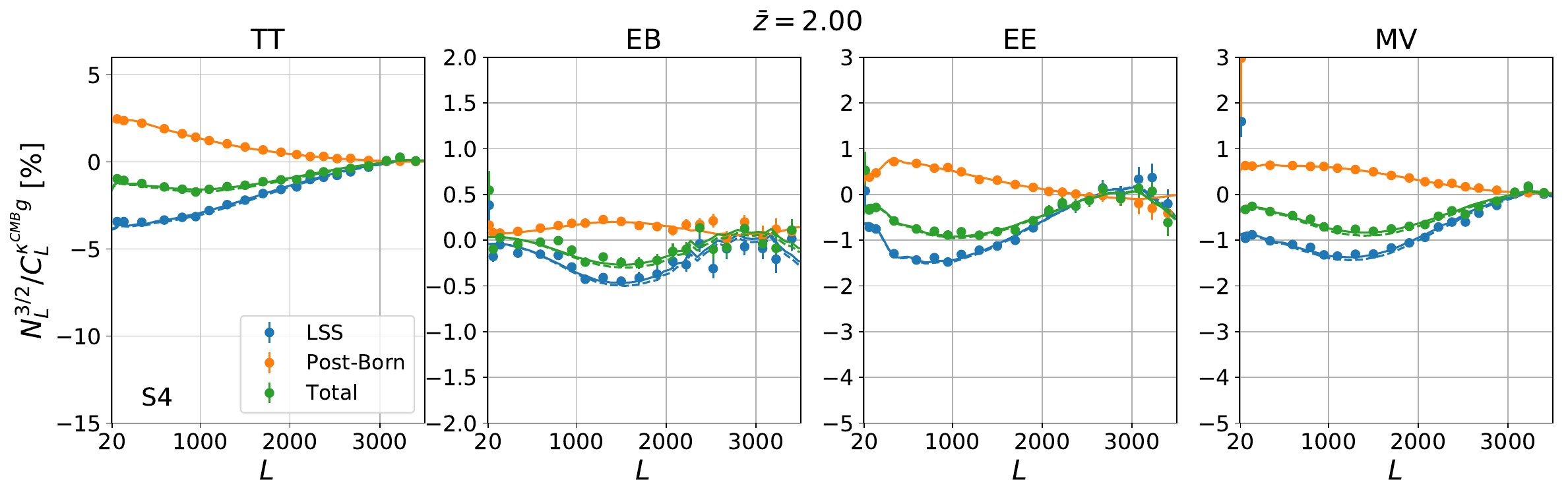}\\
\caption{Same as Fig.~\ref{fig:results_countsSO} for S4.}
\label{fig:results_countsS4}
\end{figure}

\begin{figure}[!htbp]
\centering
\includegraphics[width=\textwidth]{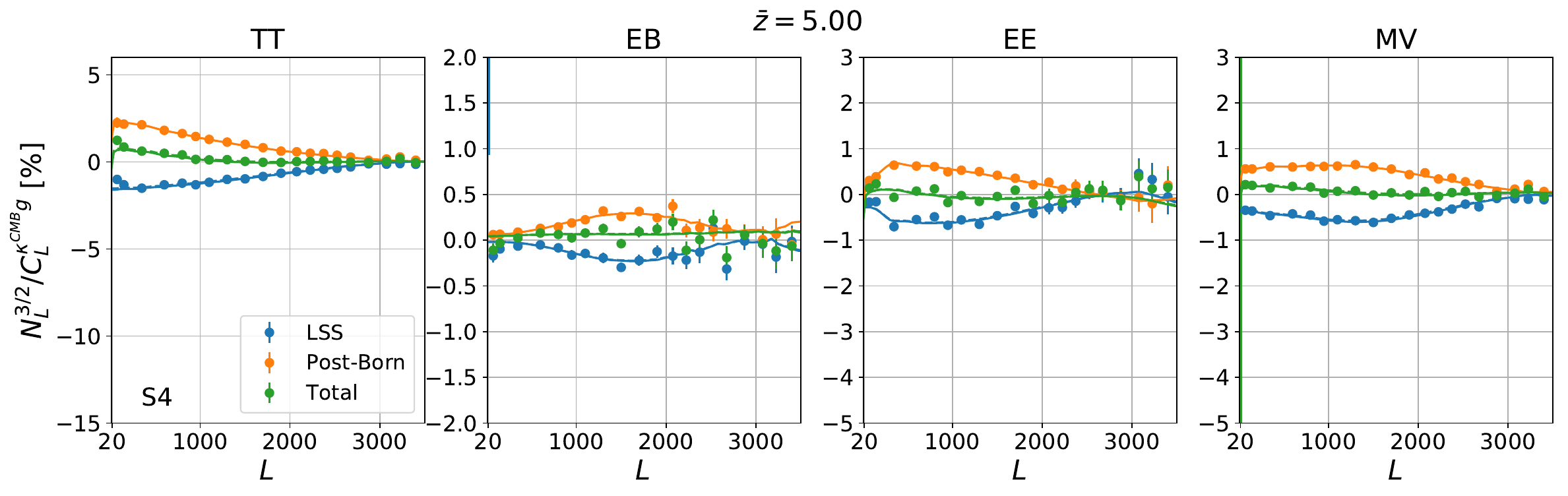}\\
\includegraphics[width=\textwidth]{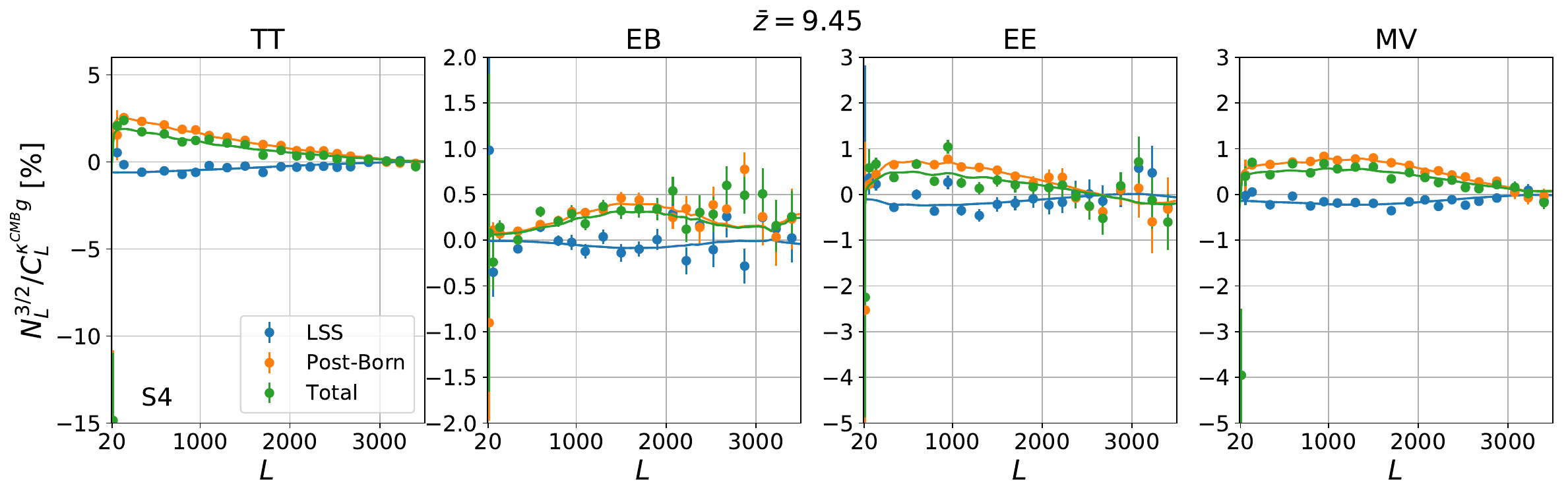}
\caption{Results for S4 lensing potential correlated with a density tracer, same as Fig.~\ref{fig:results_countsS4} but for high-redshift bins. The theoretical predictions model the non-linear matter bispectrum using tree-level prediction for $z>5$. At these redshifts the structure formation is mainly in the linear regime on relevant scales, so the main source of \nlth is post-Born effects. }
\label{results_countsS4highz}
\end{figure}

\noindent
 The increase in sensitivity between S4 and SO will boost the amplitude of the combined MV \nlth by a factor of 3 for all the reconstruction channels. However, this boost in the signal amplitude is mainly driven by the increased $\ell_{\textrm{max}}$ adopted for the lensing reconstruction in the TT estimator used for S4. As for the cross-correlation with $\phi_{\rm input}$ (and discussed further in Appendix~\ref{sec:foregrounds}), the temperature lensing estimator picks up contributions from very small-scale temperature modes which are sensitive to lensing on more non-linear scales, so inclusion of progressively smaller angular scales translates into an increase in amplitude of the \nlth. In the case of galaxy lensing at $z=1$ the \nlth bias on TT increases by a factor of 5 between SO and S4. The amplitude of the EB estimator conversely does not increase substantially even though the signal-to-noise of this channel improves significantly between SO and S4, and the EB channel also gets a larger weight in the minimum variance combination. As a result the \nlth increase in the combined MV channel is reduced.\\*
Figs.~\ref{fig:results_kappaSO} (lensing correlation) and \ref{fig:results_countsSO} (density cross-correlation) show the comparison between the theoretical predictions of \nlth discussed in Sec.~\ref{sec:quest} (and Appendix~\ref{sec:nonpert}) with simulation results for SO using the various different lensing reconstruction estimators. Figs.~\ref{fig:results_kappaS4}, \ref{fig:results_countsS4} show equivalent results for S4.
The agreement between theoretical predictions and simulations for the LSS term is influenced by the uncertainties in the theoretical modelling of the non-linear matter bispectrum, more so in the high signal-to-noise regime of S4 than for SO. There is also cosmic variance due the availability of a single realization of the non-linear matter field from the LSS simulation.
In addition to the overall lower theoretical uncertainties, the post-Born terms also agree better with the theoretical predictions because our measurement technique suppresses the bulk of the sample variance for this term.
The agreement in the worse cases is within $\simeq 25\%$ and most of the time better than 10\%. The quantitative details of the agreement do not depend strongly on the approach adopted for the non-linear modelling for SO sensitivity, in particular for the density cross-correlation. However, they are important for S4 and can generate differences in amplitude of about $50\%$ in the case of galaxy lensing cross-correlation using the very small-scale temperature. In this regime, simulation results often lie in between theoretical predictions computed with SC or GM formulae for the non-linear matter bispectrum, and lean towards one or the other approach depending on the redshift bin and on the reconstruction channel. Any improvement in accuracy in the modelling of the bispectrum could improve the accuracy of the \nlth bias prediction.
At $z\geq5$ the LSS contribution is strongly suppressed, as structures are mainly in the linear regime\footnote{We note that in the computation of \nlth we used the results of the tree-level LSS bispectrum for contributions at $z>5$.}, so the leading source of bias comes from post-Born effects alone as shown in Fig.~\ref{results_countsS4highz}.

\subsection{Detectability for future surveys}
To assess the importance of \nlth for future observations we compared the amplitude of the signal to the sensitivity expected from future date. We considered the cross-correlation power spectra between lensing from SO and S4 with an idealized galaxy survey covering the same sky fraction $ f_\mathrm{sky}=40\%$, such as Euclid or LSST. As our theoretical predictions did not incorporate details of systematic or theoretical uncertainties (such as the galaxy redshift distribution or their bias) we provide curves of the achievable detection significance $\sigma$
as a function of statistical uncertainties in galaxy lensing and galaxy density observations. For this purpose we assumed for each redshift bin a noise power spectrum for $C_L^{\kappa\kappa, z}$ and $C_L^{gg, \bar{z}}$ given by \cite{kilbinger2015}
\be
N_{\ell}^{\kappa\kappa} = \frac{\sigma_{\epsilon}^2}{2\bar{n}}, \qquad N_{\ell}^{gg} = \frac{1}{\bar{n}},
\ee
\noindent
where $\sigma_{\epsilon}\approx0.3$ is the standard deviation of the galaxy intrinsic ellipticity and $\bar{n}$ the average number of objects per arcmin$^2$ detected on the sky used to construct the galaxy density or the galaxy convergence maps in a given redshift bin. For LSST, assuming the redshift distribution of the so-called golden sample given in Refs.~\cite{lsst-science2, modi2017}, there is a total density of 50 (26) gal/arcmin$^2$ for galaxy density (lensing). Assuming a uniform distribution of objects over the 10 redshift bins as a baseline analysis (see e.g. \cite{ccl2019}), we obtain  $\bar{n} \approx 3\ (5)$ gal/arcmin$^2$ in each redshift bin. This roughly matches the shot noise expected for Euclid \cite{euclid}. We therefore take 4 gal/arcmin$^2$ as an indicative number for the shot-noise in those surveys at each redshift. This number depends on the width of the redshift bins, for example taking wider redshift bins such as those in Table \ref{table:deltabounds} would give $\bar{n} \approx 10$ for both Euclid and LSST at $z\geq 0.6$.
For a given LSS tracer, $X$, in a specific redshift bin we take
\be
    \sigma=\sqrt{\sum_b  \frac{(2 L_b +1) f_\mathrm{sky}\Delta L_b (N_{L_b}^{3/2,\kappacmb X})^2}{( C_{L_b}^{\kappacmb\kappacmb} +N_{L_b}^{\kappacmb\kappacmb}) \cdot ( C_{L_b}^{XX} + N_{L_b}^{XX} ) + (C_{L_b}^{\kappacmb X})^2}},
\label{eq:snr}
\ee
where $N_{L_b}^{3/2,\kappacmb X}$ are the binned measurements of \nlth, $\Delta L_b$ the width of a given bin $b$ of multipoles, and $L_b$ the centre of the multipole bin. We use \nlth and power spectra measured on the simulations described in Secs.~\ref{sec:sims} and \ref{sec:simcomparison}, including both LSS and post-Born contributions. In addition, we computed $C_{L_b}^{\kappacmb\kappacmb} +N_{L_b}^{\kappacmb\kappacmb}$ as the average of the autospectrum of the reconstructed CMB lensing maps of the $\kappa^G$ simulation described in Sec.~\ref{sec:sim-reconstruction} after applying the MC correction of Eq.~\eqref{eq:mcbias} in order to roughly account the major contributions to the variance \cite{kesden2003, Hu:2001kj}.
In writing Eq.~\eqref{eq:snr} we assumed a Gaussian covariance for all the auto and cross-spectra. This approximation is not expected to be accurate due to non-Gaussian contributions to the covariance of the galaxy lensing, density and small-scale $C_{L_b}^{\kappacmb\kappacmb}$ induced by the non-linear matter evolution. The reconstructed CMB lensing is also not Gaussian as the reconstruction noise is quadratic in the data. However, it is sufficient to get an indicative estimate of how important the signal could be. For low CMB noise levels it should be possible to improve on quadratic estimators sensitivity using more optimal maximum-likelihood iterative estimators \cite{hirata2003,carron2017} and, as such, the detection significance could also increase due to an improved signal-to-noise ratio. However, the behaviour of \nlth for this class of estimators might be different from the analysis presented here\footnote{To the extent that the iterative estimators provide more accurate estimates of the unlensed fields at each step to use in the quadratic estimate at that step, they could also be reducing the amplitude of the expected \nlth at each step.} and we thus leave further investigation for future work. \\*

\begin{figure}[!htbp]
\centering
\includegraphics[width=\textwidth]{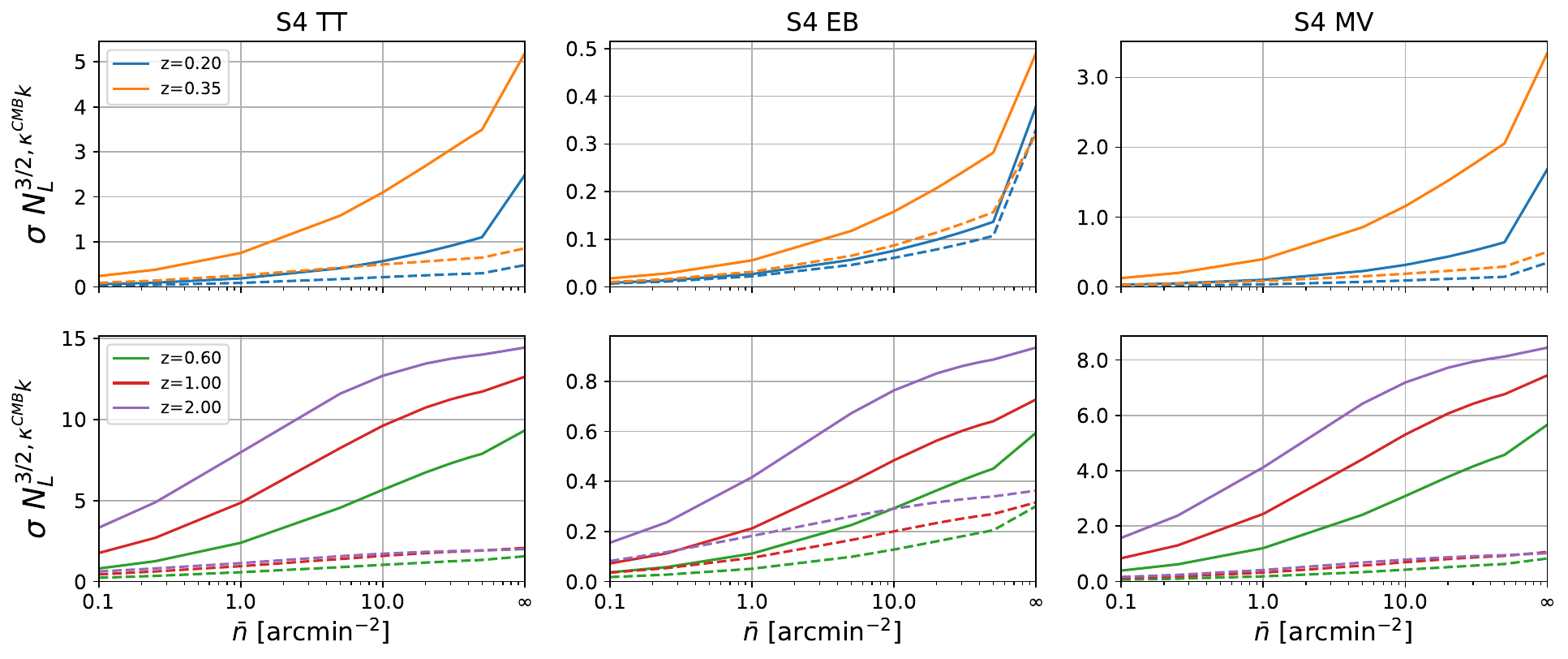}
\includegraphics[width=\textwidth]{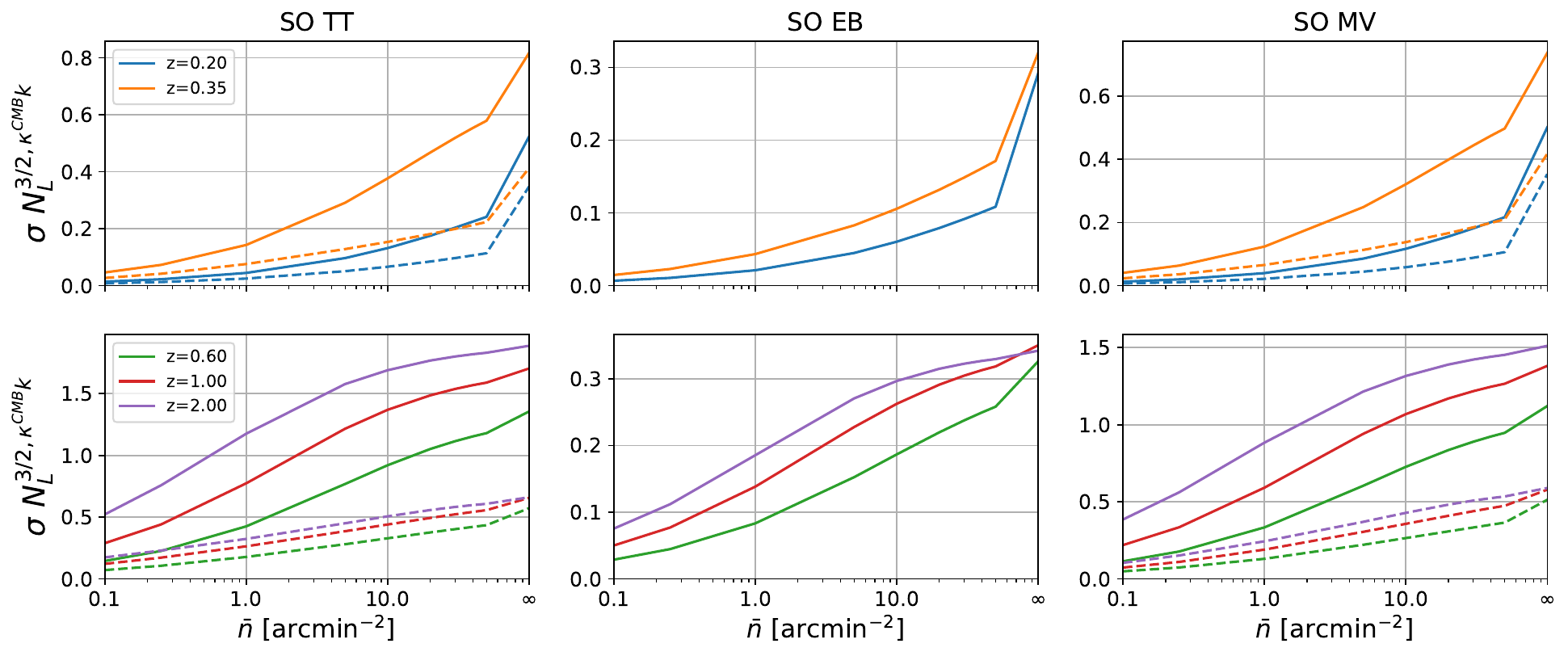}
\caption{Detection significance of \nlth measured in simulations for cross-correlation between the reconstructed CMB lensing and galaxy lensing as a function of the shot noise in an LSS survey (solid). Results for S4 (SO) are shown in the upper (lower) panels. LSST/Euclid-like surveys have $\bar{n}\approx 3$, depending on the bin thickness. Different reconstruction channels are shown from left to right, while different redshift bins are shown in different colours. The dashed lines show the detection significance $\sigma$ of the residual \nlth bias after subtraction of the analytical prediction of this work (using GM fitting formulae gives consistent results).}
\label{fig:s4snr-lensing}
\end{figure}

\begin{figure}[!htbp]
\centering
\includegraphics[width=\textwidth]{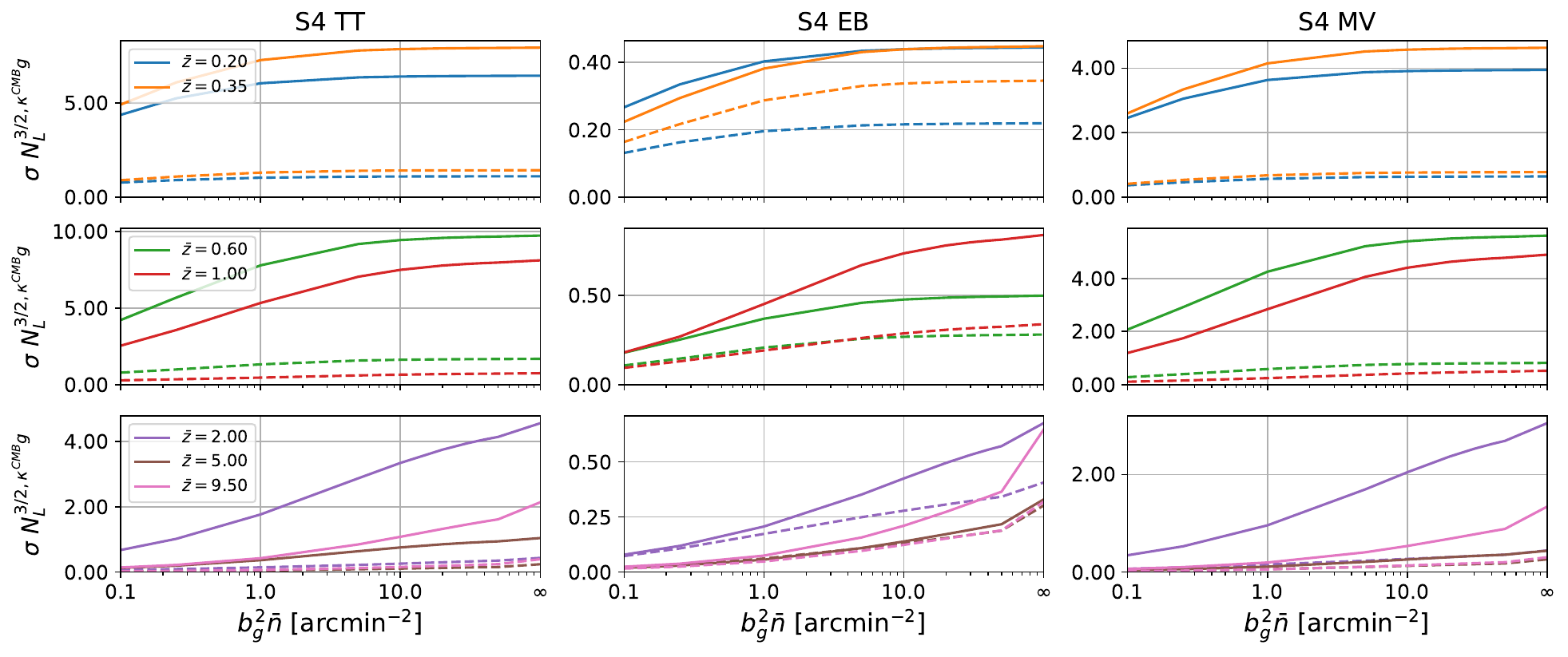}
\includegraphics[width=\textwidth]{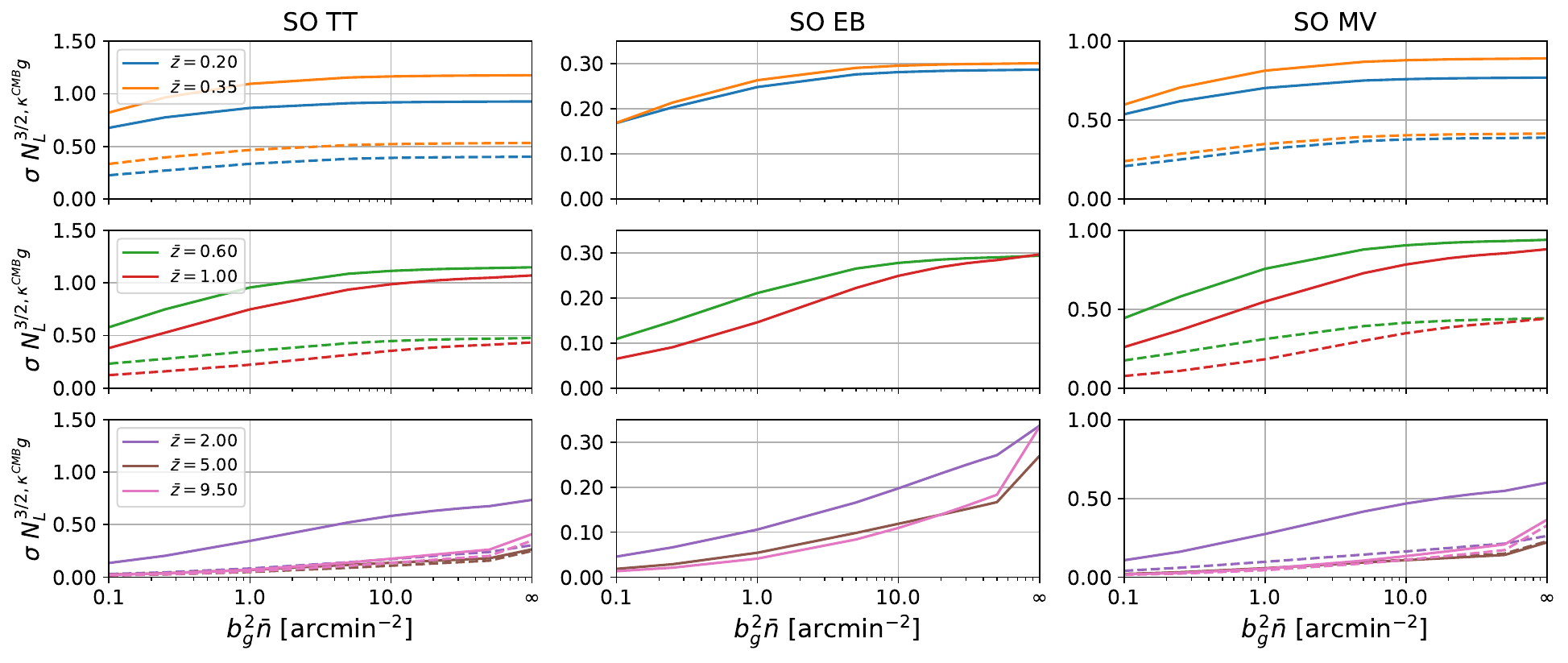}
\caption{Detection significance of \nlth measured in simulations for cross-correlation between the reconstructed CMB lensing and galaxy density as a function of the shot noise in an LSS survey (solid), assuming scale-independent constant galaxy bias $b_g$ in each bin. Results for S4 (SO) are shown in the upper (lower) panels. LSST/Euclid-like surveys have $\bar{n}\approx 4$, depending on the bin thickness. Different reconstruction channels are shown from left to right, while different redshift bins are shown in different colours. The dashed lines show the detection significance $\sigma$ of the residual \nlth bias after subtraction of the analytical predictions of this work using SC fitting formula (using GM fitting formulae gives consistent results).
}
\label{fig:s4snr-counts}
\end{figure}

In Figs.~\ref{fig:s4snr-lensing}, \ref{fig:s4snr-counts}  we show the expected detection significant of \nlth  from S4 and SO as a function of the number density in the survey, for cross-correlation with galaxy lensing and galaxy density respectively. In the case of SO MV reconstruction cross-correlated with galaxy lensing, the detection significance of \nlth will always stay below 1$\sigma$ for redshift bins at $z\lesssim 0.6$, but it would become significant for galaxy lensing at higher redshift, though with a strong dependence on the exact number density of the galaxy population used in the analysis. For for galaxy density cross-correlation the impact of the shot noise also depends on the bias of the galaxy sample; in Fig. \ref{fig:s4snr-counts} we show the significance as a function of $b_g^2 \bar n$, which is the combination of parameters that determines the detectability.
For $b_g^2 \bar{n}=4$ gal/arcmin$^2$, we have $\sigma\approx 1$ for all redshift bins at $z<2$. At S4 level of sensitivity and for the same shot noise level, the effect will be detected at $\approx 4\sigma$ in each redshift bin for cross-correlation with galaxy density at $z\leq 1$. These numbers reduce to $\approx 2\sigma$ in the case of cross-correlation with galaxy lensing, but could quickly rise if a lower shot noise level can be achieved in each redshift bin, in particular at $z\gtrsim 0.6$. Recently Ref.~\cite{wilson2019} suggested that Lyman break galaxies (LBG) at $z\gg2$ could be effectively used in the context of cosmological applications of cross-correlation with CMB lensing. Even considering an integrated redshift bin of these objects at $z\geq 3$ or, alternatively, radio-continuum surveys (e.g. SKA \cite{bonaldi2019}) \nlth effects at $z\geq 5$ would remain undetectable.
For both tracers and surveys \nlth will be unobservable for EB reconstruction, while the detection significance is enhanced for TT lensing reconstruction, where \nlth is higher, especially in the S4 case where we observe an increase of significance by a factor 2.
Differences between reconstruction estimators might therefore be an effective way to detect (and isolate) \nlth effects.\\*
The significance numbers quoted above represent an approximate lower-bound on the importance of \nlth at fixed cosmological and nuisance parameters. Even in the case of SO, where the detection significance is low in each redshift bin when considered independently, the significance of \nlth in a full combined analysis including multiple redshift bins would quickly increase above $2\sigma$. Moreover, for a full combined analysis including all the tracers in multiple redshift bins, as well as their cross-correlation with CMB lensing, the significance of \nlth could be substantially larger due to the effective large-scale cosmic-variance cancellation~\cite{cv-cancellation, schmittfull2018}.

\noindent
If \nlth is large enough to be detectable with an optimal estimator, it may also significantly contaminate other measurements if not carefully modelled.
In addition to potentially biasing cosmological parameters and bias measurements, it could also impact efforts to use cross correlations to tighten constraints on nuisance parameters describing instrumental systematics and contaminating signals. In the specific case of galaxy lensing measurements, the use of cross-correlations is emerging as standard practice to improve determination of nuisance parameters and minimize the impact of systematics \cite{chisari2015, desXC,Harnois-Deraps:2017kfd}. As a concrete example, we consider the case of intrinsic alignments (IA): in addition to the correlation with observed galaxy shapes due to gravitational lensing, the galaxy shapes can also be intrinsically aligned with the same local tidal field that contributes to the CMB lensing signal~\cite{Hirata:2004gc}. This manifests itself in an additive bias in the CMB lensing convergence cross-correlation spectra estimated from CMB lensing and galaxy shapes \cite{troxel2014,Hall:2014nja,larsen2016},
\begin{equation}
    \Delta \langle \hat{C}_L^{\kappacmb \kappa} \rangle =
     C_L^{\kappacmb I}.
\end{equation}
IA forms a non-zero bispectrum with CMB lensing that gives rise to an additional term to \nlth, but we neglect this effect in this work as it is a small correction to the main signal (see Appendix~\ref{app:postborn}).
$C_L^{\kappacmb I}$ can be modelled with the non-linear intrinsic alignment model (NLA) of \cite{Bridle:2007ft,larsen2016} using the non-linear matter power spectrum in Eq.~\eqref{eq:limbercls} with
\begin{equation}
    W_I(\chi) =  -\Aia \frac{2 C_1}{\chi^2 (1+z) D(z)} \frac{1}{n} \frac{dn}{dz} \frac{dz}{d\chi},
\label{eq:ia}
\end{equation}
where $D(z)$ is the linear growth function, $C_1 = 5 \cdot 10^{-14} h^{-2} M_\odot^{-1} \textrm{Mpc}^3$ following \cite{Bridle:2007ft} and $\Aia$ is an amplitude parameter usually constrained from the data through a joint fit with $C_L^{\kappa\kappa}$. As a working example, we derived the bias in the $\Aia$ parameter due to misidentification of the $N^{3/2}$-bias as intrinsic alignment using the likelihood
\begin{equation}
    - 2 \ln \mathcal{L} = \sum_L  \frac{(2 L +1) f_\mathrm{sky} \left(\hat{C}_L-C_L\right)^2}{\left( C_L^{\kappacmb\kappacmb} +N_L^{\kappacmb\kappacmb}\right) \cdot \left( C_L^{\kappa\kappa} + N_L^{\kappa\kappa}  \right) + \left( C_L^{\kappacmb \kappa} \right)^2} +\text{const}.
\end{equation}
To quantify the importance of modelling \nlth, we include it in the measured signal but neglect it in the assumed model, so that\footnote{As our theoretical predictions do not include any $dn/dz$ we assumed a Gaussian window function centred on our redshift bin with a width in redshift $\sigma_z = 0.01$ in the computation of Eq.~\eqref{eq:ia}. The results are practically insensitive to this parameter. }
\begin{align}
    &\text{Signal:} &
    \hat{C}_L&=C_L^{\kappacmb \kappa} + C_L^{\kappacmb I}\big\rvert_{\Aia=1} + N_L^{3/2} \\
    &\text{Model:} & C_L&=C_L^{\kappacmb \kappa} + C_L^{\kappacmb I}.
\end{align}
In Table~\ref{tab:ia-fit} we show the shift $\Delta \Aia$  in the best fit values of the intrinsic alignment amplitude $\Aia$ induced by \nlth for S4 sensitivity and a LSST/Euclid-like survey shot noise. For this setup we found that $\Delta \Aia$ is detected at $2$ to 4$\sigma$ significance when using minimum-variance lensing reconstruction for galaxy convergence at $z\gtrsim 0.6$. Going to higher redshifts, the intrinsic alignment power spectrum decreases more rapidly than the $N^{3/2}$-bias, so the apparent relative bias on the intrinsic alignment parameter increases. For polarization estimators, e.g. EB, \nlth is too low to be confused with an intrinsic alignment amplitude and thus the overall bias in the minimum-variance is hence driven by the temperature estimator. For SO sensitivity we never detect $\Delta \Aia$ even assuming no galaxy survey noise. Recent measurements of intrinsic alignment from DES constrained $\Aia$ for different galaxy populations and in different redshift bins between $0.2<z<1.3$ \cite{des-ia}. For a mixed sample of early and late type galaxies $\Aia\approx 0.5$ and constant in redshift, while using their best-fit model for the redshift evolution of $\Aia$ for early-type galaxies $\Aia \approx 3$--0.5 for the redshift bins we considered in our analysis. In light of these numbers the $\Delta \Aia$ we found could introduce significant errors in the fitting.

\begin{table}
\begin{center}
\begin{tabular}{ c|c|c|c|c|c}
\toprule
$\Delta \Aia$& $z=0.20$ & $z=0.35$ & $z=0.60$ & $z=1.00$ & $z=2.00$ \\
\hline
$TT$ & $ 0.02 \pm 0.07 $ &
$ 0.09 \pm 0.07 $ &
$ 0.27 \pm 0.07 $ &
$ 0.57 \pm 0.08 $ &
$ 1.08 \pm 0.11 $  \\
$EB$ & $ 0.00 \pm 0.7 $ &
$ 0.00 \pm 0.07 $ &
$ 0.01 \pm 0.07 $ &
$ 0.02 \pm 0.08 $ &
$ 0.05 \pm 0.10 $ \\
$MV$ & $ 0.01 \pm 0.06 $ &
$ 0.03 \pm 0.06 $ &
$ 0.10 \pm 0.06 $ &
$ 0.21 \pm 0.07 $ &
$ 0.42 \pm 0.08 $  \\
\bottomrule
\end{tabular}
\end{center}
\caption{Values of the bias in terms of galaxy intrinsic alignment amplitude $\Aia$ induced by an unmodelled \nlth assuming S4-like noise in the CMB and 3 gal/$\textrm{arcmin}^2$ in each redshift bin. The theory expectation including \nlth is calculated using the SC fit to the non-linear matter bispectrum, neglecting IA contributions to the bispectrum.}
\label{tab:ia-fit}
\end{table}
\noindent
Although the quantitative details of the detection significance and $\Aia$ estimates presented here depend on the choice of redshift bins of the LSS surveys and the approximations assumed, they clearly show that \nlth cannot be neglected in the data analysis of future experiments and will have to be taken into account. In addition, it will also be mandatory to correctly include post-Born corrections in simulations and analytical models, as neglecting the post-Born contribution would lead to an important misestimation of the size of \nlth. At S4 sensitivity neglecting post-Born contributions would increase the \nlth detection significance by 2$\sigma$ to 4$\sigma$ for cross-correlation with galaxy lensing, and by 3$\sigma$ for correlation with galaxy density if small-scale temperature is used. However, at SO sensitivity, neglecting the post-Born contribution to \nlth seems to be a good-enough approximation for tracers at $z\lesssim 0.6$ since the increase of detection significance would be marginal (less than $0.5\sigma$). Nevertheless, the analytic predictions we developed allow \nlth to be modelled with sufficient precision. The detection significance of the residual between our theoretical predictions and the simulation measurements is in fact always below $1\sigma$ for both S4 and SO, for all the tracers and at all redshifts even assuming a galaxy survey with no shot-noise (see dashed lines in Figs. \ref{fig:s4snr-counts}, \ref{fig:s4snr-lensing}\footnote{For both these figures we did not show the residual for the SO EB estimator since simulation results are noisy and consistent with no \nlth detection.}). As such no statistically significant effect of \nlth should be observed in a consistent analysis of future experiments. However, in the case where multiple redshift bins are combined using the full covariance for all the tracers, partial cosmic variance cancellation could make the residual more important and would deserve more detailed investigation.

\section{Conclusions}\label{sec:conclusions}

Forthcoming CMB observations will make CMB lensing maps at high signal to noise on large scales, and allow
powerful statistical measurements of the power spectrum and cross-power spectra to small scales.
For Gaussian fields, lensing cross-correlation estimators can be constructed that are exactly unbiased in a given fiducial model. However, the cross-correlation spectra are also sensitive to contributions from the bispectrum of large-scale structure and post-Born lensing, which we have calculated and simulated in detail. We have shown that the bispectrum contribution will be important for future data, and can potentially be detected at many sigma, depending on exactly what range of small-scale CMB temperature data are used. We showed that simple analytic models fit the results from full ray-traced simulations rather well, so that the error in modelling the signal analytically should be negligible for the foreseeable future. Future analyses must however include the \nlth bias model to obtain unbiased results.\\*
The size of the \nlth bias for cross-correlations depends on the redshift and tracer that is used.
For high-redshift lensing tracers there is a partial cancellation between the bispectrum bias from post-Born lensing and large-scale structure growth. Including both terms is therefore important to obtain correct results. If the effect is modelled via simulations, the simulations should self-consistently include the post-Born lensing effect, otherwise the bias may be overestimated.
At lower redshift the post-Born signal decreases due to the decreased path length and the large-scale structure bispectrum grows, so there is relatively less cancellation. However, the fractional bias decreases at very low redshift due to suppression by the CMB lensing window function, so the bias remains relatively small ($\alt 10\%$) for all redshifts, and therefore does not need to be modelled to high precision.\\*
In this first analysis we have made various unrealistic simplifying assumptions. Future work should study in more detail the contributions from non-linear bias, intrinsic alignments, and other contributions to the expected signals. The small-scale temperature will be dominated by foregrounds, and various kinds of foreground-mitigation strategies can be employed in addition to $\ell_{\rm max}$ cuts, so the bias will need to be calculated for the particular scheme that is adopted. In Appendix~\ref{app:fg} our first simple investigation suggests that foreground projection is also likely to remove a significant part of the contribution of very small scales to the \nlth bias, leaving a residual that must still be modelled but may be rather smaller.\\*
The presence of the \nth bias is also specific to the use of simple quadratic estimators. More optimal estimators, for example iterative maximum-likelihood~\cite{hirata2003,carron2017} or sampling-based methods~\cite{millea2017}, would behave differently: for low noise levels, optimal reconstruction methods are capable of reconstructing the realization of the lensing potential at high signal to noise almost independently of its statistics. Future work should however investigate the extent to which bispectrum biases remain in realistic situations with finite noise.

%%%%%%%%%%%%%%%%%%%%%%%%% Acknowledgements %%%%%%%%%%%%%%%%%%%%%%%%%
\section*{Acknowledgments}
We thank Julien Carron, Vanessa B\"ohm, Blake Sherwin, Emmanuel Schaan, Simone Ferraro, Emanuele Castorina and Mark Mirmelstein for discussion and comments. We thank Matteo Calabrese for his help in the lightcone reconstruction. We thank Carmelita Carbone for providing the DEMNUni simulations.
GF and AL acknowledge support from the European Research Council under
the European Union's Seventh Framework Programme (FP/2007-2013) / ERC Grant Agreement No. [616170] and
support by the UK STFC grant ST/P000525/1. GF also acknowledges the support of the CNES postdoctoral program during the early stage of the project. This research used resources of the National Energy Research Scientific Computing Center (NERSC), a DOE Office of Science User Facility supported by the Office of Science of the U.S. Department of Energy under Contract No. DE-AC02-05CH11231. The DEMNUni simulations were carried out on the CINECA Fermi supercomputer using the 5M CPUh budget awarded by the Italian SuperComputing Resource
Allocation (ISCRA) to the ``Dark Energy and Massive-Neutrino
Universe'' class-A proposal.
\appendix
\section{Perturbative \nlth calculation for cross-correlation spectra}\label{sec:Nthreetwo}

In the perturbative regime, the lensed CMB fields in harmonic space $\tilde X(\vl)$ can be obtained by series expansion in the deflection angle, giving
\begin{equation}
  \label{eq:20}
\tilde X(\vl) = X(\vl) + \delta X(\vl)+\delta^2 X(\vl) + \mathcal{O}(\phi^3),
\end{equation}
where in absence of primordial tensor modes \cite{hu2000}
\begin{eqnarray}
  \label{eq:deltaX}
  \delta X(\vl) &=& -\int_{\vl_1}\bar{X}(\vl_1)\phi(\vl-\vl_1)h_X(\vl_1,\vl)(\vl-\vl_1)\cdot \vl_1 \\
  \delta^2 X(\vl) &=& -\frac{1}{2}\int_{\vl_1,\vl_2}\bar{X}(\vl_1)\phi(\vl_2)\phi(\vl-\vl_1-\vl_2)h_X(\vl_1,\vl)(\vl_1\cdot\vl_2)\left[(\vl_1+\vl_2-\vl)\cdot\vl_1\right].
\end{eqnarray}
Here
\begin{equation}
  \label{eq:22}
  \bar{T}\equiv T,\qquad\bar{E}\equiv E,\qquad \bar{B} \equiv E,
\end{equation}
and
\begin{equation}
  \label{eq:hXDef}
  h_X(\vl_1,\vl) \equiv
  \begin{cases}
    1 & \mbox{if } X=T, \\
    \cos(2(\varphi_{\vl_1}-\varphi_\vl)) &  \mbox{if } X=E, \\
    \sin(2(\varphi_{\vl_1}-\varphi_\vl)) & \mbox{if } X=B,
  \end{cases}
\end{equation}
which satisfies $h_X(\vl_1,\vl)=h_X(-\vl_1,-\vl)$ and $h_X(\vl_1,\vl)=h_X(\vl_1,-\vl)=h_X(-\vl_1,\vl)$.\\*
The cross-correlation $\la\hat\phi\phi_\ext\ra$ between the reconstructed CMB lensing potential $\hat\phi$ and an external LSS tracer $\phi_\ext$ (as well as its auto-correlation) is sensitive to the presence of a non-zero bispectrum generated by nonlinear structure formation and post-Born lensing. Hereafter we assume that the external LSS tracer and CMB lensing potentials are uncorrelated with the unlensed CMB (i.e. we neglect correlation induced by the Integrated Sachs-Wolfe (ISW), SZ and other effects).
This should be a good approximation if the CMB maps have been adequately cleaned of non-primordial contributions.
%This is a good approximation as the effect involves mainly large scale that have a marginal contribution in the signal-to-noise of the lensing reconstruction.
We will also assume that there are no primordial B-modes, i.e. $B\(\vl\)=0$. The detailed shape of \nlth bias in cross-correlation depends on the quadratic combination used to reconstruct CMB lensing. Ref.~\cite{bohm2016} derived expressions for the TT estimator and we extend the calculation to the general reconstruction estimators below. We use the flat sky approximation for analytic results, which should be adequate on scales where there is enough signal to noise for the bias to be important. The leading effect of non-Gaussianity is expected to come from the bispectrum, and we neglect other higher-order correlators.\\*
The \nlth bias on the cross-power spectrum $C_L^{\hat\phi^{XY}\phi_\ext}$ will depend on terms of the form
\begin{eqnarray}
  \label{eq:17}
  \la\tilde X\tilde Y \phi_\ext\ra_{\mathcal{O}[(C^{\phi\phi})^{3/2}]} &=&
\la \delta X\delta Y\phi_\ext\ra + \la X\delta^2 Y\phi_\ext\ra + \la \delta^2 X Y\phi_\ext\ra.
\end{eqnarray}
Following Ref.~\cite{bohm2016} the first term on the right-hand side is called ``A1'', and the second two terms ``C1''.
These are all contractions allowed for the cross-spectrum, so that the full expression for the cross-spectrum including bispectrum corrections is
\begin{align}
  \label{eq:29}
\la C^{\hat\phi^{XY}\phi_\ext}_L\ra \approx C^{\phi\phi_\ext}_L + N^{(3/2)}_{\rm{A1}, XY}(L)
+ N^{(3/2)}_{\rm{C1}, XY}(L),
\end{align}
where the bispectrum-induced biases are
\begin{multline}
N^{(3/2)}_{\rm{A1}, XY}(L) = -A^{XY}_L  \int_{\vl_1}
B^{\phi_{\rm ext}\phi\phi}(L, l_1,l_3) \\
\times
\int_{\vl_2}
g_{XY}(\vl_2,\vL) C^{\bar X\bar Y}_{l_5} \vl_5\cdot \vl_1 \vl_5\cdot \vl_3
h_X(\vl_5,\vl_2)h_Y(\vl_5,\vl_4)
\end{multline}
and
\begin{multline}
N^{(3/2)}_{\rm{C1}, XY}(L) =\frac{A^{XY}_L}{2} \int_{\vl_1}
B^{\phi_{\rm ext}\phi\phi}(L, l_1,l_3) \\
\times
\int_{\vl_2}
\left(g_{XY}(\vl_2,\vL) C^{X\bar Y}_{l_2}h_Y(\vl_2,\vl_4) + g_{YX}(\vl_2,\vL) C^{\bar X Y}_{l_2} h_X(\vl_2,\vl_4)\right)  \vl_2\cdot \vl_1 \vl_2\cdot \vl_3.
\end{multline}
Here we defined $\vl_3 \equiv \vL-\vl_1$, $\vl_4\equiv\vL-\vl_2$, $\vl_5 \equiv \vl_1-\vl_2$, and
 $\bi_{\phi_\ext\phi\phi}$ is the mixed bispectrum between two CMB lensing modes and one external LSS tracer.
Fast methods for evaluating integrals of these forms were developed by Ref.~\cite{bohm2016}; however, since the inner integral is fast to evaluate, they can also be calculated directly quite easily.

\section{Non-perturbative \nlth calculation}
\label{sec:nonpert}

The bispectrum response expectation value in Eq.~\eqref{eq:bispectrumresponse} can be evaluated
using $\frac{\delta}{\delta\phi(\vl)} = \frac{\delta\valpha_a}{\delta\phi(\vl)}\nabla^a$, giving
\be
\frac{\delta \tilde{X}(\vl)}{\delta\phi(\vl_1)} = \frac{i}{(2\pi)^2}l_1^a
\widetilde{\nabla_a\! X}(\vl-\vl_1),
\qquad
\frac{\delta^2 \tilde{X}(\vl)}{\delta\phi(\vl_1)\delta\phi(\vl_3)}
= -\frac{1}{(2\pi)^4}l_1^a l_3^b \widetilde{\nabla_a\!\nabla_b\! X}(\vl-\vl_1-\vl_3),
\ee
where the tilde denotes the lensed quantity (i.e. the Fourier transform of the gradients at lensed positions $\vx + \valpha$).
For simplicity, we only consider the temperature in detail since that gives the largest bias for high-resolution observations. Defining
\be
\left\la\widetilde{\nabla_a\! T}(\vl) \widetilde{\nabla_b\! T}(\vl')\right\ra =
-\left\la\widetilde{\nabla_a\!\nabla_b\! T}(\vl) \widetilde{T}(\vl')\right\ra=
 (2\pi)^2\delta(\vl+\vl')\left[ l^2 \frac{\delta_{ab}}{2}\Cgg_l
+ l_{\la a}l_{ b\ra} \Cggtwo_l\right],
\ee
where angle brackets around indices denote the symmetric trace free part,
we have
\begin{multline}
 \left\la \frac{\delta^2 \left(\tilde{T}(\vl)\tilde{T}^*(\vl-\vL)\right)}
{ \delta\phi(\vl_1)\delta\phi(\vl_3)} \right\ra_G
= \frac{\delta(\vL+\vL')}{(2\pi)^2}\biggl[ \\
-\half\Cgg_{|\vl-\vl_1|}\vl_1\cdot \vl_3(\vl-\vl_1)^2 -
\Cggtwo_{|\vl-\vl_1|}\left( [(\vl-\vl_1)\cdot\vl_1 (\vl-\vl_1)\cdot \vl_3] -\half[\vl_1\cdot \vl_3  (\vl-\vl_1)^2]\right)
+ (1\leftrightarrow 3) \\
+\frac{l^2}{2}\Cgg_{l}\vl_1\cdot \vl_3  +
\Cggtwo_{l}\left( [\vl\cdot\vl_1 \vl\cdot \vl_3] -\frac{l^2}{2}[\vl_1\cdot \vl_3  ]\right) + (\vl \leftrightarrow \vL-\vl)
\biggr],
\end{multline}
where $\vl_3=-\vL'-\vl_1=\vL-\vl_1$.
From Eq.~\eqref{eq:bispectrumresponse} this then gives two non-perturbative $N^{3/2}$ contributions
\begin{multline}
N^{(3/2)}_{\rm{A1}, TT}(L) = -A^{TT}_L  \int_{\vl_1}
B^{\phi_{\rm ext}\phi\phi}(L, l_1,l_3)
  \int_{\vl_2} g_{TT}(\vl_2,\vL) \biggl(\Cggtwo_{l_5} \vl_5\cdot \vl_1 \vl_5\cdot \vl_3 \\
  +
  \frac{l_5^2}{2}[\Cgg_{l_5}-\Cggtwo_{l_5}] \vl_1\cdot\vl_3
  \biggr)
  \label{NBone}
\end{multline}
and
\begin{multline}
N^{(3/2)}_{\rm{C1}, TT}(L) = A^{TT}_L \int_{\vl_1}
B^{\phi_{\rm ext}\phi\phi}(L, l_1,l_3)
\int_{\vl_2}
 g_{TT}(\vl_2,\vL)\biggl( \Cggtwo_{l_2} \vl_2\cdot \vl_1 \vl_2\cdot \vl_3 \\
 +\frac{l_2^2}{2}[\Cgg_{l_2}-\Cggtwo_{l_2}]  \vl_1\cdot \vl_3
\biggr),
  \label{NBtwo}
\end{multline}
where (as in Appendix~\ref{sec:Nthreetwo}) $\vl_5 = \vl_1-\vl_2$.
\begin{figure}[t]
\centering
\includegraphics[width=.6\textwidth]{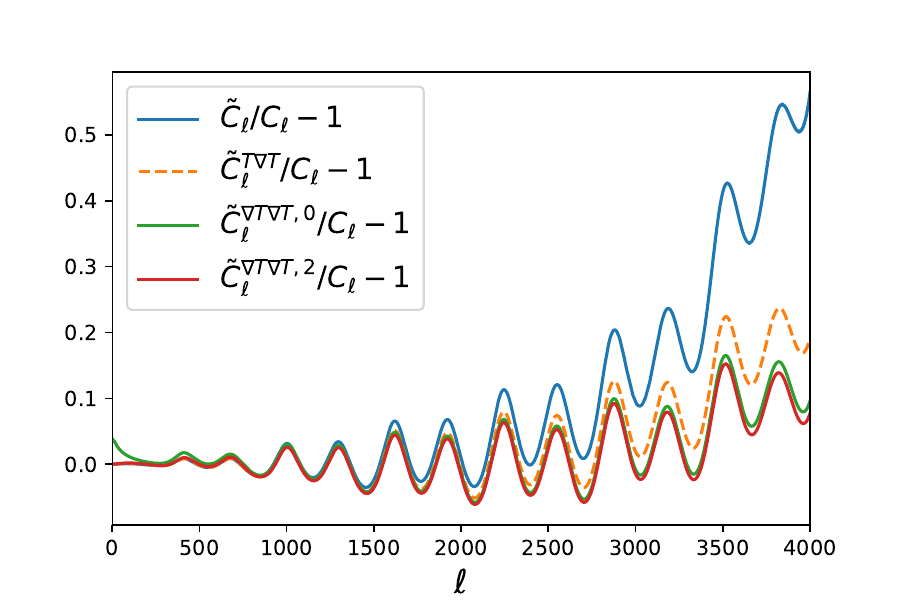}
\caption{The fractional difference between various flat-sky lensed temperature spectra compared to the unlensed spectrum: $\tilde C_\ell$ is the standard lensed spectrum; $\Cgrads_l$ is the gradient spectrum that appears in the non-perturbative mode responses; and $\Cgg_l$ and $\Cggtwo_l$ are the spectra that appear in the non-perturbative result for $N^{(3/2)}$.
The transfer of power from large scales to small scales due to lensing is substantially smaller for the various gradient spectra than for the temperature spectrum.}
\label{fig:Cgrads}
\end{figure}
\noindent
The corresponding power spectra can be evaluated in terms of correlation functions following Refs.~\cite{Seljak:1995ve,lewis2011}:
\begin{eqnarray}
l^2\Cgg_l &=&  2\pi\int r\ud r J_0(lr)\chi_0(r), \\
l^2\Cggtwo_l &=&  2\pi\int r\ud r J_2(lr)\chi_2(r),
\end{eqnarray}
where
\begin{eqnarray}
  \chi_0(r) &=& \int \frac{\ud l}{l} \frac{l^4 C_l^{TT}}{2\pi} e^{-l^2\sigma^2(r)/2}\left[J_0(lr) + \half l^2 \Cgltwo(r) J_2(lr) + \dots\right] \\
  \chi_2(r) &=&
  \int \frac{\ud l}{l} \frac{l^4 C_l^{TT}}{2\pi} e^{-l^2\sigma^2(r)/2}\left[J_2(lr) + \frac{1}{4}l^2 \Cgltwo(r) [J_0(lr) +
  J_4(lr)] +  \dots\right],
\end{eqnarray}
where the lensing correlation functions $\sigma^2(r)$ and $\Cgltwo(r)$ are defined as in~\cite{Seljak:1995ve,lewis2006}.
These results for the lensed gradient spectra have a similar form to the lensed power spectrum, but the lensing smoothing is effectively operating on $l^2C_l$ rather than $C_l$, which reduces the transfer of power to small scales: the lensed gradient spectra are larger than the unlensed spectrum on small scales, but somewhat smaller than the lensed spectrum (but well approximated by the lensed spectrum at $l\alt 2000$). See Fig.~\ref{fig:Cgrads}.
\noindent
In the leading perturbative limit in which $\Cgg_l, \Cggtwo_l \rightarrow C^{TT}_l$ the results of Eqs.~\eqref{NBone},~\eqref{NBtwo}
reduce to those in Sec.~\ref{sec:Nthreetwo} for $X=Y=T$. Numerically, the main difference to the leading perturbative result (where unlensed spectra appear in the integrals), is a $\alt 4\%$ correction to $N^{(3/2)}$. This is well captured by making the approximation $\Cgg_l\approx \Cggtwo_l$, in which limit the full result is the same as the perturbative result of Sec.~\ref{sec:Nthreetwo} with the substitution $C^{TT}_l \rightarrow \Cggtwo_l$. An almost equally good approximation is obtained by further taking $\Cggtwo_l \rightarrow \Cgrads_l$, the same lensed gradient spectrum that appears in the non-perturbative mode response. Since $N^{(3/2)}$ itself is quite small, it is not needed to high accuracy, and anyway there is significant uncertainty in the non-linear bispectrum modelling, so in practice it is likely to be a good enough approximation to simply replace all the unlensed spectra with the lensed gradient spectra.

\section{Cross-bispectrum}
\label{sec:bispectrum}
The cross-bispectrum between an external tracer and lensing has two contributions, from large-scale structure growth and post-Born lensing.

\subsection{LSS bispectrum}\label{appendix:lss}
The LSS bispectrum arises from non-linear evolution of the density perturbation $\delta$. Between three tracers $a$, $b$, $c$ with window functions $W^A$ so that
\be
A(\vnhat) = \int \ud \chi W^A(\chi) \delta(\vnhat\chi, z(\chi)),
\ee
the bispectrum is given in terms of the density bispectrum to lowest-order in the Limber approximation as~\cite{Takada:2003ef}
\be
\bi^{abc}(L_1,L_2,L_3) = \int \frac{\ud \chi}{\chi^4} W^a(\chi)W^b(\chi)W^c(\chi) B^{\delta\delta\delta}(k_1,k_2,k_3;z(\chi)).
\ee
For the tracer-lensing potential bispectrum that enters \nlth, we have  $\bi^{a \,\phi\, \phi}(L_1,L_2,L_3) = 4B^{a\, \kappacmb\, \kappacmb}(L_1,L_2,L_3)/(L_2^2L_3^2)$.\\*
The density bispectrum can be approximated by the tree-level result~\cite{Bernardeau:2001qr}:
\be
\bi^{\delta\delta\delta}(k_1, k_2, k_3;z) = 2F_2(\vk_1,\vk_2;z) P_{\delta\delta}(k_1,z(\chi))P_{\delta\delta}(k_2,z) + \text{cyc. perm.} ,
\label{eq:bdelta}
\ee
where
\be
F_2(\vk_1,\vk_2;z) = \frac{5}{7}A(k_1,k_2;z) + B(k_1,k_2;z)\frac{\vk_1\cdot\vk_2}{2k_1k_2}\left(\frac{k_1}{k_2}+\frac{k_2}{k_1}\right)
+ C(k_1,k_2;z)\frac{2}{7} \frac{(\vk_1\cdot\vk_2)^2}{k_1^2 k_2^2}.
\label{eq:Ftwo}
\ee
The baseline tree-level result has $A=B=C=1$ in Eq.~\eqref{eq:Ftwo}, so that $F_2$ is independent of redshift. We use the more accurate result by using the non-linear $P_{\delta\delta}$ from Halofit rather than the linear one~\cite{Scoccimarro:2000ee}.
We also use the
extended fitting formulae for $A$, $B$ and $C$ from Ref.~\cite{GilMarin:2011ik} (``GM''), which are calibrated from simulations at low redshift relevant for most cross-correlation tracers.

\subsection{Post-Born bispectrum}
\label{app:postborn}
We follow the derivation and results of Ref.~\cite{pratten2016}, which provides the second-order post-Born correction to the lensing convergence from a source at $\chi_s$ given by
\be
\kappa^{(2)}(\vL) = -2 \int_0^{\chi_s}  d\chi W(\chi,\chi_s)\int_0^\chi d\chi' W(\chi',\chi) \int_{\vL'}
%\frac{d^2 {\vL'}}{(2\pi)^2}
\,\vL'\cdot \vL \, \vL'\cdot(\vL-\vL') \Psi(\vL',\chi)\Psi(\vL-\vL',\chi'),
\label{eq:distortiontwo}
\ee
where $W(\chi,\chi_s) = W_\kappa(\chi,\chi_s)/(\gamma\chi^2) = (1/\chi-1/\chi_s)\Theta(\chi_s-\chi)$.
In the lowest-order Limber approximation the power spectrum of the two fields is given by \cite{Kaiser92}
\begin{align}
\label{eqn:LimberPS}
\left\langle {A} \left( \vL ; \chi \right)  {B} \left( \vL^{\prime} ; \chi^{\prime} \right) \right\rangle &= \left( 2 \pi \right)^2 \, \delta \left( \vL + \vL^{\prime} \right) \, \frac{\delta_D \left( \chi - \chi^{\prime} \right)}{\chi^2} \, P_{AB} \left( \frac{L}{\chi} , z(\chi) \right) .
\end{align}
Contracting Eq.~\eqref{eq:distortiontwo} with the linear expression for two other sources $a_1$ and $a_2$ then gives
\begin{align}
\bi^{a_1^{(1)}a_2^{(1)} \kappa_s^{(2)} }_{L_1 L_2 L_3}  &=  2 \, \left[ \vL_1 \cdot \vL_2 \right] \int d\chi \frac{W(\chi,\chi_s)}{\chi^2} \int_0^{\chi}d\chi' \frac{W(\chi',\chi)}{{\chi'}^2} \nonumber \\
&\,\times \left[  \left( \vL_1 \cdot \vL_3 \right) W^{a_1}(\chi)W^{a_2}(\chi') P_{\delta\Psi}\left(\frac{L_1}{\chi},z(\chi)\right) P_{\delta\Psi}\left(\frac{L_2}{\chi'},z(\chi')\right) + (1 \leftrightarrow 2) \right]\\
&=
 2 \frac{\vL_1 \cdot \vL_2 }{L_1^2 L_2^2} \left[\vL_1\cdot \vL_3 \Mkappa^{a_1a_2\kappa_s}(L_1,L_2) + \vL_2 \cdot \vL_3 \Mkappa^{a_2a_1\kappa_s}(L_2,L_1)\right],
\label{eq:kkz}
\end{align}
where we defined
\begin{align}
C^{a\kappa}_L(\chi)=  \int d \chi' \, \frac{W^a (\chi') W_\kappa(\chi',\chi)}{(\chi')^2} \, P_{\delta \delta} \left( \frac{L}{\chi'} , z(\chi') \right),
\end{align}
\be
\Mkappa^{a_1a_2\kappa_s}(L,L') \equiv  \int d\chi \frac{W^{a_1}(\chi)W_\kappa(\chi, \chi_s)}{\chi^2} P_{\delta\delta}\left(\frac{L}{\chi}, z(\chi)\right) C_{L'}^{a_2\kappa}(\chi).
\ee
Hence the post-Born bispectrum for three lensing sources is
\be
\bi^{\kappa_1\kappa_2\kappa_3}_{L_1,L_2,L_3} =  2 \frac{\vL_1 \cdot \vL_2 }{L_1^2 L_2^2} \left[\vL_1\cdot \vL_3 \Mkappa^{\kappa_1\kappa_2\kappa_3}_s(L_1,L_2) + \vL_2 \cdot \vL_3 \Mkappa_s^{\kappa_2\kappa_1\kappa_3}(L_2,L_1)\right]  + (2\leftrightarrow 3) + (1\leftrightarrow 3).
\ee
If $\kappa_2=\kappa_3 =\kappa_{\rm CMB}=\kappa$ and $\kappa_1$ is low redshift then $\Mkappa^{\kappa \kappa_1 \kappa}_s(L,L') \gg \Mkappa^{\kappa_1\kappa  \kappa}(L,L') = \Mkappa^{\kappa \kappa  \kappa_1}(L,L')$ since the integral range is much larger (corresponding to e.g. high-$\ell$ CMB lens - low-$\ell$ density coupling).
If the convergence is measured by galaxy shear, there will be additional contributions to the shear-CMB lensing bispectrum from intrinsic alignments, however in the simplest models this is expected to be a small correction~\cite{merkel2017} and we do not consider it further in this paper.\\*
% the bispectrum can then be approximated by only keeping these terms (but doesn't look great at $z=2$).
%probably only the $2 \leftrightarrow 3$ perm is needed, since PB is small on low-redshift galaxy lensing.
%
In general the observed angular density of galaxies has many contributions, e.g. from magnification bias and redshift distortions, plus velocity and GR effects. For simplicity we only consider the dominant density source so that lensing only affects the observed density of galaxies via the deflection to the observed galaxy positions. To second order the observed density perturbation is then
\be
\delta(\vtheta) = \delta(\vtheta+\nabla \phi) \approx   \delta(\vtheta) + \nabla_a \phi \nabla^a \delta(\vtheta).
\ee
The second-order correction to the density from lensing is then
\be
\delta^{(2)}(\vL;\chi) = 2 \int_0^\chi \ud \chi' W(\chi',\chi) \int_{\vL'} \vL'\cdot(\vL-\vL') \Psi(\vL',\chi')\delta(\vL-\vL',\chi).
\ee
Integrating over a source redshift window $W^{g}(\chi)$ and contracting with two linear sources then gives
\begin{align}
\bi^{g^{(2)}a_2^{(1)} a_3^{(1)} }_{L_1 L_2 L_3}  &=
2 [\vL_2\cdot \vL_3]\int \ud \chi \frac{W^g(\chi)}{\chi^2}\int_0^{\chi} \ud \chi' \frac{W(\chi',\chi)}{(\chi')^2}
\nonumber\\
&\times \left[
W^{a_2}(\chi')W^{a_3}(\chi) P_{\delta\Psi}\left(\frac{L_2}{\chi'},z(\chi')\right) P_{\delta\delta}\left(\frac{L_3}{\chi},z(\chi)\right) + (2\leftrightarrow 3)
\right] \\
&= -2 \frac{\vL_2 \cdot \vL_3 }{L_2^2 L_3^2} \left[L_3^2\Mkappa^{ga_2a_3}(L_3,L_2) + L_2^2 \Mkappa^{ga_3a_2}(L_2,L_3)\right].
\end{align}
The corresponding post-Born bispectrum for a non-lensing tracer with two $\kappa$ is then
\begin{multline}
\bi^{g\kappa_2\kappa_3}_{L_1,L_2,L_3} =  2 \frac{\vL_1 \cdot \vL_2 }{L_1^2 L_2^2} \left[\vL_1\cdot \vL_3 \Mkappa^{g\kappa_2\kappa_3}(L_1,L_2) + \vL_2 \cdot \vL_3 \Mkappa^{\kappa_2g\kappa_3}(L_2,L_1)\right] + (2\leftrightarrow 3) \\
 -2 \frac{\vL_2 \cdot \vL_3 }{L_2^2 L_3^2} \left[L_3^2\Mkappa^{g\kappa_2\kappa_3}(L_3,L_2) + L_2^2 \Mkappa^{g\kappa_3\kappa_2}(L_2,L_3)\right].
\end{multline}
The last term isn't really ``post-Born'' in the lensing sense, since it comes from Born lensing of the density distribution. However, it should be included for consistency, so that the observed bispectrum does not have sensitivity to large unobservable constant lensing deflections between us and the source plane.

\section{Foreground and bias-mitigated estimators}\label{app:fg}

The \nlth bias is most important for lensing reconstruction using the very small-scale CMB temperature. However these modes are also strongly contaminated by a variety of foregrounds, so lensing reconstruction requires very thorough foreground cleaning, detailed foreground modelling in the simulation and fiducial model, or use of estimators that are less sensitive to foregrounds and hence more robust. One of the latter options is ``bias-hardening''~\cite{namikawa2013}: using a lensing estimator that projects out the response to a variety of possible sources of contamination, for example point sources~\cite{namikawa2013} or a spatially modulating amplitude of a foreground with a known spectrum~\cite{planck-lensing2018}. If the source being projected only has a small overlap with lensing, the signal to noise is not substantially reduced, however bias hardening does depend on knowing what needs projecting. Ref.~\cite{Schaan:2018tup} instead consider a multipole decomposition on the angular dependence of the modes used for lensing reconstruction, equivalent to a
magnification/shear decomposition for reconstruction of low-$L$ lensing modes. This is similar to bias-hardening to the monopole at each scale, at the cost of increasing the reconstruction noise relative to bias hardening to foregrounds with a particular spectrum. We summarize these in turn, and then present the results for the \nlth bias, showing that it is substantially reduced after foreground mitigation. Since foregrounds are most important for the temperature, for simplicity we only consider temperature lensing reconstruction here.

\begin{figure}[t]
\centering
\includegraphics[width=.9\textwidth]{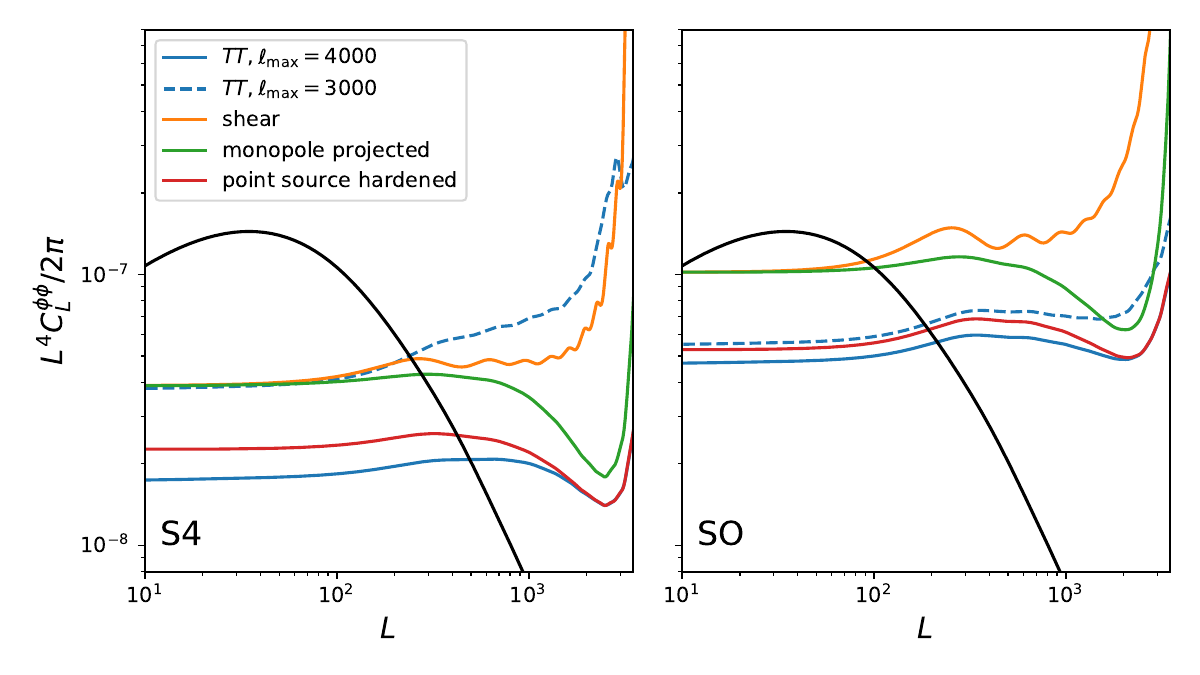}
\caption{Flat-sky temperature lensing reconstruction noise for S4-like observations (left) and Simons Observatory (right, including foreground residual noise) for $\lmax=3000$ and $\lmax=4000$ compared to the various foreground bias-mitigated estimators discussed in the text with $\lmax=4000$. The black line shows the fiducial theory curve for comparison. }
\label{fig:foregroundN0}
\end{figure}

\subsection{Monopole-projected and shear estimators}
Consider circularly-symmetric sources with a normalized radial profiles $I(r)$ (or a large number of sources which average to give this same radial profile). On the flat sky the total signal is
\be
T_f(\vl) = \int \ud c A(\vc) I(l) e^{-i\vl\cdot \vc},
\ee
%at from position $\vc$ has $A_j(\vc) I_j(\vl) = e^{-i\vl\cdot \vc} A_j(\vc)I_i(l)$
where $I(l)$ is the Bessel transform of $I(r)$ and $A(\vc)$ is the amplitude of the source at position $\vc$.
If the sources are uncorrelated and spatially unclustered on the sky, we then have
\be
\label{sourceresponse}
\la T_f(\vl_2)T_f(\vl_3)\ra \propto f(\vL) I(l_2)I(l_3),
\ee
where $f(\vx) = A^2(\vx)$ measures the relevant spatial variation in source power and $\vL=\vl_2+\vl_3$.
%(c.f.~\cite{Osborne:2013nna}).
The sources therefore contaminate the lensing quadratic estimator with
  \begin{multline}
    \Delta \hat\phi (\VL) = A_L^{TT} \int_{\vl_2} g_{TT}(\vl_2,\VL) \la T_f(\vl_2) T^*_f(\vl_2-\vL)\ra
    \propto  f(\vL) \int_{\vl_2} g_{TT}(\vl_2,\vL)  I(l_2)I(|\vL-\vl_2|).
  \end{multline}
\noindent
The foregrounds are mainly important at high CMB multipoles $l\agt 2000$, and for lensing reconstruction, most of the lensing signal is at lensing multipoles $L\alt 2000$. This motivates considering $L \ll l_2, l_3$, so that if $I(l)$ is a smooth function of $l$,
we have $I(l_2)I(l_3)\approx I(\ell)^2 + \clo(L^2/\ell^2)$ (and is exact in the point source limit, where $I(l)=\text{const}$). Here we have defined $\vl \equiv (\vl_2-\vl_3)/2$ (where $\vl_2+\vl_3=\vL$). We can further define the angle $\phi\equiv \phi_{\vl L}$ as the angle between $\vl$ and $L$, which determines the relative orientation of the small-scale temperature modes and the larger-scale lensing modes, so that to quadratic order in
$L/\ell$
%Note that we \emph{cannot} use this series approximation for terms involving $C_l$ at $l\gg 100$, because the $C_l$ are not smooth over the scale of the acoustic peaks (except possibly in the far damping tail).
%\approx f(\vL) \int l\ud l I(l)^2 \int \d\phi g(\vl_2,\vl_3).
%;for a review see~\cite{Lewis:2011au})
\be
  \Delta \hat\phi (\VL) \propto f(\vL) \int l\ud l I(l)^2 \int \d\phi g_{TT}(\vl+\vL/2,\vL).
  \label{deltaPhi}
\ee
The weight function $g$ can be expanded in multipole moments
\be
g_{TT}(\vl+\vL/2,\vL) = \sum_n g^{(n)}(l,L)\cos(2n\phi)
\ee
for integer $n$, so that the foreground contamination in Eq.~\eqref{deltaPhi} only comes from the monopole ($n=0$) term (for a review of this kind of decomposition for bispectrum triangles see Ref.~\cite{Lewis:2011au}).
We can therefore define a monopole-projected quadratic estimator that is fairly immune to the foregrounds of the assumed form by using a modified weight function
\be
g^{\rm proj.}(\vl+\vL/2,\vL) \equiv g_{TT}(\vl+\vL/2,\vL) - g^{(0)}(l,L).
\ee
We call the corresponding lensing quadratic estimator the monopole-projected estimator.
%\int \frac{\ud \phi}{2\pi} g(\vl_2,\vl_3)
In the squeezed limit $L\ll l_2,l_3$ and $L \alt 200$, this amounts to using the shear-only estimator with~\cite{Schaan:2018tup,Prince:2017sms,Bucher:2010iv}
\be
g^{\rm shear}(\vl,\vL) = \cos(2\phi_{\vL\vl})
\frac{\Cgrads_l}{2 \Cexpt_l \Cexpt_{|\vL-\vl|} } \frac{\ud \ln \Cgrads_l}{\ud \ln l}.
\ee
The shear estimator has the advantage of being separable, and hence easy to evaluate in practice\footnote{Ref.~\cite{Schaan:2018tup} use $(\Cexpt_l)^2$ in the denominator, however on real data the first step is usually to inverse-variance filter the data (applying multipole cuts, and accounting for masks and possibly anisotropic noise), so we use a form consistent with that. We also use the gradient power spectrum $\Cgrads_l$ that appears in non-perturbative response as described in Sec.~\ref{sec:nonpert}.}, though out of the squeezed limit it is no longer equivalent to monopole projection and can become significantly suboptimal.
Example reconstruction noise curves are shown in Fig.~\ref{fig:foregroundN0}.

\subsection{Point source bias-hardened estimator}
The un-normalized estimator
\begin{equation}
\hat{g}^a(\VL) \equiv \int_\vl g^a(\vl,\VL)\tilde T_\expt(\vl)\tilde{T}^*_\expt(\vl-\VL),
\end{equation}
for some weight function $g^a(\vl,\VL)$, has an estimator response to signal $b(\vL)$ given by
\be
\response^{ab}_L = \int_\vl g^a(\vl,\VL) f^b(\vl,\vL-\vl),
\ee
where the normalization is $A_L^a = (\response_L^{aa})^{-1}$ and the signal response is given by
\be
\left\la\frac{\delta}{\delta b(\vL)}\left(\tilde T(\vl_1) \tilde{T}(\vl_2)\right)\right\ra = \delta(\vl_1+\vl_2-\vL) f^b(\vl_1,\vl_2).
\ee
The bias-hardened lensing estimator is then constructed using weight functions~\cite{namikawa2013}:
\be
g^\hard(\vl,\VL) \equiv g^{\phi}(\vl,\VL) - \frac{\response^{\phi b}_L}{\response^{bb}_L}g^{b}(\vl,\VL),
\ee
so that taking the expectation for fixed $b$ gives $\la \hat{g}^\hard(\VL)\ra_b = 0$.
For unclustered point sources $I_l =\text{const}$ and hence from Eq.~\eqref{sourceresponse}
the minimum variance weight function is then simply
$$
g^{\rm pt.src.}(\vl,\vL) = \frac{1}{\Cexpt_l \Cexpt_{|\vL-\vl|} }.
$$
This bias-hardened estimator is also separable and easy to apply in practice. It has substantially lower reconstruction noise than the monopole-projected estimator (see Fig.~\ref{fig:foregroundN0}), but only projects unclustered point sources (though the generalization to other spectra is straightforward).

\subsection{\nlth for foreground-mitigated estimators}
\label{sec:foregrounds}

\begin{figure}[t]
\centering
\includegraphics[width=.9\textwidth]{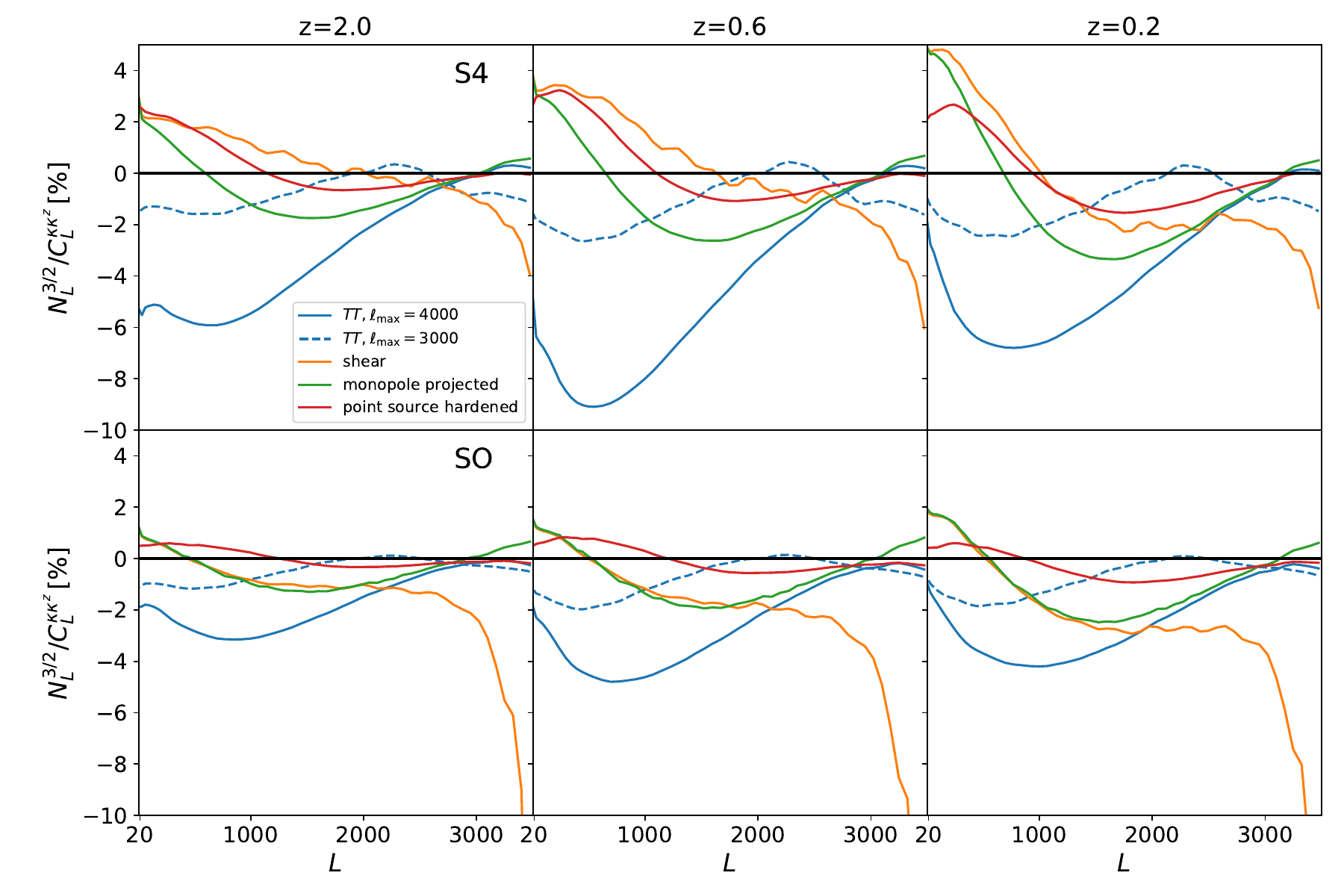}
\caption{Fractional \nlth biases for the cross-correlation of CMB lensing $TT$ reconstruction with lensing at redshifts, $z=2$, $0.6$, and $0.2$ for an idealized S4-like observation (top panel) and Simons Observatory (lower panel, foreground-cleaned maps). Blue lines are the standard minimum variance quadratic estimators with different $\ell_{\rm max}$, and other lines show the foreground-mitigated estimators described in the text with $\ell_{\rm max}=4000$. The shear estimator bias blows up because its response crosses zero on small scales. }
\label{fig:foregrounds}
\end{figure}

Fig.~\ref{fig:foregrounds} shows the fractional \nlth bias for the cross correlation of lensing with the CMB temperature lensing quadratic estimators described above, compared to the standard estimator with two $\ell_{\rm max}$ cuts (neglecting any foreground residual contribution to the effective noise for S4). The estimators designed to project out foregrounds also project out some of the signal responsible for the \nlth bias, significantly reducing its relative amplitude, suggesting that the bias in the temperature estimators is in practice not substantially more problematic than the bias from polarization.

\section{Comparison of convergence skewness}
\label{app:skewness}
\begin{figure}[htb]
\centering
\includegraphics[width=\textwidth]{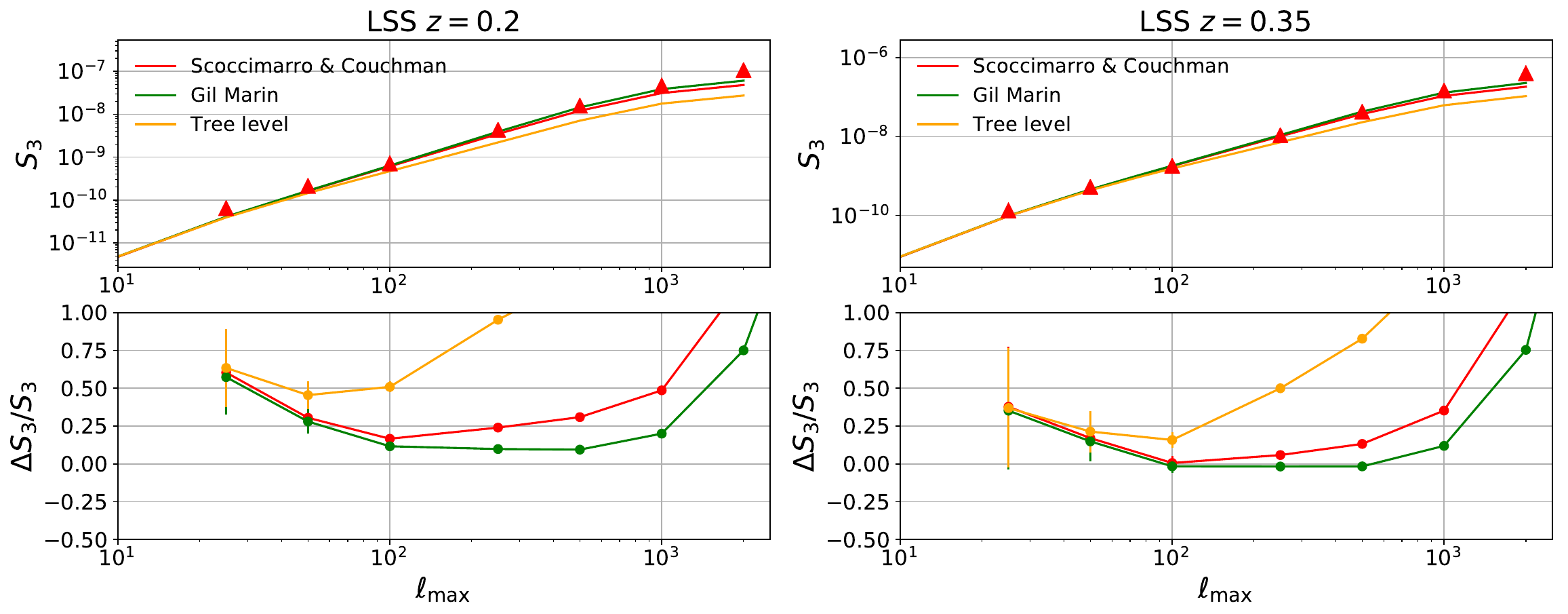}
\caption{Top: Skewness of the simulated $\kappa_z$ fields induced by non-linear evolution (triangles) compared to analytic predictions using the different fitting formulae for the non-linear matter bispectrum (solid lines, as described in the legends) for $z=0.2$ (left) and $z=0.35$ (right). Bottom: relative difference between simulation and analytic results. The error bar includes the sample variance as in Eq~\ref{eq:vars3}.}
\label{fig:skewness-lowz}
\end{figure}

\begin{figure}[htb]
\centering
\includegraphics[width=\textwidth]{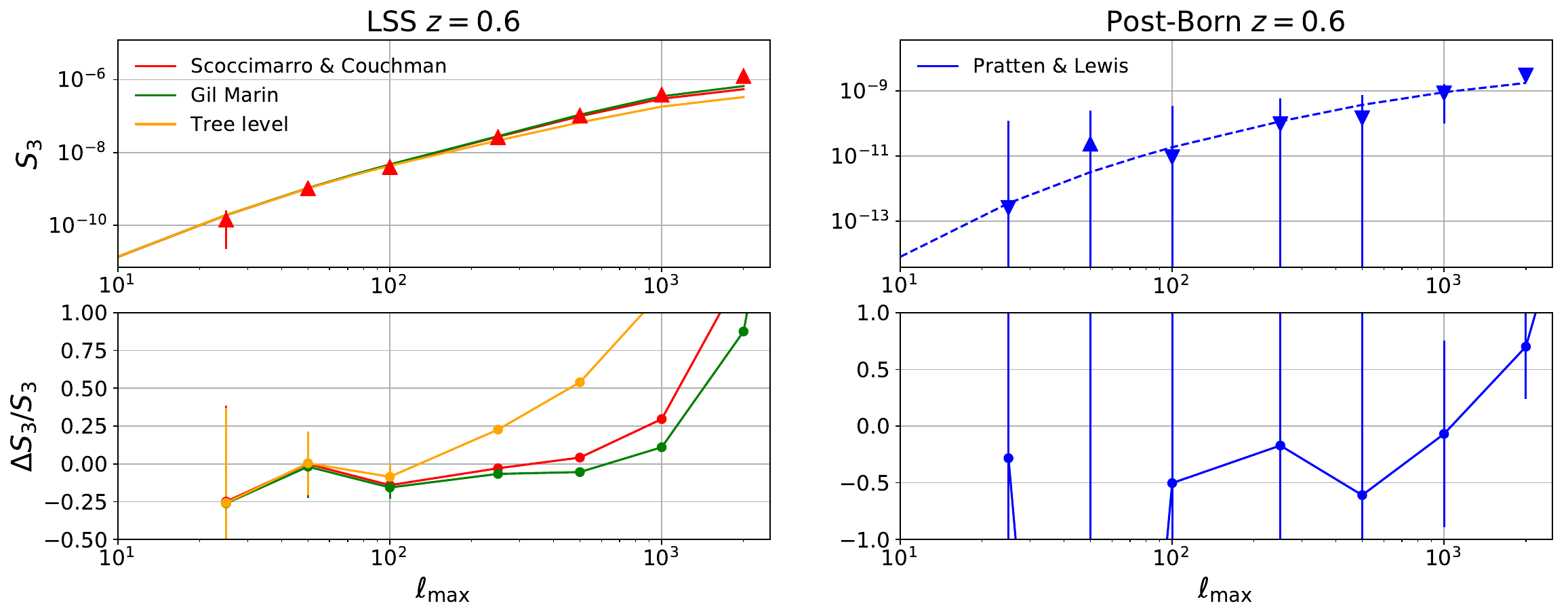}\\
\includegraphics[width=\textwidth]{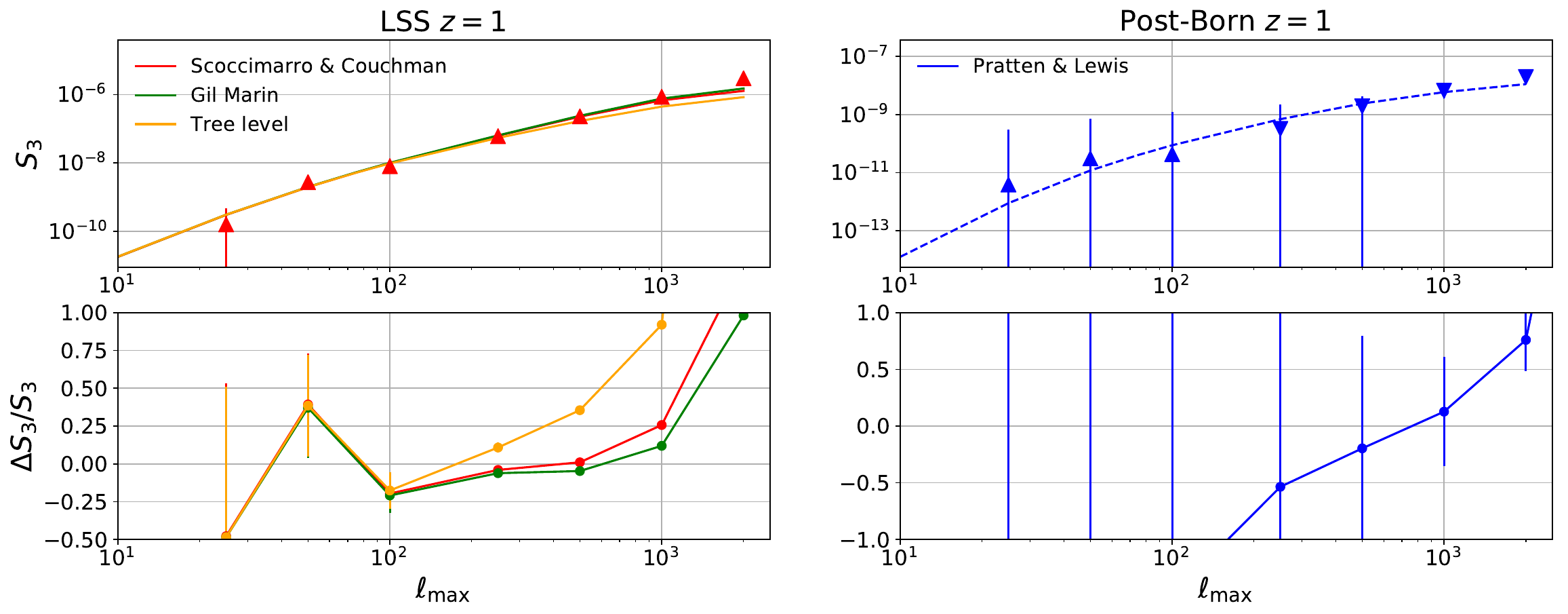}\\
\includegraphics[width=\textwidth]{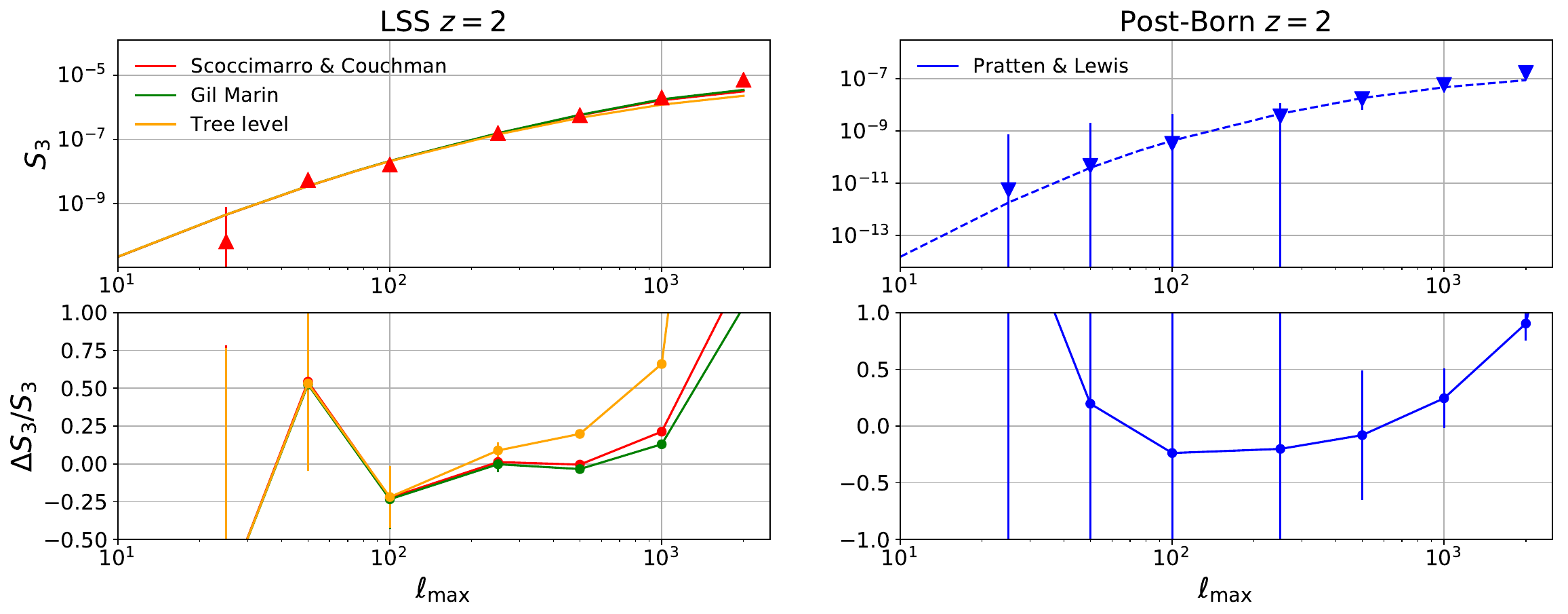}
\caption{Left: same as Fig.~\ref{fig:skewness-lowz} for $z=0.6$ (top), $z=1$ (middle) and $z=2$ (bottom). Right: post-Born induced skewness (top) on $\kappa^{z,{\rm ML}}$ compared with analytical predictions of \cite{pratten2016} (solid line). Reverse triangle and dashed lines denote negative values. The fractional difference between simulation and analytic results is shown in the bottom panel. }
\label{fig:skewness-highz}
\end{figure}

Since \nlth effect we want to investigate with the simulations depends on the non-Gaussian statistics of the observables, we tested if the overall level of non-Gaussianity present in the simulated convergence maps was consistent with theoretical expectations by focussing on the skewness of the convergence maps. The appearance of a skewness in the pdf of tracers of the matter fluctuations is a direct consequence of the non-linear gravitational evolution, as densities can grow arbitrarily large, while under dense regions can never have less than zero mass.\\*
The skewness $S_3[X]=\left\langle XXX \right\rangle$ of a pixelized map of a generic scalar field $X$ can be estimated using
\begin{equation}
\hat S_3[X]= \frac{1}{N_\textrm{pix}}\sum_p^{N_\textrm{pix}} X_p^3,
\end{equation}
where $p$ is the pixel index and $N_\textrm{pix}$ the total number of pixels in the map.
Following \cite{Srednicki,komatsu2001} the skewness can be expressed in terms of the reduced bispectrum $b^{XXX}_{L_1 L_2 L_3}$ of the $X$ field as
\begin{align}
S_3[b^{XXX}_{L_1 L_2 L_3}]&=
\sum_{L_1 L_2 L_3}^{L_\textrm{max}}  \frac{(2 L_1+1)(2L_2+1)(2L_3+1)}{(4\pi)^2}
\begin{pmatrix}
L_1 & L_2 & L_3 \\
0 & 0 & 0
\end{pmatrix}^2
b^{XXX}_{L_1 L_2 L_3},
\end{align}
\noindent
with the variance of the estimator dominated by the disconnected six-point function
\begin{align}
\sigma_{S_3}^2\simeq \frac{6}{4\pi} \sum_{L_1 L_2 L_3}^{L_\textrm{max}} \frac{(2 L_1+1)(2 L_2+1)(2 L_3+1)}{(4\pi)^2}
\begin{pmatrix}
L_1 & L_2 & L_3 \\
0 & 0 & 0
\end{pmatrix}^2
C^{XX}_{L_1}C^{XX}_{L_2}C^{XX}_{L_3}.
\label{eq:vars3}
\end{align}
Here $L_\textrm{max}$ represents the band-limit of the $X$ field in harmonic domain. The skewness of the Born-approximated convergence, $\kappa_{z}$, provides a measurement of the LSS-induced bispectrum since the convergence is directly proportional to the surface mass densities. Post-Born corrections however induce a characteristic change in the $\langle\kappa_{z}\kappa_{z}\kappa_{z}\rangle$ bispectrum due to the coupling of the second-order post-Born correction with two linear perturbation as first noted in  Ref.~\cite{pratten2016}. The post-Born bispectrum to the total bispectrum contributes to suppress squeezed configurations and enhance equilateral ones. The quantity $S_3^{\rm post-Born} = S[\kappa_{z}^{\rm ML}] - S[\kappa_{z}]$ provides a straightforward way to measure the skewness due to the post-Born correction alone. For post-Born corrections to the bispectrum become more important at higher redshifts as there are more independent lenses along the line of sight and the impact of multiple-lens couplings increases. In Figs.~\ref{fig:skewness-lowz}, \ref{fig:skewness-highz} we show a comparison between  $S_3^{\rm post-Born}$ and $S_3[\kappa_{z}]$ and the theoretical prediction of these quantities derived as a function of the map band limits $L_\textrm{max}$. We use the results of Ref.~\cite{pratten2016} to estimate $S_3^{\rm post-Born}$ analytically and different fitting formulae to evaluate the LSS bispectrum $b^{\rm LSS}_{L_1L_2L_3}$: we compare the simulation results with prediction based on tree-level perturbation theory~\cite{Bernardeau:2001qr} or fitting formulae of Refs.~\cite{GilMarin:2011ik} (GM) and \cite{Scoccimarro:2000ee} (SC) for the matter bispectrum. More details can be found in Appendix \ref{appendix:lss}.
\noindent
We found that GM fitting formulae provide a very good description of the skewness with better than $\simeq 10\%$ error at all redshift up to angular scales $L\approx 1000$. This scale corresponds to $k\simeq 1.8,1,0.7, 0.3, 0.4\ h/\textrm{Mpc}$ going from the lowest to the highest redshift bin we considered. This is consistent with the expected accuracy of the fitting formulae, which were calibrated for scales $k\lesssim 0.4 h/\textrm{Mpc}$ for $0\leq z\leq 1.5$, and perform quite well even outside these $z$ and $k$ ranges. SC results give similar results, but fit less well than GM at low redshifts. The tree-level perturbation theory only provides satisfactory results for high-redshift bins.

\noindent
The post-Born contribution $S_3^{\rm post-Born}$ becomes visible only at $z\geq 0.6$, and in this regime matches well the analytic predictions of Ref.~\cite{pratten2016}, though the statistical error remains large. All these tests show that the analytic models that we will use to model the relevant pieces of  \nlth are consistent with the simulations to reasonable accuracy.

\bibliographystyle{JHEP}
\bibliography{xcbib,texbase/antony,texbase/cosmomc}
\end{document}